\documentclass{article}

\usepackage[vmargin=1.18in,hmargin=1.1in]{geometry}
\usepackage{amsmath,amssymb,amsfonts,graphicx,amsthm,nicefrac,mathtools,bm}
\usepackage{hyperref}
\usepackage[numbers]{natbib}
\usepackage[english]{babel}
\usepackage{times}
\usepackage[T1]{fontenc}
\usepackage{bbm,mdframed}
\usepackage[dvipsnames]{xcolor}
\usepackage{xspace}
\usepackage{rotating,multirow}
\usepackage{enumerate,graphics,enumitem}
\usepackage[algo2e]{algorithm2e}
\usepackage{subcaption}

\setlist[itemize]{noitemsep, topsep=0pt}
\setlist[enumerate]{itemsep=5pt, topsep=5pt, leftmargin=25pt}

\newtheorem{theorem}{Theorem}

\definecolor{verylightblue}{rgb}{0.7,0.8,1}
  {\begin{mdframed}[backgroundcolor=verylightblue]\begin{theorem}}%
  {\end{theorem}\end{mdframed}}

\definecolor{verylightgray}{gray}{0.95}
  {\begin{mdframed}[backgroundcolor=verylightgray]\begin{proof}}%
  {\end{proof}\end{mdframed}}

\newtheorem{lemma}{Lemma}
\definecolor{verylightred}{rgb}{1,0.8,0.8}
  {\begin{mdframed}[backgroundcolor=verylightred]\begin{lemma}}%
  {\end{lemma}\end{mdframed}}

\newtheorem{proposition}{Proposition}
  {\begin{mdframed}[backgroundcolor=verylightblue]\begin{proposition}}%
  {\end{proposition}\end{mdframed}}

\theoremstyle{definition}
\newtheorem{definition}{Definition}
\theoremstyle{remark}

\makeatletter

\newtheorem*{rep@theorem}{\rep@title}
\newcommand{\newreptheorem}[2]
{\newenvironment{rep#1}[1]
{\def\rep@title{#2 \ref{##1}} \begin{rep@theorem}}%
 {\end{rep@theorem}}}
\makeatother
\newreptheorem{theorem}{Theorem}
\newreptheorem{lemma}{Lemma}
\newreptheorem{corollary}{Corollary}
\newreptheorem{proposition}{Proposition}

\newcommand{\figref}[1]{Figure~\ref{fig:#1}}

\newcommand{\secref}[1]{Section~\ref{sec:#1}}

\newcommand{\lemref}[1]{Lemma~\ref{lem:#1}}

\newcommand{\propref}[1]{Proposition~\ref{prop:#1}}

\newcommand{\thmref}[1]{Theorem~\ref{thm:#1}}

\newcommand{\eqnref}[1]{\eqref{eqn:#1}}

\newcommand{\PP}[1]{\textnormal{Pr}\!\left\{{#1}\right\}} 
\newcommand{\EE}[1]{\mathbb{E}\left[{#1}\right]} 

\newcommand{\EEst}[2]{\mathbb{E}\left[{#1}\ \middle| \ {#2}\right]} 
\newcommand{\PPst}[2]{\text{Pr}\!\left\{{#1}\ \middle| \ {#2}\right\}} 

\def\R{\mathbb{R}}

\def\LL{\mathcal{L}}

\newcommand{\ignore}[1]{}

\usepackage{tikz}
\usetikzlibrary{arrows}
\usetikzlibrary{shapes}

\newcommand{\thedate}{\today}
\newcommand{\theauthor}{Tijana Zrnic\textsuperscript{1}, ~ ~ Aaditya Ramdas\textsuperscript{2}, ~ ~ Michael I. Jordan\textsuperscript{1,3}\\
Departments of EECS\textsuperscript{1} and Statistics\textsuperscript{3}, University of California, Berkeley\\
Department of Statistics and Data Science\textsuperscript{2}, Carnegie Mellon University\\
{\small \texttt{\{tijana.zrnic,jordan\}@berkeley.edu, aramdas@cmu.edu} }
}
\newcommand{\thetitle}{Asynchronous Online Testing of Multiple Hypotheses}

\date{\thedate}
\author{\theauthor}
\title{\thetitle}

\usepackage[mmddyy]{datetime}

\newcommand{\nulls}{\mathcal{H}^0}

\newcommand{\fdp}{\textnormal{FDP}}
\newcommand{\fdr}{\textnormal{FDR}}

\newcommand{\mfdr}{\textnormal{mFDR}}

\newcommand{\saffasync}{\text{SAFFRON}_{\text{async}}}
\newcommand{\lordasync}{\text{LORD}_{\text{async}}}
\newcommand{\saffmarkov}{\text{SAFFRON}_{\text{dep}}}
\newcommand{\lordmarkov}{\text{LORD}_{\text{dep}}}
\newcommand{\lordmini}{\text{LORD}_{\text{mini}}}
\newcommand{\saffmini}{\text{SAFFRON}_{\text{mini}}}

\newcommand{\talpha}{\widetilde{\alpha}}

\newcommand{\One}[1]{{\bf{1}}\left\{{#1}\right\}}

\def\N{\mathbb N}
\def\F{\mathcal{F}}

\def\Ss{\mathcal{S}}

\def\X{\mathcal{X}}
\def\V{\mathcal{V}}
\def\C{\mathcal{C}}

\def\cR{\mathcal{R}}

\makeatletter
\newcommand{\dotfrac}[2]{
\mathchoice
{\ooalign{$\genfrac{}{}{0pt}{0}{#1}{#2}$\cr\leavevmode\cleaders\hb@xt@ .22em{\hss $\displaystyle\cdot$\hss}\hfill\kern\z@\cr}}
{\ooalign{$\genfrac{}{}{0pt}{1}{#1}{#2}$\cr\leavevmode\cleaders\hb@xt@ .22em{\hss $\textstyle\cdot$\hss}\hfill\kern\z@\cr}}
{\ooalign{$\genfrac{}{}{0pt}{2}{#1}{#2}$\cr\leavevmode\cleaders\hb@xt@ .22em{\hss $\scriptstyle\cdot$\hss}\hfill\kern\z@\cr}}
{\ooalign{$\genfrac{}{}{0pt}{3}{#1}{#2}$\cr\leavevmode\cleaders\hb@xt@ .22em{\hss $\scriptscriptstyle\cdot$\hss}\hfill\kern\z@\cr}}
}
\makeatother

\usepackage{algorithm,algorithmic}

\newcommand{\defn}{\ensuremath{:\, =}}

\makeatletter
\long\def\@makecaption#1#2{
        \vskip 0.8ex
        \setbox\@tempboxa\hbox{\small {\bf #1:} #2}
        \parindent 1.5em  
        \dimen0=\hsize
        \advance\dimen0 by -3em
        \ifdim \wd\@tempboxa >\dimen0
                \hbox to \hsize{
                        \parindent 0em
                        \hfil 
                        \parbox{\dimen0}{\def\baselinestretch{0.96}\small
                                {\bf #1.} #2
                                } 
                        \hfil}
        \else \hbox to \hsize{\hfil \box\@tempboxa \hfil}
        \fi
        }
\makeatother

\begin{document}

 \maketitle


\begin{abstract}
We consider the problem of \emph{asynchronous online testing}, aimed at
providing control of the false discovery rate (FDR) during a continual
stream of data collection and testing, where each test may be a sequential test
that can start and stop at arbitrary times.  This setting increasingly characterizes
real-world applications in science and industry, where teams of researchers
across large organizations may conduct tests of hypotheses in a decentralized
manner.  The overlap in time and space also tends to induce dependencies
among test statistics, a challenge for classical methodology, which either
assumes (overly optimistically) independence or (overly pessimistically) arbitrary
dependence between test statistics.  We present a general framework that
addresses both of these issues via a unified computational abstraction that
we refer to as ``conflict sets.''  We show how this framework yields algorithms
with formal FDR guarantees under a more intermediate, local notion of dependence.  
We illustrate our algorithms in simulations
by comparing to existing algorithms for online FDR control.
\end{abstract}

\section{Introduction}
\label{sec:intro}

As applications of machine learning expand in scope beyond the classical setting of a single decision-maker
and a single dataset, the decision-making side of the field has become increasingly important.
Unfortunately, research on the decision-making side of the field has lagged relative to the pattern-recognition side, often focusing only on the validity of single decisions. Arguably, however, deployed machine learning models witness large collections of decisions, typically occurring in an extended asynchronous stream. In such settings, it is essential to consider error rates over sets of decisions, and not merely over single decisions.

Although it is not a focus of research in machine learning, multiple decision-making
has been prominent during the past two decades in statistics, in the wake of
seminal research by \citet{BH95} on \emph{false discovery rate}
(FDR) control in multiple testing.  That literature has, however, principally
focused on batch data analysis and relatively small-scale problems.  Modern applications
in domains such as medicine, commerce, finance, and transportation are increasingly
of planetary scale, with statistical analysis and decision-making tools being used
to evaluate hundreds or thousands of related hypotheses in small windows of
time~(see, e.g.,~\cite{tang2010overlapping,xu2015infrastructure}).
These testing processes are often sequential, conducted in the context of a continuing stream
of data analysis.  The sequentiality is at two levels---each individual test is often a
sequential procedure, terminating at a random time when a stopping criterion is
satisfied, and also the overall set of tests is carried out sequentially, with
possible overlap in time.  In this setting---which we refer to as \emph{asynchronous online testing}---the goal is to control a criterion such as the FDR not
merely at the end of a batch of tests, but at any moment in time, and to do so
while recognizing that the decision for a given test must generally be made while
other tests are ongoing.

\begin{figure}[h]
\centering
\begin{subfigure}[b]{\textwidth}
   \includegraphics[width=\linewidth]{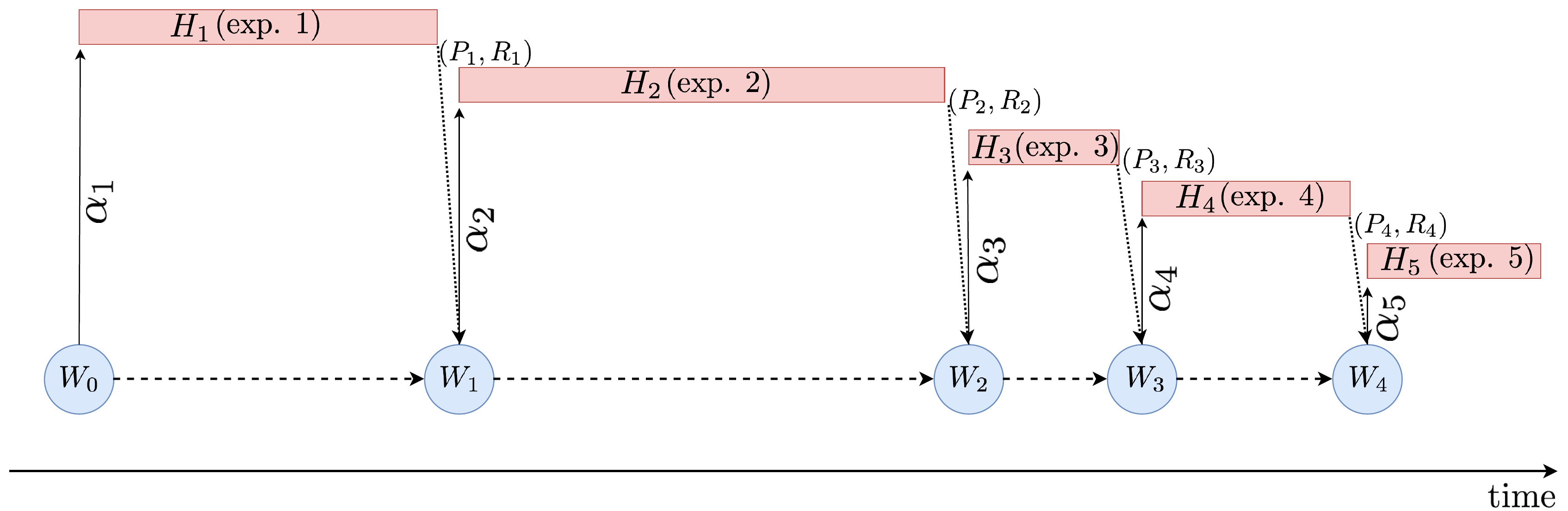}
\end{subfigure}

\begin{subfigure}[b]{\textwidth}
   \includegraphics[width=0.62\linewidth]{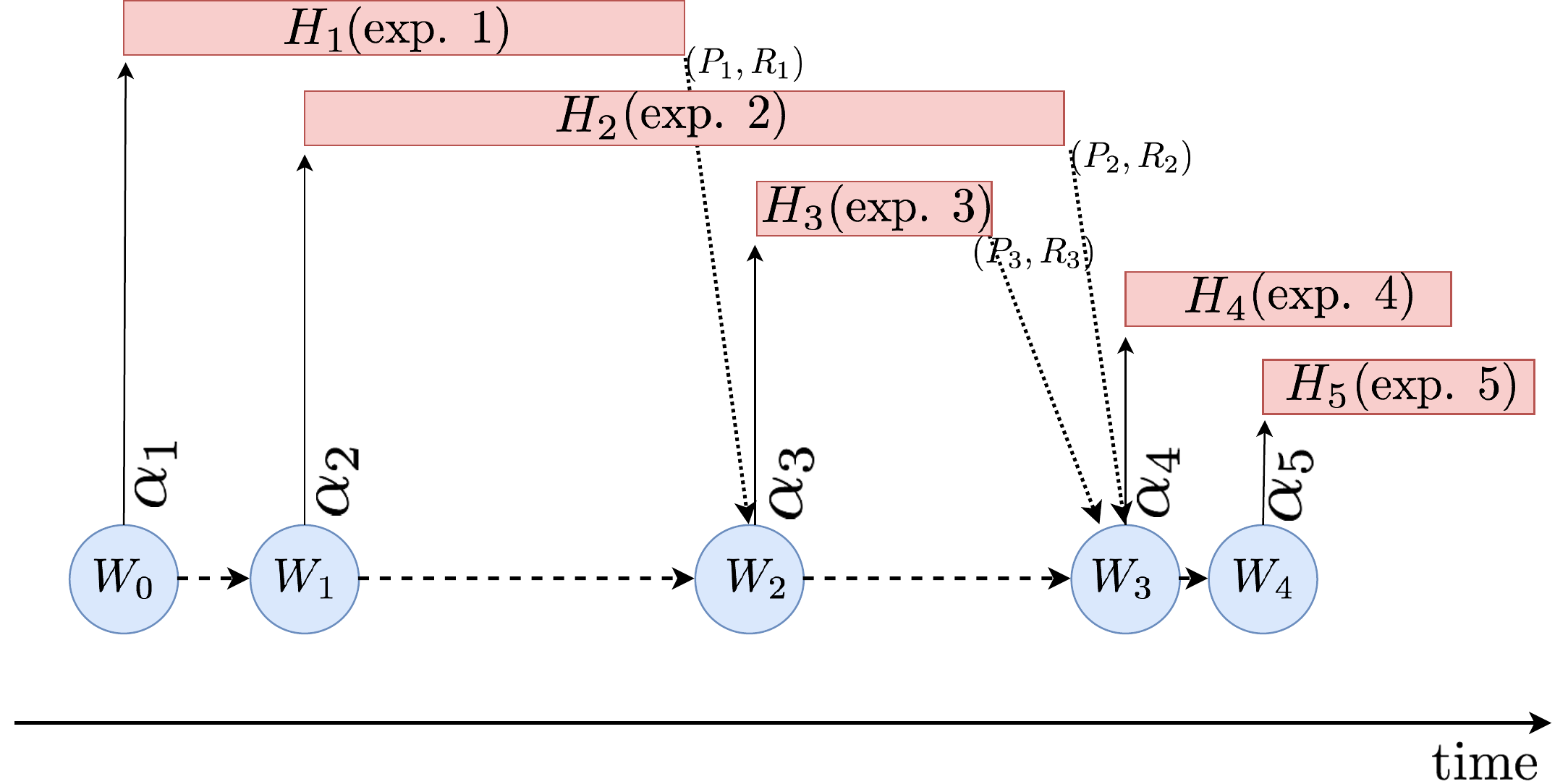}
\end{subfigure}

\caption{Testing five hypotheses synchronously (top) and asynchronously (bottom). In both cases, the test levels $\alpha_t$ depend on the outcomes of previously completed tests, which in the synchronous case includes all previously started tests. At the start time of experiment $t$, $W_{t-1}$ is used to denote the remaining ``wealth'' for making false discoveries. At the end of experiment $t$, a $p$-value $P_t$ and its corresponding decision $R_t \defn \One{P_t\leq\alpha_t}$ are known, which is used to adjust the available wealth at the start time of the next new test.}
\label{fig:async-comparison}
\end{figure}

A recent literature on ``online FDR control'' has responded to one aspect of this
problem, namely the problem of providing FDR control during a sequence of tests,
and not merely at the end, by adaptively setting the test levels for the
tests~\cite{foster2008alpha, javanmard2016online,RYWJ17, ramdas2018saffron}.
These methods are \emph{synchronous}, meaning that each test can only start when
the previous test has finished.  Our goal is to consider the more realistic
setting in which each test is itself a sequential process and where tests can
overlap in time.  This is done in real applications to gain time efficiency, and because
of the difficulties of coordination in a large-scale, decentralized setting.
To illustrate this point, \figref{async-comparison} compares the testing of five hypotheses
within an asychronous setting and a synchronous setting. In the asynchronous setting,
the test level $\alpha_t$ used to test hypothesis $H_t$ is allowed to depend only
on the outcomes of the previously \emph{completed} tests---for example, $\alpha_3$
can depend on the outcome of $H_1$, however not on the outcome of $H_2$.  In the
synchronous setting, on the other hand, the test level $\alpha_t$ can depend on
all previously started (hence also completed) tests. To account for the uncertainty
about the tests in progress, the test levels assigned by asynchronous online procedures
must be more conservative.  Thus, there is a tradeoff---although asynchronous procedures
take less time to perform a given number of tests they are necessarily less powerful
than their synchronous counterparts.  The management of this tradeoff involves
consideration of the overall power achieved per unit of real time, and consideration
of the complexity of the coordination required in the synchronous setting.

Another limitation of existing work on online multiple testing is that the dependence assumptions
on the tested sequence of test statistics, under which the formal false discovery rate guarantees
hold, are usually at one of two extremes---they are either assumed to be independent,
or arbitrarily dependent. From a practical perspective, independence seems overly optimistic as new tests may use previously collected data to formulate hypotheses, or to form a prior, or as evidence while testing. On the other hand, arbitrary dependence is likely too pessimistic, as older data and test
outcomes with time become ``stale,'' and no longer have direct influence on newly created
tests.  We see that a reconsideration of dependence is natural in the setting of online
FDR control, and is particularly natural in the asynchronous setting, given that tests
that are being conducted concurrently are often likely to be dependent, since they may use the same or highly correlated data during their overlap.

We therefore define and study a notion of \emph{local dependence}, and place it within
the context of asynchronous multiple testing. Working with $p$-values for simplicity, and letting $P_t$ denote the $t$-th tested $p$-value, we say that a sequence of $p$-values $\{P_t\}$ satisfies local dependence if the following condition holds:
\begin{equation}
\label{eqn:localdep}
     \text{ for all $t>0$, there exists $L_t \in \N$ such that } P_t \perp P_{t-L_t-1},P_{t-L_t-2},\dots,P_1,
\end{equation}
where $\{L_t\}$ is a fixed
sequence of parameters which we will refer to as ``lags.''  Clearly, when $L_t=0$ for all $t$, we
obtain the independent setting, and when $L_t=t$, we recover the arbitrarily dependent
setting. If $L_t\equiv L$ for all $t$, condition \eqnref{localdep} captures a lagged
dependence of order $L$.

\begin{figure}[h]
\centerline{\includegraphics[width=0.7\textwidth]{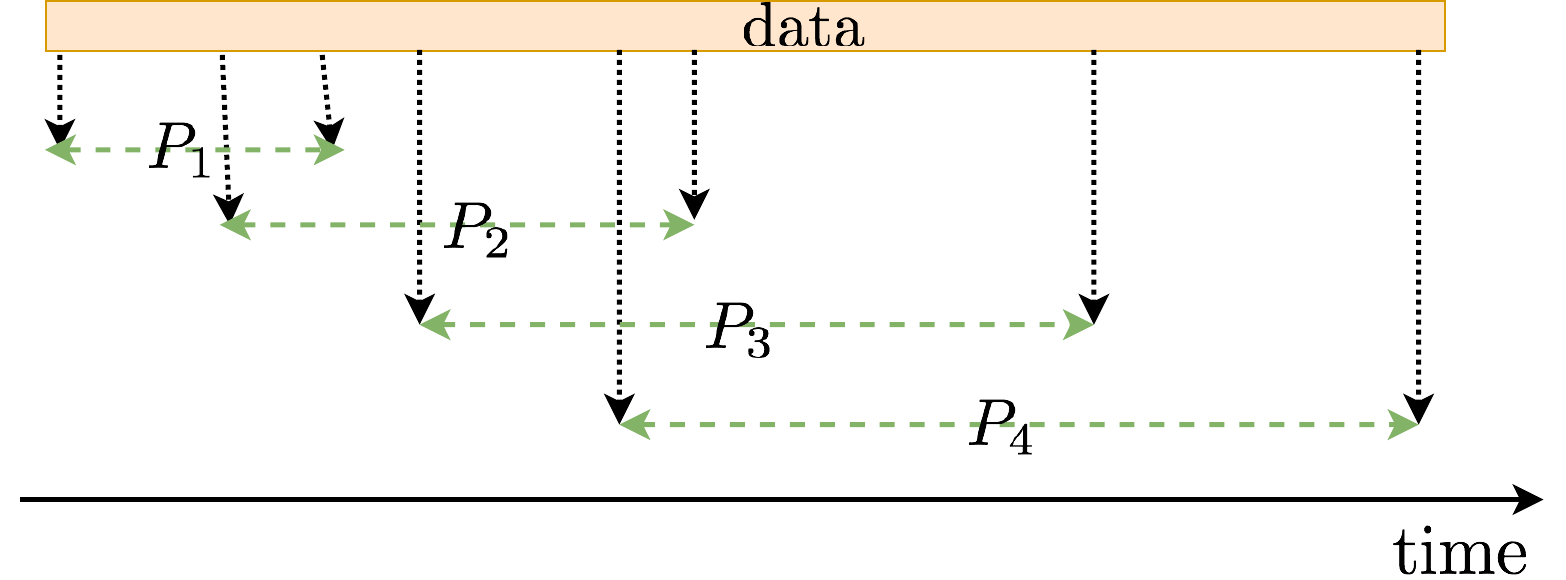}}
\caption{Example of $p$-values within a short interval computed on overlapping data. They exhibit local dependence; for example, $P_3$ and $P_4$ are independent of $P_1$.} 
\label{fig:markov-diag}
\end{figure}

To further emphasize the natural connection between asynchrony and local dependence,
consider the simple setting in \figref{markov-diag}.  This diagram captures the setting
in which a research team is collecting data over time, and decides to run multiple tests
in a relatively short time interval. For example, such a situation might arise when testing multiple treatments against a common control \citep{robertson2018online}, or in large-scale A/B testing by internet companies \citep{xu2015infrastructure}. Since there is overlap in the data these tests use to compute
their test statistics, the corresponding $p$-values could be arbitrarily dependent.  In general several tests might share data with the first test. 
Thus the $p$-values are locally dependent, with the lag parameter being equal to
the number of consecutive tests that share data.

In this work, we reinforce this connection between asynchronous online testing and
dependence by developing a general abstract framework in which, from an algorithmic
point of view, these two issues are treated with a single formal structure.  We do so
by associating with each test a \emph{conflict set}, which consists of other tests
that have a potentially adversarial relationship with the test in question. Within
this framework, we develop algorithms with provable guarantees on the rate of false discoveries.  The core
idea is to enforce a notion of pessimism with regard to the conflict set---when computing
a new test level, the algorithm ``hallucinates'' the worst-case outcomes of the conflicting
tests.

We derive procedures that handle conflict sets as strict generalizations of current
state-of-the-art online FDR procedures; indeed, when there are no conflicts, for
example when there is no asynchrony and when the $p$-values are independent,
our solutions recover LORD \citep{javanmard2016online}, LOND \citep{javanmard2015online},
and SAFFRON \citep{ramdas2018saffron}, the latter of which recovers alpha-investing
\citep{foster2008alpha} as a special case for a particular choice of parameters.  On the other hand,
if the conflict sets are as large as possible---for example, if the parameter $L_t$
or the number of tests run in parallel tend to infinity---our algorithms behave like
alpha-spending,\footnote{Alpha-spending is a generalization of the Bonferroni correction
in which the assigned test levels do not have to be equal. In other words, the Bonferroni
correction suggests testing $n$ hypotheses under level $\alpha/n$, while
alpha-spending merely requires $\sum_{i=1}^n \alpha_i \leq \alpha$, where $\alpha_i$
is the test level for the $i$-th hypothesis.} which was designed to control a more
stringent criterion called the family-wise error rate (FWER), under any dependence
structure. Independently, we also prove that the original LOND procedure controls the FDR even 
under positive dependence (PRDS), the first online procedure to provably have this guarantee
under the PRDS condition that is popular in the offline FDR literature \citep{BY01,ramdas2017unified}.

\paragraph{Organization.} The rest of this paper is organized as follows.  After a presentation of the general problem formulation and related work, \secref{conflict} presents the key notion of conflict
sets.  We present two general procedures based on conflict sets, deferring their formal FDR guarantees to \secref{unify}.  In \secref{async}, we couch asynchronous testing in terms of conflict sets. In a
similar fashion, in \secref{markov}, we describe synchronous testing of locally
dependent $p$-values using conflict sets, and present procedures having FDR guarantees
within this environment. \secref{minibatch} then combines the ideas of local dependence
and asynchronous testing into an overall framework designed for testing asynchronous
batches of dependent $p$-values. \secref{unify} provides additional, stronger guarantees of the presented algorithms, which hold under more stringent assumptions on the $p$-value sequence. 
In \secref{sims} we present simulations designed to explore our methods, comparing them to existing procedures
that handle dependent $p$-values. Finally, we conclude the paper with a short discussion
in \secref{summary}.
All proofs are deferred to the Appendix.

\subsection{Technical preliminaries}

We briefly overview the technical background upon which our work builds.  Recall that the \emph{false discovery rate} (FDR) \citep{BH95} is defined as follows:
 \begin{align*}
 \fdr \equiv \EE{\fdp} = \EE{\frac{|\nulls \cap \cR|}{|\cR|\vee 1}},
 \end{align*}
 where $\nulls$ is the unknown set of true null hypotheses and $\cR$ is the set of hypotheses rejected by some procedure.
Formally we have $\nulls = \{i: H_i \text{ is true}\},~\cR = \{i: H_i \text{ is rejected}\}.$
The random ratio appearing inside the expectation is called the \emph{false discovery proportion} (FDP). It is also of theoretical and practical interest to study a related metric called the \emph{modified false discovery rate}
(mFDR):
\begin{align*}
\mfdr \equiv \frac{\EE{|\nulls \cap \cR|}}{\EE{|\cR|\vee 1}}.
\end{align*}
\citet{foster2008alpha} show that in the long run the mFDR behaves similarly to the FDR in an online environment. Similarly, \citet{genovese2002operating} prove that the $\mfdr$ and $\fdr$ achieved by the celebrated Benjamini-Hochberg procedure \citep{BH95} become equivalent as the number of hypotheses tends to infinity. In this work, we mainly focus on the control of $\mfdr$, as we can provide simple proofs under less restrictive assumptions. Importantly, in the Appendix we provide a side-by-side comparison of the $\mfdr$ and $\fdr$ for all of the experiments in this paper; as we show there, the plots for $\mfdr$ and $\fdr$ are visually indistinguishable when the number of non-nulls is non-negligible, and mFDR dominates the FDR when non-nulls are sparse. Thus, our experiments suggest that mFDR control suffices for FDR control as well.

In addition, we point out one advantage of mFDR over FDR which is especially relevant in the online context. Suppose that different sequences of hypotheses are tested with different algorithms controlling the mFDR. Then, one can retroactively group the set of discoveries resulting from these different algorithms, all the while knowing that the mFDR is still controlled. If the original sets of discoveries come with FDR guarantees only, one cannot argue FDR control over the overall batch of discoveries. This decentralized testing of different sequences using different algorithms is particularly aligned with the online FDR setup, where not all hypotheses are known in advance. In fact, this "online" property of the mFDR was recognized even within offline FDR control \citep{van2008controlling}.

To simplify our presentation, we will often suppress the distinction, referring to both of these metrics as ``FDR.''

In online FDR control, the set of rejections possibly changes at each time step, implying changes in $\mfdr$ and $\fdr$. Therefore, in online settings, we have to consider $\cR(t)$, which is the set of rejections up to time $t$, and the naturally implied $\mfdr(t)$ and $\fdr(t)$. We will also use the symbol $\V(t)\defn \cR(t)\cap\nulls$ to denote the set of false rejections made up to time $t$. The main objective of online FDR algorithms is to ensure $\mfdr(t)\leq\alpha$ or $\fdr(t)\leq\alpha$, for a chosen level $\alpha$ and \emph{for all times $t$}.

Many of the online FDR algorithms that have been proposed to date in the literature are special cases of the generalized alpha-investing (GAI) framework \citep{aharoni2014generalized}.  The initial interpretation of these algorithms, as put forward by \citet{foster2008alpha}, relied on a notion of dynamically changing ``alpha-wealth.'' \citet{RYWJ17,ramdas2018saffron} subsequently presented an alternative perspective on GAI algorithms. In this view, GAI algorithms are viewed as keeping track of an empirical estimate of the true false discovery proportion, denoted $\widehat{\fdp}(t)$, and they assign test levels $\alpha_t$ in a way that ensures $\widehat{\fdp}(t)\leq\alpha$ for all time steps $t$, where $\alpha$ is the pre-specified FDR level. In the earlier paper \citep{RYWJ17}, they show that such control of FDP estimates also yields FDR control.  This perspective---which is equivalent to the earlier, wealth characterization of GAI algorithms---will provide the mathematical framework upon which we build in this paper. 

Finally, we recap the typical assumptions made for null $p$-values in the FDR literature. If a hypothesis $H_i$ is truly null, then the corresponding $p$-value $P_i$ is stochastically larger than the uniform distribution (``super-uniformly distributed,'' or ``super-uniform'' for short),
meaning that:
\begin{equation*}
\text{If the null hypothesis $H_i$ is true, then } \PP{P_i \leq u}
\leq u \text{ for all } u\in[0,1].
\end{equation*}
This condition is sometimes generalized to the online FDR setting by incorporating a filtration $\F^{i-1}$, resulting in the following assumption:
\begin{equation}
\label{eqn:superunifclassical}
\text{If the null hypothesis $H_i$ is true, then } \PPst{P_i \leq u}{\F^{i-1}}
\leq u \text{ for all } u\in[0,1],
\end{equation}
Here, $\F^i$ captures all relevant information about the first $i$ tests. As we discuss in later sections, however, this condition can be overly stringent when there are interactions between $p$-values, and we will accordingly introduce weaker super-uniformity assumptions.


\subsection{Problem formulation and contribution}

We now give a formal introduction to the problem setting, at the same time introducing the necessary notation for the sections to follow.

At time step $t\in\N$, the test of hypothesis $H_t$ begins, and the $p$-value resulting from this test is denoted $P_t$. In contradistinction to the standard online FDR paradigm, $P_t$ is not required to be known at time $t$; indeed, this test is not fully \emph{performed} at time $t$, but is only initiated at time $t$.  The \emph{decision time} for $H_t$ is denoted $E_t$; this is the time of possible rejection. Fully synchronous testing is thus an instance of this setting in which $E_t=t$, as assumed in classical online FDR work. In general, however, $E_t\neq t$. Note also that, unlike in the classical online FDR problem, the set of rejections $\cR(t)$ and false rejections $\V(t)$ at time $t$ now consider not all $\{P_i:i\leq t\}$, but only $\{P_i:E_i\leq t\}$:
$$\cR(t) = \{i\in[t]: E_i\leq t, H_i \text{ is rejected}\},~ \V(t) = \cR(t)\cap \nulls.$$
In addition, to capture the desideratum of statistical validity in the face of data reuse, we allow the possibility of the $p$-values not being completely independent; in particular, we allow \emph{local dependence}. Here, we envision $P_t$ having arbitrary, possibly adversarial dependence on $P_{t-1},\dots,P_{t-L_t}$, while the dependence between $P_t$ and $P_j$, $j<t-L_t$ is limited. For simplicity the reader can assume $P_t \perp P_{t-L_t-1},P_{t-L_t-2},\dots,P_1$, however in later sections we will discuss some restricted forms of dependence between $P_t$ and $P_j$, for $j<t-L_t$, handled by our results.

We treat $E_t$ as fixed but unknown before time $E_t$ itself. While the $p$-value and the duration of a test could indeed be dependent random quantities---for example, when the duration is a reasonably good proxy for sample size---here $E_t$ is \emph{not} the absolute duration on a meaningful time scale, but it merely captures how many tests have started before the decision for the $t$-th test. Thus, treating $E_t$ as fixed roughly corresponds to asserting independence between $P_t$ and the number of newly created tests before test $t$ finishes. As we envision a highly decentralized setting with little between-test coordination, we deem this assumption reasonable. We do, however, acknowledge the possibility of a more coordinated setting with $P_t$ and $E_t$ randomly coupled, and this is an important avenue for future work.

Under the setup described above, the goal is to  produce test levels $\alpha_t$ dynamically at the beginning of the $t$-th test, such that, \emph{despite arbitrary local dependence and regardless of the decision times $E_t$}, the false discovery rate is controlled at any given moment under a pre-specified level $\alpha$. In this work, we provide procedures which achieve this goal, both at all fixed times $t\in\N$, as well as adaptively chosen stopping times.

It is important to remark that, even under independence of $p$-values, we cannot simply ignore the asynchronous aspects of the problem and naively apply an existing online FDR algorithm. We discuss two such plausible but naive applications of online methodology, and discuss why they are invalid.

 One natural adjustment could be to apply an online FDR algorithm whenever each test finishes (that is, whichever test is the $t$-th one to finish, test it at level $\alpha_t$). This scheme would only assign $\alpha_t$ to a test at the end of that test, which is unrealistic because sequential hypothesis tests (parametric tests such as Wald's sequential probability ratio test (SPRT), and nonparametric tests as well \citep{balsubramani_sequential_2016}) typically require specification of the target type I error level in advance because it is an important component of their stopping rule. The same holds for more recent, multi-armed bandit approaches to testing \citep{yang2017multi, jamieson2018bandit}. Thus, we need to specify $\alpha_t$ at the \emph{start} of test $t$. That said, there do exist tests which only require the test level at decision time. If \emph{all} tests in the sequence are of the latter kind, existing online methodology is indeed sufficient; however, we find this assumption too strong, especially given the popularity of bandit approaches in modern testing applications.

\begin{figure}[t]
\centerline{\includegraphics[width=0.35\textwidth]{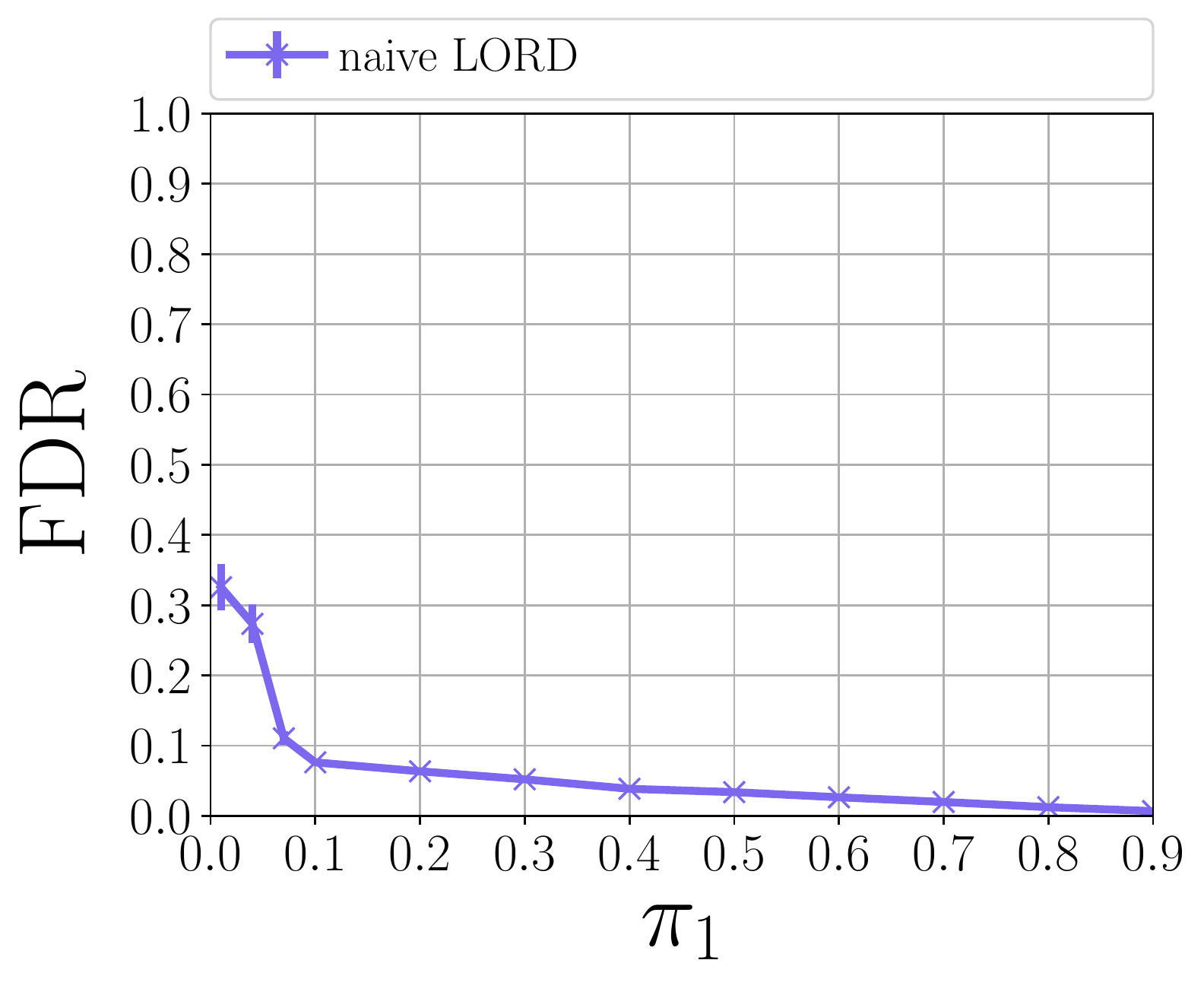}
\includegraphics[width=0.35\textwidth]{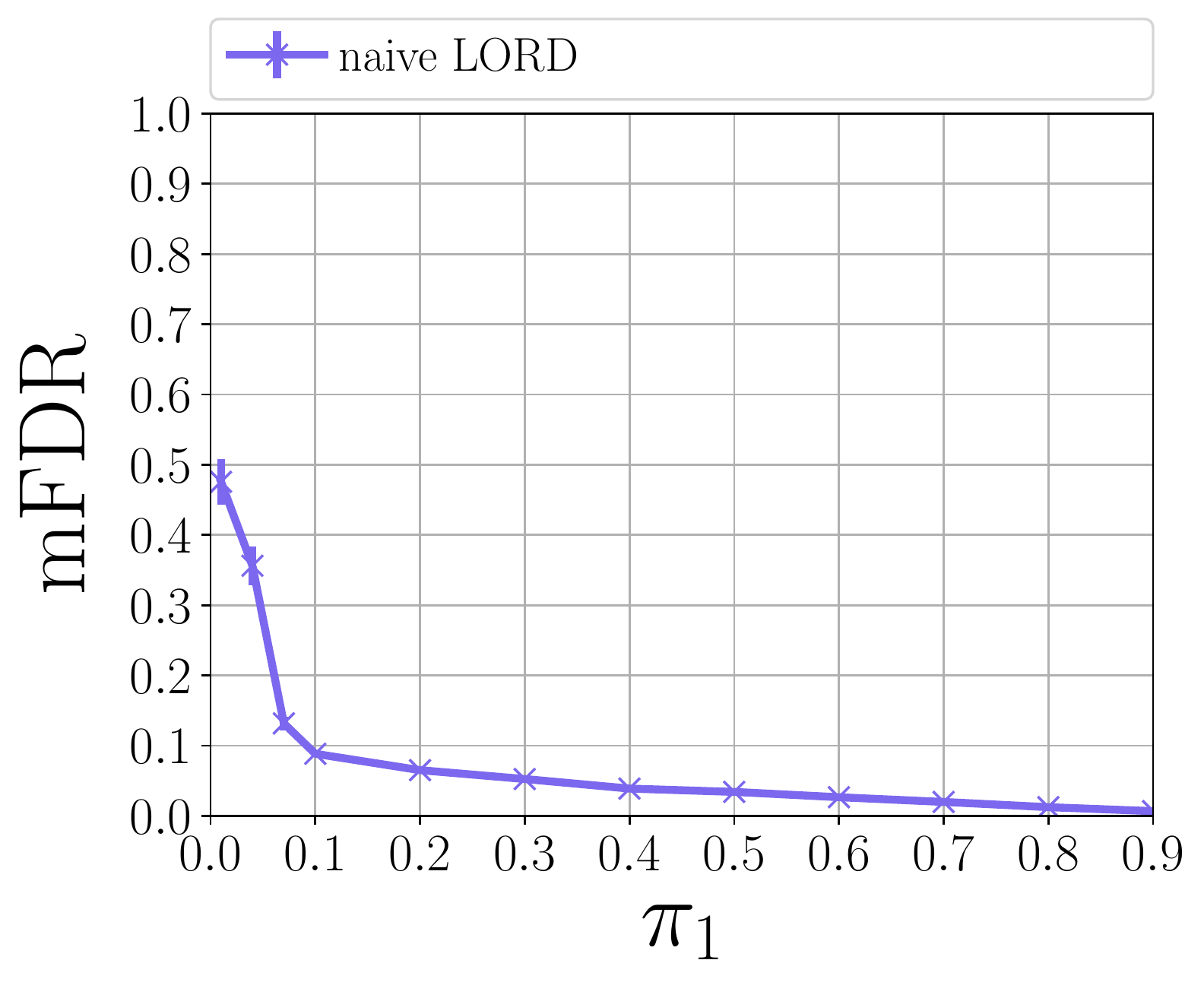}}
\caption{FDR and mFDR achieved by a naive application of the LORD algorithm with target FDR level $\alpha=0.05$ \citep{javanmard2016online} in an asynchronous environment. We adopt the experimental setting from \secref{sims}; we set the asynchrony parameter to $p=1/150$, and the mean of observations under the alternative to $\mu_c=3$. Here, $\pi_1$ is the proportion of tested hypotheses which are non-null. Both FDR and mFDR are controlled only for $\pi_1\geq 0.4$. \vspace{-0.5cm}} 
\label{fig:naive}
\end{figure}

Alternatively, one could imagine computing $\alpha_t$ at the start of test $t$ by applying an online FDR algorithm to the completed tests only, and ignoring those that have not finished. From the theory perspective, this clearly comes with no formal guarantee; for example, one could imagine starting $N$ tests, and all $N$ tests finishing at once, after the $N$-th test has started. Given that there are no completed tests at the time of test level assignment, all tests would receive the same level, which would violate the FDR requirement under any natural setting (we ignore trivial special cases such as all hypotheses being non-null). Somewhat less trivially, \figref{naive} plots the FDR and mFDR achieved by this naive heuristic in a simulation setting from Section \ref{sec:sims}. When the proportion of non-null $p$-values is relatively small---as one would generally expect in practice---this heuristic severely violates the FDR requirement.

\subsection{Related work}

There is a vast literature on sequential testing~\cite[see, e.g.,][]{wald1945sequential,
chernoff1959sequential, albert1961sequential, naghshvar2013active}.
We do not aim to contribute to that literature per se; rather, our goal is to consider
multiple testing through a more realistic lens as an outer sequential process, one that
acknowledges the existence of inner sequential processes that are based on sequential testing.

Likewise, there is a large and growing literature on false discoveries in multiple testing, aimed at solving a range of problems, often addressing issues of scientific reproducibility in research \citep{ioannidis2005most}.
  Here we focus on work whose methods or objectives have the most overlap with ours.
 In particular, we focus on literature on ``online'' methods in multiple testing, and compare and contrast those solutions to the ones we propose.
 
 The most salient difference is that we address the general problem of asynchrony; when there is no asynchrony, meaning $E_t  = t$, our approach recovers a slew of existing methods, including work  by \citet{foster2008alpha}, \citet{aharoni2014generalized}, \citet{javanmard2015online,javanmard2016online}, \citet{RYWJ17,ramdas2018saffron}.

Most previous work also differs from ours in that it assumes that condition \eqnref{superunifclassical} holds.  This condition is too strong for the notion of local dependence this paper considers; indeed, in \secref{markov} we present a simple toy example in which this assumption fails. An exception is the work of \citet{javanmard2015online,javanmard2016online}, who discuss sufficient conditions for achieving FDR control under arbitrary dependence within the $p$-value sequence. However, these conditions essentially imply an alpha-spending-like correction for the test levels, making their proposed procedure overly conservative. We elaborate on this argument and empirically demonstrate this observation in \secref{sims}.

\citet{robertson2018online} have investigated the performance of several online FDR algorithms empirically, including all of those listed above, when the $p$-value sequence is positively dependent. They do not, however, provide any formal guarantees for those procedures that have thus far been shown to work only under independence. We make partial progress to justifying their empirical observations by proving that LOND provably controls FDR under positive dependence.

Recently, there has also been some work specifically motivated by controlling false discoveries in A/B testing in the tech industry \citep{yang2017multi}. However, their setup was again fully synchronous, and assume that the observations are independent across all experiments, which are the two assumptions this paper deems too strong and circumvents.

The vast literature on adaptive data analysis \citep{dwork2015preserving, dwork2015reusable, bassily2016algorithmic, blum2015ladder} focuses on an online setting where a distribution is adaptively queried for a chosen functional, and at each step these queries are answered by making use of a single data set coming from that distribution. This line of work also has the goal of preventing false discovery, however by proving generalization bounds, rather than controlling the FDR in online multiple testing.

Ordered hypothesis testing considers tests for which additional prior information is available, and allows sorting null hypotheses from least to most promising \citep{li2017accumulation, lei2016power, lynch2017control, g2013sequential}. In these papers, however, the word ``sequential'' or ``ordered'' does not refer to online testing; these methods are set in an offline environment, requiring access to all $p$-values at once.  In our approach, we allow testing a possibly infinite number of hypotheses with no available knowledge of the future $p$-values.

\section{Conflict sets: the unifying approach}
\label{sec:conflict}

In this section we describe a general, abstract formulation of multiple testing under asynchrony and dependence, which unifies the seemingly disparate solutions of this paper and provides the point of departure for deriving specific algorithms. We describe two such procedures, which we will refer to as LORD* and SAFFRON*, that control $\mfdr$ within this framework.

LORD* and SAFFRON* build off the LORD \citep{javanmard2016online} and SAFFRON \citep{ramdas2018saffron} algorithms. Like SAFFRON, SAFFRON* allows the user to choose a parameter $\lambda_t \geq \alpha_t$, which is the ``candidacy threshold'' at time $t$, meaning that, if $P_t\leq\lambda_t$, then $P_t$ is referred to as a \emph{candidate} for rejection. We will discuss this extension introduced in the SAFFRON procedure further below; for now, we simply note that it is an analog of the notion of ``null-proportion adaptivity'' in the offline multiple testing literature. Indeed, \citet{ramdas2018saffron} argue that LORD can be seen as the online analog of the BH procedure~\citep{BH95}, while SAFFRON can be seen as the online analog of the adaptive Storey-BH procedure~\citep{Storey02, Storey04}.

Throughout we let $R_t\defn \One{P_t\leq\alpha_t}$ denote the \emph{indicator for rejection}, and $C_t\defn \One{P_t\leq\lambda_t}$ denote the \emph{indicator for candidacy}.

We now define several filtrations, which capture the increasing information available to the experimenter as well as the FDR algorithm.


By $\LL^t$, we denote a filtration that captures all relevant information about the tests that started up to, and including, time $t$, for the LORD* procedure. Formally, $\LL^t\defn \sigma(\{R_1,\dots,R_t,\})$. For SAFFRON*, we also incorporate candidates in the filtration: $\Ss^t\defn \sigma(\{R_1,C_1,\dots,R_t,C_t\})$. Many of our arguments will apply to both algorithms; we accordingly use $\F^t$ to indicate a generic filtration that can be either $\LL^t$ or $\Ss^t$.

With each test and its corresponding hypothesis, we associate a \emph{conflict set}. For the test starting at step $t$, we denote this set $\X^t$; it consists of a (not necessarily strict) subset of $\{1,\dots,t-1\}$. For example, $\X^5$ could be $\{3,4\}$. The reason why we refer to this set as \emph{conflicting} for test $t$ is because it contains the indices of tests that interact with the $t$-th test in some unknown way. This could mean that, at time $t$, there is missing information about these tests, or that there potentially exists some arbitrary dependence between those tests and the upcoming one. More explicitly, we let
$$\X^t = \{i \in [t-1]~:~E_i\geq t\}\cup \{t-L_t,\dots,t-1\},$$
where $L_t$ is the sequence of dependence lags. In words, $\X^t$ consists of all tests that have not finished running or are locally dependent with test $t$.

We require the conflict sets to be \emph{monotone}: each index $t$ has to be in a continuous ``block'' of conflict sets. More formally, if there exists $j$ such that $t\in\X^j$, then $t\in\X^i$, for all $i\in\{t+1,\dots,j\}$. Without any constraint on the sequence $\{L_t\}$, the conflict sets need not be monotone. Therefore, we translate the condition of monotonicity of conflict sets into a constraint on the sequence $\{L_t\}$ as: $L_{t+1}\leq L_t + 1$. Informally, this is just a requirement that the ``non-conflicting information'' does not decrease with time. This will ensure that the test level $\alpha_t$ and candidacy threshold $\lambda_t$ have at least as much knowledge about prior tests as $\alpha_{t-1}$ and $\lambda_{t-1}$. Moreover, this requirement is indeed a natural one, and usual testing practices satisfy it; for example, this condition holds if dependent $p$-values come in disjoint blocks.

 We define the \emph{last-conflict time} of test $t$ as $\tau_t \defn \max\{j:t\in\X^j\}$. If test $t$ never appears in a conflict set, we take $\tau_t = t$.
 

Consider again the filtration $\F^t$. A subtlety we initially ignored is that the superscript $t$ does not correspond to the physical quantity of time. In particular, different tests may run for different lengths of time and the decision time for each test may even be random; therefore, $R_t$ might be known before $R_{t-1}$. This motivates us to define a filtration as a counterpart of $\F^t$ whose increase at each step corresponds to the real increase in knowledge with time. We introduce $\F^{-\X^t}$ as the \emph{non-conflicting filtration}; the sigma-algebra $\F^{-\X^t}$ contains information about the tests that started before time $t$ which are \emph{not} in the conflict set of test $t$. In particular, $\LL^{-\X^t}\defn\sigma(\{R_i: i\leq t-1, i\not\in \X^t\})$ for LORD*, $\Ss^{-\X^t}\defn\sigma(\{R_i,C_i: i\leq t-1, i\not\in \X^t\})$ for SAFFRON*, and again we use $\F^{-\X^t}$ to generically denote either $\LL^{-\X^t}$ or $\Ss^{-\X^t}$. We have that $\F^{-\X^t}\subseteq \F^{t-1}$. Notice that we promised to make this set a \emph{filtration}; if $\X^t$ was an arbitrary set of indices, this would not in general be satisfied. However, it is straightforward to verify that the monotonicity property of conflict sets ensures that $\F^{-\X^t}$ indeed forms a filtration.

We will design $\alpha_t$ and $\lambda_t$ to be $\F^{-\X^t}$-measurable. This is essentially the idea of pessimism mentioned earlier---among all tests that finished before the $t$-th one starts, $\alpha_t$ and $\lambda_t$ have to ignore the ones conflicting with test $t$ in order to guard against unknown interactions that the conflicting tests have with the upcoming one.

Finally, we will generally require the following super-uniformity condition for null $p$-values:
\begin{align}
\label{eqn:superuniformity-conflict}
    \text{If the null hypothesis $H_t$ is true, then } \PPst{P_t \leq
  u}{\F^{-\X^{E_t}}} \leq u, \text{ for all } u\in[0,1].
\end{align}
This is a condition that requires validity of null $p$-values: given the knowledge one has before making a decision, if a hypothesis is truly null, it has to be well-behaved. However, unlike in classical online FDR work, we do not have $\F^{-\X^{E_t}}=\F^{t-1}$. As we discuss further in later sections, assumption \eqnref{superuniformity-conflict} will allow arbitrary local dependence, as well some limited, but nevertheless important, forms of dependence between distant $p$-values. Note that, if the distant $p$-values are independent---a setting we study in \secref{markov} and \secref{minibatch}---this condition is automatically satisfied.

\subsection{The LORD* algorithm}

Following a recently proposed framework \citep{RYWJ17}, we define LORD* and SAFFRON* as arbitrary update rules which control a certain estimate of the false discovery proportion under a pre-specified level $\alpha$; the two algorithms differ in their choice of estimate. In Subsection \ref{subs:oracle}, we introduce additional analysis tools which will justify the choice of these estimates.

LORD* is defined as any update rule for $\alpha_t$ that ensures that the estimate
$$\widehat \fdp_{\text{LORD*}}(t)\defn \frac{\sum_{j\leq t}\alpha_j}{(\sum_{j\leq t, j\not\in \X^t}R_j)\vee 1}.$$
is at most $\alpha$ for all $t\in\N$. 



Below we state two different versions of LORD*, using two different test level updates. Algorithm 1 generalizes the LORD++ procedure \citep{javanmard2016online,RYWJ17}, while Algorithm 2 generalizes its predecessor, the LOND procedure \citep{javanmard2015online}. These are not the only ways of assigning $\alpha_j$ that are consistent with the assumptions and satisfy the definition of LORD* in their control of $\widehat \fdp_{\text{LORD*}}$, but they are our focus in the remainder of the paper. Other rules can be developed as extensions of the rules in the LORD paper \citep{javanmard2016online}.

To state the algorithms in this paper, we will make use of the variable $r_k$, which refers to the first time that $k$ rejections are non-conflicting, meaning that there exist $k$ rejected hypotheses which are no longer in the conflict set at that time. That is, we define $r_k$ as:\footnote{Here, as well as in the rest of this paper, we define the minimum of an empty set to be $-\infty$.}
\begin{equation}
\label{eqn:rk}
    r_k\defn\min\{ i\in [t] : \sum_{j=1}^i R_j\One{\tau_j\leq i}\geq k\}.
\end{equation}

\label{algs1to4}
\begin{algorithm}[H]
\footnotesize
\SetAlgoLined
\SetKwInOut{Input}{input}
\Input{FDR level $\alpha$, non-negative non-increasing sequence $\{\gamma_j\}_{j=1}^\infty$ such that $\sum_j \gamma_j=1$, initial wealth $W_0\leq\alpha$}
Set $\alpha_1 = \gamma_1 W_0$\newline
 \For{$t=1,2,\dots$}{
 start $t$-th test with level $\alpha_t$\newline
  $\alpha_{t+1} = \gamma_{t+1} W_0 + \gamma_{t+1-r_1} (\alpha - W_0) +  \left(\sum_{j\geq 2}\gamma_{t+1-r_j}\right)\alpha$
 }
 \caption{The LORD++ algorithm under general conflict sets (a special case of LORD*)}
\end{algorithm}

\begin{algorithm}[H]
\SetAlgoLined
\footnotesize
\SetKwInOut{Input}{input}
\Input{FDR level $\alpha$, non-negative non-increasing sequence $\{\gamma_j\}_{j=1}^\infty$ such that $\sum_j \gamma_j=1$}
Set  $\alpha_1 = \gamma_1 \alpha$\newline
 \For{$t=1,2,\dots$}{
 start $t$-th test with level $\alpha_t$\newline
  $\alpha_{t+1} = \alpha\gamma_{t+1}\left((\sum_{j=1}^t \One{P_j\leq\alpha_j, \tau_j \leq t}) \vee 1\right)$
 }
 \caption{The LOND algorithm under general conflict sets (a special case of LORD*)}
\end{algorithm}


It is a simple algebraic exercise to verify that the two update rules given for $\alpha_t$ indeed guarantee that $\widehat \fdp_{\text{LORD*}}(t)\leq\alpha$ for all $t\in\N$.

\subsection{The SAFFRON* algorithm}

In response to LORD and LOND's FDP estimate, the SAFFRON method was derived after observing that the former might be overly conservative estimates of the FDP. Indeed, if the tested sequence contains a significant fraction of non-nulls, and if the non-nulls yield strong signals for rejection, the realized FDP and the estimated FDP might be very far apart. Motivated by this observation, SAFFRON was developed as the adaptive counterpart of LORD which keeps track of an empirical estimate of the null proportion, similar to the way in which Storey et al.\ \citep{Storey02, Storey04} improved upon the BH procedure \citep{BH95}. We thus propose the SAFFRON* algorithm to maintain control over the following estimate:
$$\widehat \fdp_{\text{SAFFRON*}}(t)\defn \frac{\sum_{j< t, j\not\in \X^t}\frac{\alpha_j}{1-\lambda_j}\One{P_j>\lambda_j} + \sum_{j\in\{\X^t\cup \{t\}\}}\frac{\alpha_j}{1-\lambda_j}}{(\sum_{j\leq t,j\not\in \X^t}R_j) \vee 1}.$$
Any update rule for $\alpha_t$ and $\lambda_t$ ensuring $\widehat \fdp_{\text{SAFFRON*}}(t)\leq\alpha$ for all $t\in\N$ satisfies the definition of SAFFRON*. Algorithm 3 and Algorithm 4 describe two particular instances of SAFFRON*, obtained for specific choices of the sequence $\{\lambda_j\}$. We present an algorithmic specification of SAFFRON* for the constant sequence $\{\lambda_j\}\equiv \lambda$ in Algorithm 3. A different case of SAFFRON* is presented in Algorithm 4, where we use the alpha-investing strategy $\lambda_j=\alpha_j$ \citep{foster2008alpha, ramdas2018saffron}.


For the updates below, recall the definition of $r_k$ from equation \eqnref{rk}.


\begin{algorithm}[H]
\SetAlgoLined
\footnotesize
\SetKwInOut{Input}{input}
\Input{FDR level $\alpha$, non-negative non-increasing sequence $\{\gamma_j\}_{j=1}^\infty$ such that $\sum_j \gamma_j=1$, candidate threshold $\lambda\in(0,1)$, initial wealth $W_0\leq\alpha$}
$\alpha_1 =  (1-\lambda) \gamma_1 W_0$\newline
 \For{$t=1,2,\dots$}{
 start $t$-th test with level $\alpha_t$\newline
  $\alpha_{t+1} = \min\left\{\lambda,(1-\lambda)\left(W_0\gamma_{t+1-C_{0+}} + (\alpha - W_0)\gamma_{t+1-r_1-C_{1+}} + \sum_{j\geq 2} \alpha\gamma_{t+1-r_j-C_{j+}}\right)\right\}$, \\
  where $C_{j+}=\sum_{i=r_j+1}^t C_i\One{i\not\in\X^t}$

 }
 \caption{The SAFFRON* algorithm for constant $\lambda$ under general conflict sets}
\end{algorithm}


\begin{algorithm}[H]
\SetAlgoLined
\footnotesize
\SetKwInOut{Input}{input}
\Input{FDR level $\alpha$, non-negative non-increasing sequence $\{\gamma_j\}_{j=1}^\infty$ such that $\sum_j \gamma_j=1$, initial wealth $W_0\leq \alpha$}
$s_1 = \gamma_1 W_0$\newline
$\alpha_1 = s_1/(1 + s_1)$\newline
 \For{$t=1,2,\dots$}{
 start $t$-th test with level $\alpha_t$\newline
  $s_{t+1} = W_0\gamma_{t+1-R_{0+}} + (\alpha - W_0)\gamma_{t+1-r_1-R_{1+}} + \sum_{j\geq 2} \alpha\gamma_{t+1-r_j-R_{j+}}$, where $R_{j+}=\sum_{i=r_j+1}^t R_i\One{i\not\in\X^t}$\\
  $\alpha_{t+1} = s_{t+1}/(1+s_{t+1})$,
 }
 \caption{The alpha-investing algorithm under general conflict sets (a special case of SAFFRON*)}
\end{algorithm}


\subsection{Oracle estimate under conflict sets}
\label{subs:oracle}

Following \citet{ramdas2018saffron}, we analyze LORD* and SAFFRON* through an \emph{oracle estimate} of the false discovery proportion. This quantity serves as a good estimate of the true false discovery proportion, and controlling it under a pre-specified level guarantees that FDR is also controlled. Let the oracle estimate of the FDP be defined as:
\begin{align*}
    \fdp^*(t) \defn \frac{\sum_{j\leq t,j\in\nulls}\alpha_j}{(\sum_{E_j\leq t} R_j)\vee 1},
\end{align*}
where we recall that $\alpha_j$ is required to be $\F^{-\X^j}$-measurable, across all $j$. The following proposition gives formal justification for using $\fdp^*(t)$ as a proxy for the true FDP.

\begin{proposition}
\label{prop:oracleconflict}
Suppose that the null $p$-values are super-uniform conditional on $\F^{-\X^{E_t}}$, meaning $\PPst{P_t \leq
  u}{\F^{-\X^{E_t}}} \leq u$, for all  $u\in[0,1]$ and $t\in\nulls$. Then, for all times $t \in \N$, the condition $\fdp^*(t)\leq\alpha$ implies that $\mfdr(t)\leq\alpha$.
\end{proposition}

Note that \propref{oracleconflict} is technically true even if we only consider tests for which $E_j\leq t,j\in\nulls$ in the numerator of $\fdp^*(t)$. However, it is not clear how to achieve this without ensuring $\fdp^*(t)\leq \alpha$. For example, even if $\frac{\sum_{E_j\leq t,j\in\nulls}\alpha_j}{(\sum_{E_j\leq t} R_j)\vee 1}\leq \alpha$ at time $t$, it is possible that in subsequent rounds all tests will finish without any new rejections, thus increasing the FDP estimate. Therefore, we need to assign $\alpha_j$ conservatively, such that this estimate is provably controlled under $\alpha$, despite unknown future outcomes which might augment the FDP estimate.

The fact that $\alpha_t$ is measurable with respect to $\F^{-\X^t}$ should give us pause. Even though we are only required to guarantee $\fdp^*(t)\leq\alpha$, we cannot rely on the rejection indicators that push down the value of $\fdp^*(t)$, if they are in the current conflict set. As a consequence, $\alpha_t$ has to ensure $\fdp^*(t)\leq\alpha$ for the \emph{worst-case} configuration of conflicting rejections; that is, when $R_j=0$ for all $j\in\X^t$. This motivates us to define the oracle estimate of the FDP \emph{under conflict sets}:
\small
\begin{align}
\label{eqn:oracleconflict}
\fdp_{\text{conf}}^*(t) \defn \frac{\sum_{j\leq t,j\in\nulls}\alpha_j}{(\sum_{j\leq t, j\not\in\X^t} R_j)\vee 1}.
\end{align}
\normalsize

Since this quantity is only more conservative than the oracle estimate, controlling it under $\alpha$ will preserve the guarantees given by \propref{oracleconflict}. However, notice an unfortunate fact about both oracle estimates---they depend on the unobservable set $\nulls$. This implies that not even $\fdp_{\text{conf}}^*(t)$ can be controlled tightly. For this reason, LORD* and SAFFRON* construct \emph{empirical} estimates of $\fdp_{\text{conf}}^*(t)$, such that the properties given in \propref{oracleconflict} are retained. For LORD*, claiming $\mfdr$ control at fixed times boils down to a simple observation: for any chosen $\alpha$, $\fdp_{\text{conf}}^*(t)\leq \widehat \fdp_{\text{LORD*}}(t)\leq\alpha$, hence by \propref{oracleconflict} $\mfdr$ is controlled. SAFFRON* controls mFDR by virtue of ensuring that, on average, $\fdp_{\text{conf}}^*(t)\leq\alpha$. We make this argument formal in Section \ref{sec:unify}.


\section{Example 1: Asynchronous online FDR control}
\label{sec:async}

In this section, we look at one instantiation of the conflict-set framework, which considers arbitrary asynchrony but limits possible dependencies between $p$-values. This immediately gives two procedures for asynchronous online testing as special cases of LORD* and SAFFRON*. From here forward we will refer to these methods as $\lordasync$ and $\saffasync$, respectively. In \secref{unify}, we provide $\mfdr$ guarantees of these procedures in terms of the general conflict-set setting, as well as additional $\fdr$ guarantees for $\lordasync$ and $\saffasync$ under a strict independence assumption.


In this section, the only conflicting tests are those whose outcomes are unknown, since the allowed dependencies will be fairly restrictive. Therefore, the asynchronous conflict set at time $t$ is:
\begin{align*}
    \X_{\text{async}}^t = \{i\in[t-1]: E_i\geq t\},
\end{align*}
which is observable at time $t-1$. This simplified conflict set implies that the last-conflict time of test $t$ is, naturally, $\tau_t = E_t$.

Denote by $\cR_t$ the \emph{set of rejections} at time $t$, and similarly let $\C_t$ denote the \emph{set of candidates} at time $t$:
$$\cR_t = \{i\in[t]: E_i = t, P_i\leq\alpha_i\}, ~\C_t = \{i\in[t]: E_i = t, P_i\leq\lambda_i\}.$$
Therefore, $\cR(t) = \cup_{i=1}^t \cR_t$.
With this, we can write the non-conflicting filtrations $\LL_{\text{async}}^{-\X^t}$ and $\Ss_{\text{async}}^{-\X^t}$ compactly as:
$$\LL_{\text{async}}^{-\X^t} \defn \sigma(\cR_1,\dots,\cR_{t-1}),~\Ss_{\text{async}}^{-\X^t} \defn \sigma(\cR_1,\C_1,\dots,\cR_{t-1},\C_{t-1}).$$
Since the arguments for $\lordasync$ and $\saffasync$ have significant overlap, for brevity we write $\F_{\text{async}}^{-\X^t}$ to refer to both $\LL_{\text{async}}^{-\X^t}$ and $\Ss_{\text{async}}^{-\X^t}$, where possible. Recall from \secref{conflict} that $\alpha_t$ is designed to be measurable with respect to $\F_{\text{async}}^{-\X^t}$; here this essentially means that it is computed as a function of the outcomes known by time $t$. For $\saffasync$, additionally $\lambda_t$ is $\Ss_{\text{async}}^{-\X^t}$-measurable. More generally, for $\lordasync$, we can choose $\alpha_t = f_t(\cR_1,\dots,\cR_{t-1})$, for any deterministic function $f_t$ as long as the correct FDP estimate is controlled. The $\saffasync$ procedure also keeps track of encountered candidates, hence we can take $\alpha_t = g_t(\cR_1,\C_1,\dots,\cR_{t-1},\C_{t-1})$ and $\lambda_t = h_t(\cR_1,\C_1,\dots,\cR_{t-1},\C_{t-1})$, for deterministic functions $g_t$ and $h_t$.



Our mFDR guarantees hold under a condition which we term \emph{asynchronous super-uniformity}:
\begin{equation}
  \label{eqn:superuniformity-async}
\text{If the null hypothesis $H_t$ is true, then } \PPst{P_t \leq
  u}{\F_{\text{async}}^{-\X^{E_t}}} \leq u, \text{ for all } u\in[0,1].
 \end{equation}
This condition essentially shapes the allowed dependencies between $p$-values. It is immediately implied if the $p$-values are independent. However, it is strictly weaker. For example, it allows revisiting $p$-values which were previously not rejected. Suppose we have tested independent $p$-values thus far, and we failed to reject $H_t$, that is $P_t > \alpha_t$. If at a later time $s>t$ we have a higher error budget $\alpha_s > \alpha_t$, we can, somewhat surprisingly, test $H_t$ using the \emph{same} $p$-value $P_t$ again at time $s$. This clearly violates independence of $P_t$ and $P_s$ (as they are identical), however condition \eqnref{superuniformity-async} is nevertheless satisfied. Indeed, for all $u>\alpha_t$:
 $$\PPst{P_s \leq u}{\F_{\text{async}}^{-\X^{E_s}}} = \PPst{P_s \leq u}{P_s > \alpha_t}\leq \frac{u - \alpha_t}{1-\alpha_t} \leq u,$$
 where the equality follows because $P_s\perp \F_{\text{async}}^{-\X^{E_s}}~|~ \One{P_s > \alpha_t}$. On the other hand, if $u\leq \alpha_t$, $\PPst{P_s \leq u}{P_s > \alpha_t} = 0 \leq u$, and hence condition \eqnref{superuniformity-async} follows.

\subsection*{The $\lordasync$ and $\saffasync$ algorithms}

We turn to an analysis of how the abstract LORD* and SAFFRON* procedures translate into our asynchronous testing scenario, for the particular choice of conflict set $\X^t_{\text{async}}$. They utilize all available information; the conflict set---the tests whose outcomes the algorithms ignore---consists only of the tests about which we temporarily lack information.

Plugging in the definition of $\X^t_{\text{async}}$, we obtain the following empirical estimate of the false discovery proportion for $\lordasync$:
$$\widehat \fdp_{\lordasync}(t)= \frac{\sum_{j\leq t}\alpha_j}{(\sum_{j\leq t}\One{P_j\leq\alpha_j, E_j< t}) \vee 1}.$$
For $\saffasync$, we obtain the following estimate:
$$\widehat \fdp_{\saffasync}(t)= \frac{\sum_{j\leq t}\frac{\alpha_j}{1-\lambda_j}(\One{P_j>\lambda_j,E_j< t} + \One{E_j\geq t})}{(\sum_{j\leq t}\One{P_j\leq\alpha_j, E_j< t})\vee 1}.$$

Consider the dynamics of these two algorithms, and how their pessimism comes into play. Whenever a test starts, they increase their FDP estimate, expecting that the resulting $p$-value will have no favorable contribution. However, when the test in question ends, they readjust the FDP estimate if they see a positive outcome, namely a candidate and/or rejection. This shows that testing in parallel indeed has a cost---due to pessimistic expectations about the tests in progress, the algorithms remain conservative when assigning a new test level. For this reason, asynchronous testing should be used with caution, and the number of tests run in parallel should be monitored closely. Indeed, in the asymptotic limit where the number of parallel tests tends to infinity, the algorithm behaves like alpha-spending; i.e., the sum of all assigned test levels converges to the error budget $\alpha$.

Substituting $\X^t$ for $\X^t_{\text{async}}$ in \hyperref[algs1to4]{Algorithms 1-4} yields procedures for asynchronous online FDR control. The explicit statements of these algorithms, which correspond to asynchronous versions of LORD++, LOND, SAFFRON, and alpha-investing, are given in the Appendix.


\section{Example 2: Online FDR control under local dependence}
\label{sec:markov}

In this section, we derive online FDR procedures that handle local dependencies. We begin with the fully synchronous setting studied in classical online FDR literature, and turn to the asynchronous environment in the next section.

A standard assumption in existing work on online FDR has been independence of $p$-values, a requirement that is rarely justified in practice. Tests that cluster in time often use the same data, null hypotheses depend on the outcomes of recent tests, etc. On the other hand, arbitrary dependence between any two $p$-values in the sequence is also arguably unreasonable---very old data used for testing in the past are usually considered ``stale,'' and hypotheses tested a long time ago may bear little relevance to current hypotheses. In light of this, we consider a notion of \emph{local dependence}: 
\begin{align*}
    \text{for all $t>0$, there exists $L_t\in\N$ such that } P_t\perp P_{t-L_t-1},P_{t-L_t-2},\dots,P_1,
\end{align*}
where $\{L_t\}$ is a fixed sequence of parameters which we refer to as lags.

Since we allow $P_t$ to have arbitrary dependence on the previous $L_t$ $p$-values, some of these dependencies might be adversarial toward the statistician, and, with ``peeking'' into this adversarial set, the nulls might no longer behave super-uniformly. Suppose we observe a sample $X\sim N(\mu,1)$, and wish to test two hypotheses using this sample. Let the two hypotheses be $H_1:\mu < 0$ and $H_2:\mu \geq 0$. If, for instance, $R_1=0$, we know that $P_2\leq 1-\alpha_1$ almost surely, implying that $P_2$ is not super-uniform, given the information about past tests. On the other hand, if we were to ignore the outcome of the first test, $P_2$ would indeed be super-uniform.

This observation motivates us to define the conflict set for testing under local dependence as:
$$\X^t_{\text{dep}}\defn \{t-L_t,\dots,t-1\}.$$

The non-conflicting filtrations $\LL_{\text{dep}}^{-\X^t}$ for $\lordmarkov$ and $\Ss_{\text{dep}}^{-\X^t}$ for $\saffmarkov$ are respectively given by:
$$\LL_{\text{dep}}^{-\X^t} \defn \sigma(R_1,\dots,R_{t-L_t-1}),~\Ss_{\text{dep}}^{-\X^t} \defn \sigma(R_1,C_1,\dots,R_{t-L_t-1},C_{t-L_t-1}).$$
Since most formal arguments in this section apply to both procedures, we use $\F_{\text{dep}}^{-\X^t}$ to indicate that the filtration in question could be both $\LL_{\text{dep}}^{-\X^t}$ and $\Ss_{\text{dep}}^{-\X^t}$.

In contrast to asynchronous testing, the levels $\alpha_t$ and $\lambda_t$ under local dependence ignore some portion of available information, specifically the outcomes of the last $L_t$ tests. Notice the difference between these two settings---in the asynchronous setting, pessimism guards against \emph{unknown outcomes}, while here pessimism guards against \emph{known outcomes}. Perhaps counterintuitively, this observation means that the pessimism of $\lordmarkov$ and $\saffmarkov$ actually guards against possible disadvantageous direct impact of the last $L_t$ $p$-values on the upcoming one. In the Appendix we instantiate the test levels and candidacy thresholds according to Algorithms 1-4, however more generally we allow $\alpha_t = f_t(R_1,\dots,R_{t-L_t-1})$ for $\lordmarkov$, and $\alpha_t = g_t(R_1,C_1,\dots,R_{t-L_t-1},C_{t-L_t-1})$ and $\lambda_t = h_t(R_1,C_1,\dots,R_{t-L_t-1},C_{t-L_t-1})$ for $\saffmarkov$.


Consider some $P_t$ which is from a null hypothesis. As previously emphasized, we cannot trust $P_t$ to behave like a true null, given that we already know its last $L_t$ predecessors that have a direct impact on it. The appropriate super-uniformity condition satisfied by locally dependent $p$-values thus ignores these last $L_t$ $p$-values and is of the following form:
\begin{equation}
  \label{eqn:superuniformity-markov}
\text{If the null hypothesis $H_t$ is true, then } \PPst{P_t \leq
  u}{\F_{\text{dep}}^{-\X^t}} \leq u , \text{ for all } u\in[0,1].
 \end{equation}
This will allow setting $\alpha_t\in\F_{\text{dep}}^{-\X^t} $, while knowing $\PPst{P_t \leq
  \alpha_t}{\F_{\text{dep}}^{-\X^t}} \leq \alpha_t$. Importantly, unlike in the previous section where the appropriate super-uniformity condition implied dependence constraints on the $p$-values, condition \eqnref{superuniformity-markov} is immediately true by local dependence.

\subsection*{The $\lordmarkov$ and $\saffmarkov$ algorithms}

As in \secref{async}, we analyze the particular instances of LORD* and SAFFRON* that are obtained by taking the conflict set of \secref{conflict} to be $\X^t_{\text{dep}}=\{t-L_t,\dots,t-1\}$. Since this conflict set is deterministic, unlike $\X^t_{\text{async}}$, the estimate of the false discovery proportion that $\lordmarkov$ and $\saffmarkov$ keep track of is completely determined $L_t$ steps ahead, that is at time $t-L_t-1$.

By definition of the general estimates and the conflict set in consideration, $\lordmarkov$ controls the following quantity:
$$\widehat \fdp_{\lordmarkov}(t) = \frac{\sum_{j\leq t}\alpha_j}{(\sum_{j\leq t,j\not\in\{t-L_t,\dots,t-1\}}R_j)\vee 1}.$$
The $\saffmarkov$ method, on the other hand, controls the estimate:
$$\widehat \fdp_{\saffmarkov}(t) = \frac{\sum_{j< t-L_t}\frac{\alpha_j}{1-\lambda_j}\One{P_j>\lambda_j} + \sum_{j=t-L_t}^t \frac{\alpha_j}{1-\lambda_j}}{(\sum_{j\leq t,j\not\in\{t-L_t,\dots,t-1\}}R_j)\vee 1}.$$

In the case of running asynchronous tests, the algorithms were constructed as pessimistic; however, they had access to as much information as the statistician performing the tests. Here, that is not the case---$\lordmarkov$ and $\saffmarkov$ choose to ignore the outcomes of completed tests as long as they are in the conflict set of subsequent tests. Only after the last-conflict time $\tau_i$, positive outcomes are rewarded by readjusting the FDP estimate. On the other hand, the statistician's perspective is different---as soon as round $t$ is over, the statistician knows the outcome of the $t$-th test. Just like testing in parallel, testing locally dependent $p$-values comes at a cost---if the lags are large, the algorithm keeps increasing the FDP estimate, assigning ever smaller test levels, waiting for rewards from tests performed a long time ago. In the extreme case of $L_t=t$, the test levels steadily decrease so that their sum converges to $\alpha$, regardless of the fact that discoveries have possibly been made.

Explicit setting-specific algorithms, obtained by substituting $\X^t$ for $\X^t_{\text{dep}}$ in Algorithms 1-4, resulting in LORD++, LOND, SAFFRON, and alpha-investing under local dependence, are given in the Appendix.



\section{Example 3: Controlling FDR in asynchronous mini-batch testing}
\label{sec:minibatch}

Here we merge the ideas of the previous two sections, bringing together asynchronous testing and local dependence of $p$-values. Although there are various ways one could think of in which these two concepts intertwine, here we discuss a particularly simple and natural one.

\begin{figure}[h]
\centerline{\includegraphics[width=\textwidth]{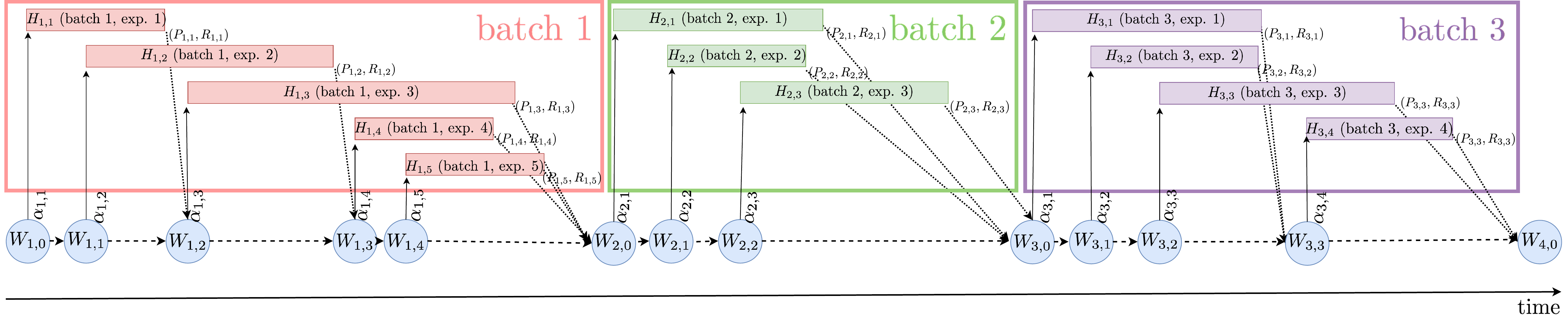}}
\caption{Running three mini-batches of tests. The batches are run synchronously, while the tests that comprise each of them are run asynchronously. We use $W_{t,j-1}$ to denote the remaining "wealth" for making false discoveries before starting the $j$-th test in the $t$-th batch.} 
\label{fig:minibatch}
\end{figure}

Let a \emph{mini-batch} represent a grouping of an arbitrary number of tests that are run asynchronously, which result in dependent $p$-values; for instance, these tests could be run on the same data. After a mini-batch of tests is fully executed, a new one can start, testing new hypotheses, independent of the previous batch, and doing so on fresh data. From the point of view of asynchrony, such a process could be thought of as a compromise between synchronous and asynchronous testing---batches are internally asynchronous, however they are globally sequential and synchronous. If all batches are of size one, one recovers classical online testing; if the batch-size tends to infinity, the usual notion of asynchronous testing is obtained. \figref{minibatch} depicts an example of a mini-batch testing process with three mini-batches.

We introduce notation that captures this setting. We will use two time indices; $P_{b,t}$ denotes the $p$-value resulting from the test that starts as the $t$-th one in the $b$-th batch, testing hypothesis $H_{b,t}$. We allow any two $p$-values in the same batch to have arbitrary dependence; however, we require any two $p$-values in different batches to be independent. This can be written compactly as:
$$P_{b_1,i}\perp P_{b_2,j}, \text{ for any } b_1,b_2,i,j, \text{ such that } b_1\neq b_2.$$

We will denote the size of the $b$-th batch as $n_b$. Thus, the first batch results in $P_{1,1},\dots,P_{1,n_1}$, the second one in $P_{2,1},\dots,P_{2,n_2}$, etc. Analogously, the test levels and candidacy thresholds will also be doubly-indexed; $\alpha_{b,t}$ and $\lambda_{b,t}$ are used for testing $P_{b,t}$. Further, we define $R_{b,t}\defn \One{P_{b,t}\leq\alpha_{b,t}}$, and $C_{b,t}\defn \One{P_{b,t}\leq\lambda_{b,t}}$ as the rejection and candidacy indicators, respectively. By $\cR_b$ we will denote the set of rejections in the $b$-th batch, and by $\C_b$ the set of candidates in the $b$-th batch.

Recall the key ideas of the previous two sections---tests running in parallel, or those resulting in dependent $p$-values, are seen as conflicting. We again pursue this approach, and let the conflict set of $P_{b,t}$ consist of all other $p$-values in the same batch. More formally, the mini-batch conflict set can be defined as:
$$\X_{\text{mini}}^{b,t} = \{(b,i):i< t\}.$$
Notice that in \secref{async}, the conflicts arise solely due to missing information, in \secref{markov} solely due to dependence, while here they are due to both.

The instances of LORD* and SAFFRON* used to test mini-batches will be referred to as $\lordmini$ and $\saffmini$. As before, we will define the past-describing filtrations for both of these algorithms. Due to local dependence, as in \secref{markov}, whole batches of tests are mutually conflicted. Only at the finish time of a batch are the discoveries taken into account. For this reason, from the perspective of any batch, all rejections in any prior batch happened at one time step. Consequently, there is no need to consider the actual finish time of any test from previous batches, and thus the respective non-conflicting filtrations for $\lordmini$ and $\saffmini$ will be of the form:
$$\LL_{\text{mini}}^{-\X^{b,t}} = \sigma(\cR_1,\dots,\cR_{b-1}),~\Ss_{\text{mini}}^{-\X^{b,t}} = \sigma(\cR_1,\C_1,\dots,\cR_{b-1},\C_{b-1}).$$
As before, we use $\F_{\text{mini}}^{-\X^{b,t}}$ to refer to both of these two filtrations simultaneously. The test levels $\{\alpha_{b,t}\}$ and candidacy thresholds $\{\lambda_{b,t}\}$ are therefore computed as functions of the outcomes of the tests in previous batches, i.e., we can write $\alpha_{b,t} = f_{b,t}(\cR_1,\dots,\cR_{b-1})$ for $\lordmini$, and similarly, $\alpha_{b,t} = g_{b,t}(\cR_1,\C_1,\dots,\cR_{b-1},\C_{b-1})$ and $\alpha_{b,t} = h_{b,t}(\cR_1,\C_1,\dots,\cR_{b-1},\C_{b-1})$ for $\saffmini$.


By analogy with the last section, we do not necessarily expect the $p$-value $P_{b,t}$ to be well-behaved, given that we have seen the outcomes of tests whose $p$-values have dependence on $P_{b,t}$. By the local dependence assumption, it is straightforward to verify that the following condition holds true:
\begin{align}
\label{eqn:superuniformity-mini}
    \text{If the null hypothesis $H_{b,t}$ is true, then } \PPst{P_{b,t} \leq
  u}{\F_{\text{mini}}^{-\X^{b,t}}} \leq u , \text{ for all } u\in[0,1].
\end{align}

\subsection*{The $\lordmini$ and $\saffmini$ algorithms}

By definition of the mini-batch conflict set and the general estimate of LORD*, $\lordmini$ is obtained as an update rule for $\alpha_{b,t}$ such that the following quantity is controlled for all $b,t\in\N$:
$$\widehat \fdp_{\lordmini}(b,t) = \frac{\sum_{i<b}\sum_{j\leq n_i}\alpha_{i,j} + \sum_{j\leq t}\alpha_{b,j}}{(\sum_{i< b} \sum_{j\leq n_i}R_{i,j})\vee 1}.$$
Similarly, $\saffmini$ controls the following adaptive estimate:
$$\widehat \fdp_{\saffmini}(b,t) = \frac{\sum_{i< b}\sum_{j\leq n_i}\frac{\alpha_{i,j}}{1-\lambda_{i,j}}\One{P_{i,j}>\lambda_{i,j}} + \sum_{j\leq t}\frac{\alpha_{b,j}}{1-\lambda_{b,j}}}{(\sum_{i< b} \sum_{j\leq n_i}R_{i,j})\vee 1}.$$

Since the set of rejections corresponding to tests that are not in the current conflict set is invariant throughout the testing of any whole batch, the FDP estimate gradually increases while a batch is being tested. Only when the batch has finished testing in its entirety does the algorithm get rewarded for every rejection it made in that batch. This implies that the batch size should be carefully chosen, as the achieved power decreases with batch size. This is numerically verified in \secref{sims}.

The LORD++, LOND, SAFFRON, and alpha-investing procedures for mini-batch testing are explicitly stated in the Appendix, obtained by substituting $\X_{\text{mini}}^{b,t}$ into Algorithms 1-4.


\section{Controlling mFDR and FDR at fixed and stopping times}
\label{sec:unify}

The previous three sections have shown that the abstract framework of conflict sets is a useful representational tool for expressing interactions across different tests, yielding three natural specific testing protocols. In this section, we return to the abstract unified framework in order to prove $\mfdr$ guarantees of LORD* and SAFFRON*, which implies $\mfdr$ control of all of the setting-specific algorithms. Additionally, we provide several results on strict FDR control under asynchrony and dependence, although under more stringent conditions.

\subsection{mFDR control}

We begin by focusing on fixed-time $\mfdr$ control. As mentioned earlier, the claim for LORD* follows trivially from \propref{oracleconflict}, so the proof of \thmref{mfdr-conflict}, given in the Appendix, focuses on providing guarantees for SAFFRON*.

\begin{theorem}\label{thm:mfdr-conflict}
Suppose that the null $p$-values are super-uniform conditional on $\F^{-\X^{E_t}}$, meaning $\PPst{P_t \leq
  u}{\F^{-\X^{E_t}}} \leq u$, for all  $u\in[0,1]$ and $t\in\nulls$. Then, LORD* and SAFFRON* with target FDR level $\alpha$ both guarantee that $\mfdr(t) \leq \alpha$ for all $t \in \N$.
\end{theorem}

Notice that the super-uniformity assumption above reduces to conditions \eqnref{superuniformity-async}, \eqnref{superuniformity-markov}, and \eqnref{superuniformity-mini}, in the three settings previously described.

The result of \thmref{mfdr-conflict} actually holds more generally; in particular, in the following theorem we show that $\mfdr$ is also controlled at certain stopping times. Our approach is based on constructing a process which behaves similarly to a submartingale, which allows us to derive a result mimicking optional stopping. This process, however, is not a submartingale in the general case. For example, it is not a submartingale in the synchronous setting under local dependence, described in \secref{markov}.

More specifically, we show that LORD* and SAFFRON* control $\mfdr$ at any stopping time $T$ which satisfies the following conditions:
\begin{enumerate}
    \item[(C1)] $T$ is defined with respect to the filtration $\F^{-\X^{t+1}}$, $\{T=t\}\in\F^{-\X^{t+1}}$;
    \item[(C2)] $T$ is almost surely bounded.
\end{enumerate}

Recall that $\F^{-\X^{t+1}}$ denotes the non-conflicting information about the first $t$ tests (in particular, not the first $t+1$), and hence the offset by 1 in indexing. Intuitively, this means that the decision to stop at time $t$ can depend on all information up to time $t$ that the algorithm is allowed to utilize.

Condition (C2) is a mild one, as in practice we primarily care about bounded stopping times. For instance, one would not wait infinitely long to observe the first rejection; if $T_{r_1}$ denotes the time of the first rejection, a natural stopping time would be $T\defn T_{r_1}\wedge t_{\max}$, where $t_{\max}$ is the fixed longest time one is willing to wait for a rejection.

When $E_t=t$ and the $p$-values are independent, we require the same conditions as \citet{foster2008alpha} in their stopping-time analysis of the mFDR. Consequently, their result can be seen as a special case of Theorem \ref{thm:stoppinglord&saff}, given that alpha-investing is a special instance of SAFFRON.

\begin{theorem}
\label{thm:stoppinglord&saff}
Suppose that the null $p$-values are super-uniform conditional on $\F^{-\X^{E_t}}$, meaning $\PPst{P_t \leq
  u}{\F^{-\X^{E_t}}} \leq u$, for all  $u\in[0,1]$ and $t\in\nulls$. Consider any stopping time $T$ that satisfies conditions (C1-C2). Then, LORD* and SAFFRON* with target FDR level $\alpha$ both control $\mfdr$ at $T$: $\mfdr(T)\leq\alpha$.
\end{theorem}


\subsection{FDR control}

Even though the main objective of the paper is to provide $\mfdr$ guarantees, one can also obtain $\fdr$ control for $\lordasync$ and $\saffasync$, provided that the $p$-values in the sequence are independent. This is in line with earlier work where (synchronous) online FDR control has only been proved under independence assumptions \citep{javanmard2016online,RYWJ17,ramdas2018saffron}. While our arguments below generalize the earlier ones, we stress that the independence assumption may not be reasonable in asynchronous settings, which is why we focused on the mFDR for most of the paper and we only present the argument below for completeness.

For FDR control, we additionally require $\alpha_t$ and $\lambda_t$ to be \emph{monotone}. In the context of $\lordasync$, this means that 
$$\alpha_t = f_t(\cR_1,\dots,\cR_i,\dots\cR_{t-1})\geq f_t(\cR_1,\dots,\cR_i',\dots\cR_{t-1}) =\alpha_t'$$
whenever $\cR_i' \subseteq \cR_i$. For $\saffasync$, we require the same condition also when $\C_i' \subseteq \C_i$, both for $\alpha_t$ and $\lambda_t$. All update rules stated in this paper are monotone by design.

First we state a technical lemma that is the key ingredient in proving $\fdr$ control of our asynchronous procedures, which generalizes several similar lemmas that have appeared in related work \citep{javanmard2016online,RYWJ17,ramdas2018saffron}.

\begin{lemma}
\label{lem:superunif}
Assume that null $p$-values are independent of each other and of
the non-nulls. Moreover, let $g: \{\N\cup\{0\}\}^M \to \R$ be any coordinate-wise non-decreasing function. Then, for any index $t \leq M$ such that $t \in \nulls$, we have:
\begin{align*}
\EEst{ \frac{\alpha_t \One{P_t >
      \lambda_t}}{(1-\lambda_t)g(|\cR|_{1:M})}}{\F_{\text{async}}^{-\X^{E_t}}} &\geq~
\EEst{\frac{\alpha_t}{g(|\cR|_{1:M})}}{\F_{\text{async}}^{-\X^{E_t}}}  \geq~
\EEst{ \frac{ \One{P_t \leq
      \alpha_t}}{g(|\cR|_{1:M})}}{\F_{\text{async}}^{-\X^{E_t}}},
\end{align*}
where $|\cR|_{1:M} = (|\cR_1|,\dots,|\cR_M|)$.
\end{lemma}

With this lemma, we directly obtain FDR guarantees of $\lordasync$ and $\saffasync$ under independence, as stated in \thmref{fdr-async}.

\begin{theorem}\label{thm:fdr-async}
Suppose that the null $p$-values are independent of each other and of
the non-nulls, and that $\alpha_t$ and $\lambda_t$ are monotone. Then, $\lordasync$ and $\saffasync$ with target FDR level $\alpha$ both guarantee $\fdr(t)\leq\alpha$ for all $t\in\N$.
\end{theorem}



Additionally, we prove that the original LOND algorithm \citep{javanmard2015online} controls FDR for an arbitrary sequence of $p$-values that satisfy positive regression dependency on a subset (PRDS) \citep{BY01}, without any correction. In other words, under the PRDS assumption, it suffices to take all conflict sets in the sequence to be empty. For convenience, we state the formal definition of PRDS in the Appendix.

Recall the setup of the LOND algorithm. Given a non-negative sequence $\{\gamma_j\}_{j=1}^\infty$ such that $\sum_{j=1}^\infty \gamma_j=1$, the test levels are set as $\alpha_t = \alpha \gamma_t (|\cR(t-1)|\vee 1)$, where $|\cR(t-1)|$ denotes the number of rejections at time $t-1$. Note that this rule is monotone, in the sense that $\alpha_t$ is coordinate-wise non-decreasing in the vector of rejection indicators $(R_1,\dots,R_{t-1})$. Below, we prove that LOND controls the FDR at any time $t \in \N$ under PRDS.

Recalling the definition of reshaping \citep{ramdas2017unified,blanchard2008two}, we will also prove that if $\{\beta_t\}$ is a sequence of reshaping functions, then using the test levels $\talpha_t := \alpha \gamma_t \beta_t(|\cR(t-1)|\vee 1)$ controls FDR under arbitrary dependence. We call this the reshaped LOND algorithm. As one example, using the Benjamini-Yekutieli reshaping yields $\talpha_t := \alpha \gamma_t (|\cR(t-1)|\vee 1)/(\sum_{i=1}^t \tfrac1i)$.

\begin{theorem}
\label{thm:positivedep}
\begin{enumerate}
  \item[(a)] The LOND algorithm satisfies $\fdr(t) \leq \alpha$ for all $t \in \N$ under positive dependence (PRDS).
  \item[(b)] Reshaped LOND satisfies $\fdr(t) \leq \alpha$ for all $t \in \N$ under arbitrary dependence.
\end{enumerate}
\end{theorem}


\section{Numerical experiments}
\label{sec:sims}

Here we present the results of several numerical simulations, which show the gradual change in performance of LORD* and SAFFRON* with the increase of asynchrony and the lags of local dependence.\footnote{The code for all experiments in this section is available at: https://github.com/tijana-zrnic/async-online-FDR-code}  We also compare these solutions to existing procedures with formal FDR guarantees under dependence. The plots in this section compare the achieved power and FDR of $\lordasync$, $\saffasync$, $\lordmarkov$, $\saffmarkov$, $\lordmini$ and $\saffmini$ for different problem parameters, in settings with $p$-values computed from Gaussian observations. We present additional experiments on real data in the Appendix.

The justification for focusing on synthetic data is two-fold. First, there is no standardized real data set for testing online FDR procedures. The quintessential applications of these methods involve testing with sensitive data, which are not publicly available due to privacy concerns. Second, even when real data are obtainable, it is unclear how one would evaluate the ground truth.

In all of the simulations we present the FDR is controlled at $\alpha=0.05$, and we estimate the FDR and power by averaging the results of 200 independent trials. The SAFFRON-type algorithms use the constant candidacy threshold sequence $\lambda=1/2$, across all tests. The LORD-type algorithms use the LORD++ update for test levels. Each figure additionally plots the performance of uncorrected testing, in which the constant test level $\alpha_t=\alpha=0.05$ is used across all $t\in\N$, and alpha-spending, whose test levels decay according to the $\{\gamma_t\}_{t=1}^\infty$ sequence of LORD* and SAFFRON*.

The experiments test for the means of $M=1000$ Gaussian observations, and each null hypothesis takes the form $H_i: \mu_i = 0$, where $\mu_i$ is the mean of the Gaussian sample. We generate samples $\{Z_i\}_{i=1}^M$, where $Z_i\sim N(\mu_i,1)$ and the parameter $\mu_i$ is chosen as $\mu_i = \xi F_1$, where $\xi\sim\text{Bern}(\pi_1)$,
for a fixed proportion of non-nulls in the sequence $\pi_1$, and some random variable $F_1$. We consider two distributions for $F_1$---a degenerate distribution with a point mass at $\mu_c$, where $\mu_c$ is a fixed constant for the whole sequence, or $N(0,2\log(M))$. The motivation for the latter is that $\sqrt{2\log(M)}$ is the minimax amplitude for estimation under the sparse Gaussian sequence model. In the case of the mean coming from a degenerate distribution, we form one-sided $p$-values as $P_i = \Phi(-Z_i)$, where $\Phi$ is the standard Gaussian CDF. If the mean has a Gaussian distribution, we form two-sided $p$-values, i.e., $P_i = 2\Phi(-|Z_i|)$.

\subsection{Varying asynchrony}

\begin{figure}[tb]
\centerline{\includegraphics[width=0.25\textwidth]{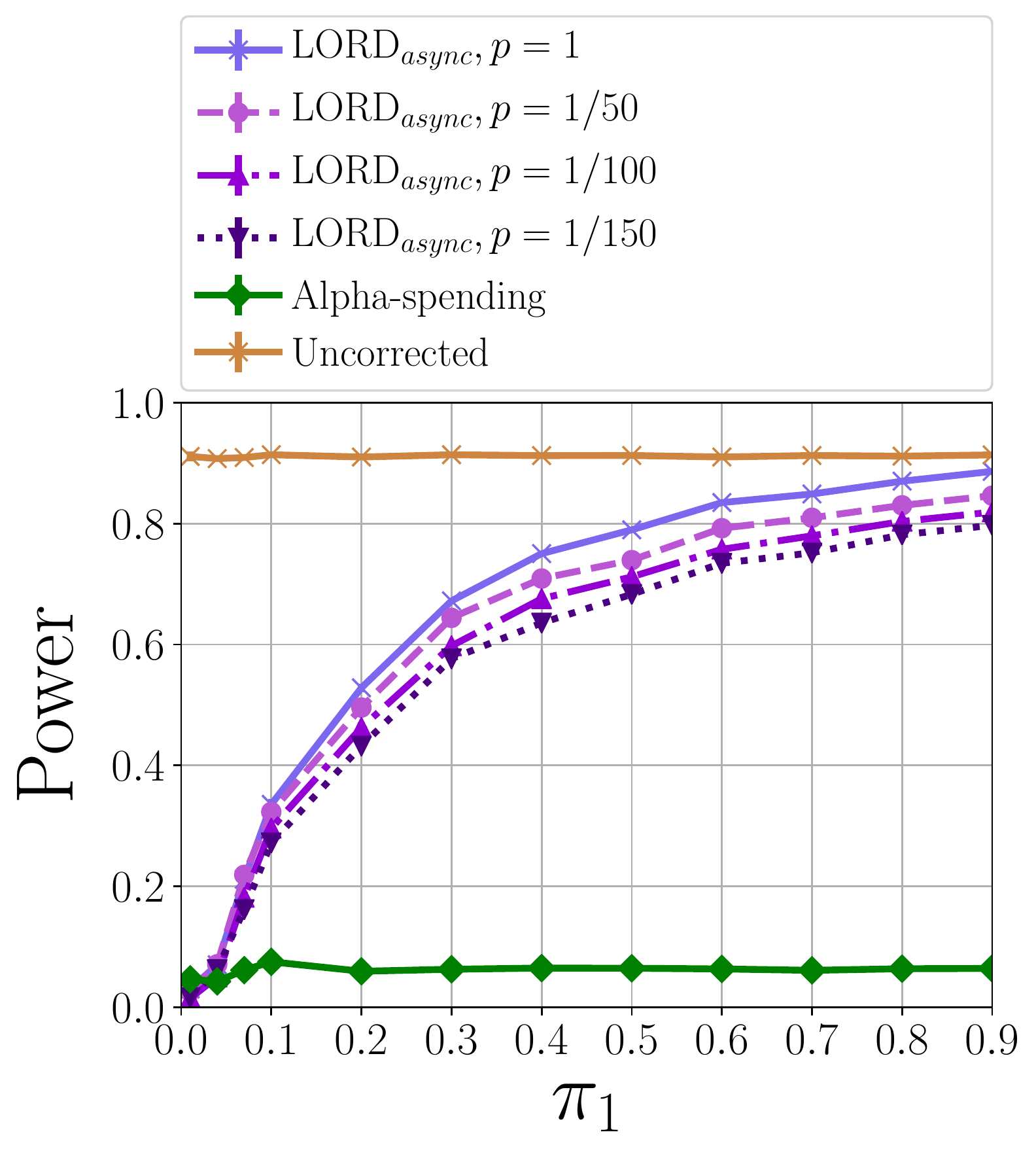}
\includegraphics[width=0.25\textwidth]{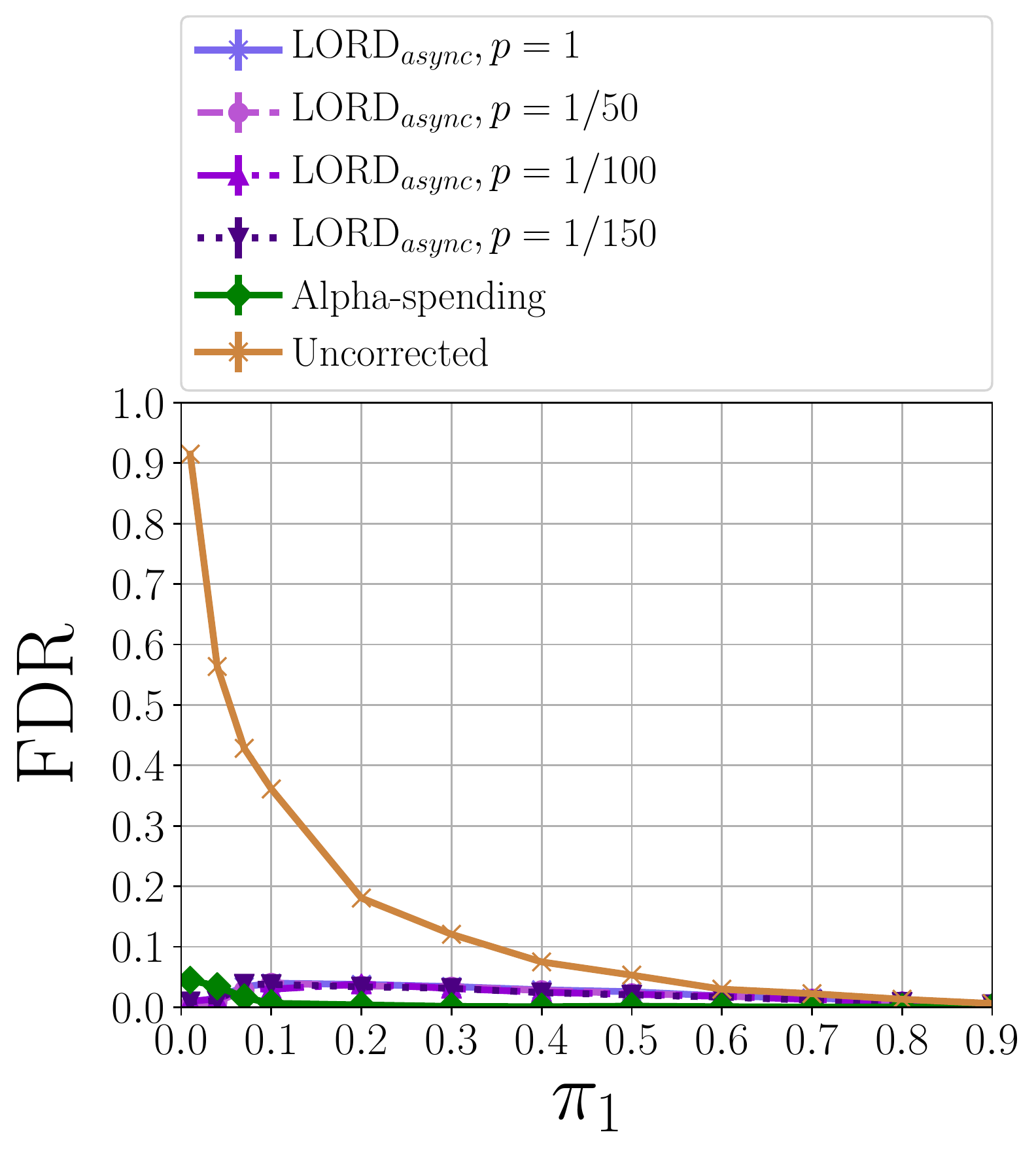}
\includegraphics[width=0.25\textwidth]{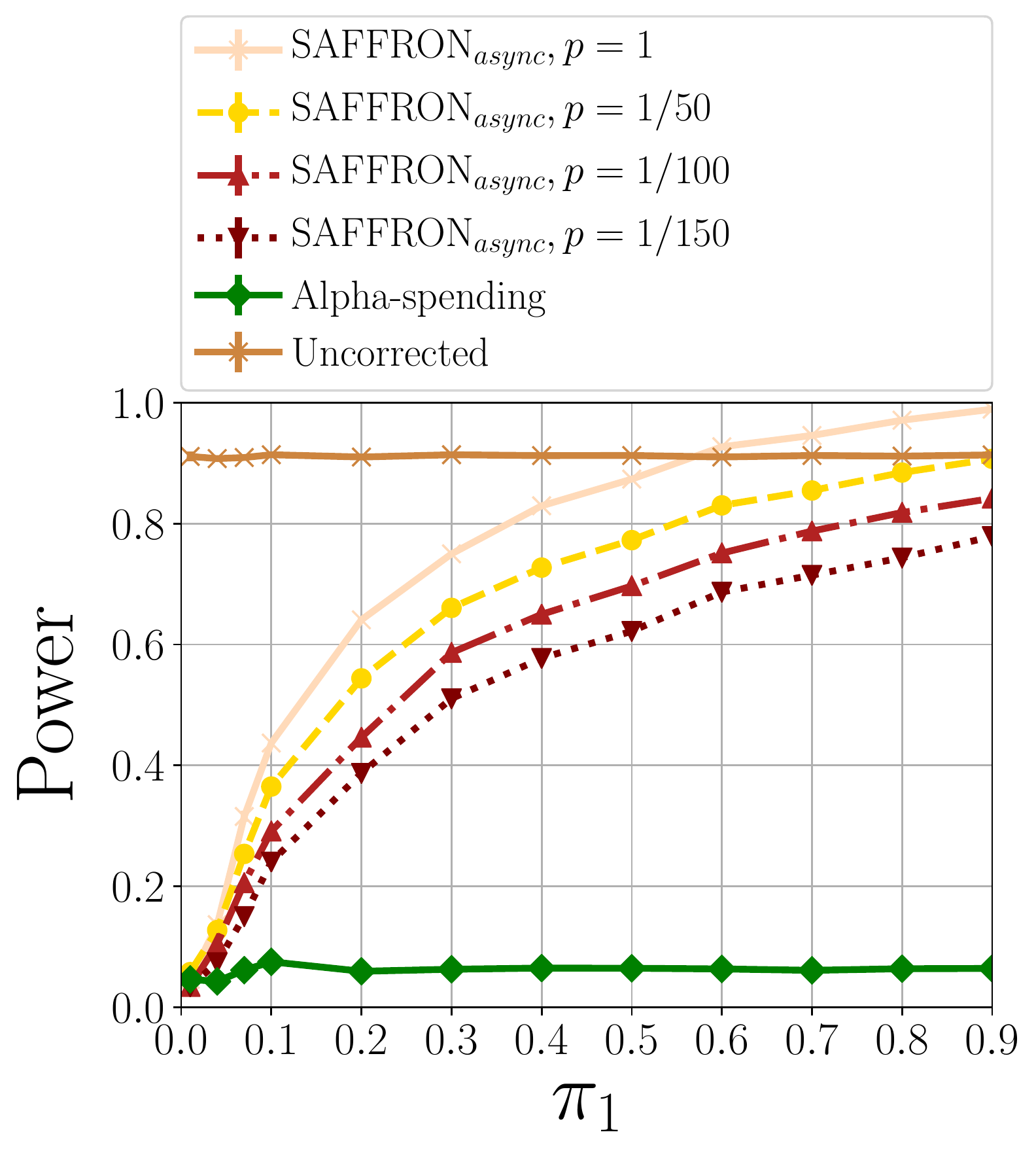}
\includegraphics[width=0.25\textwidth]{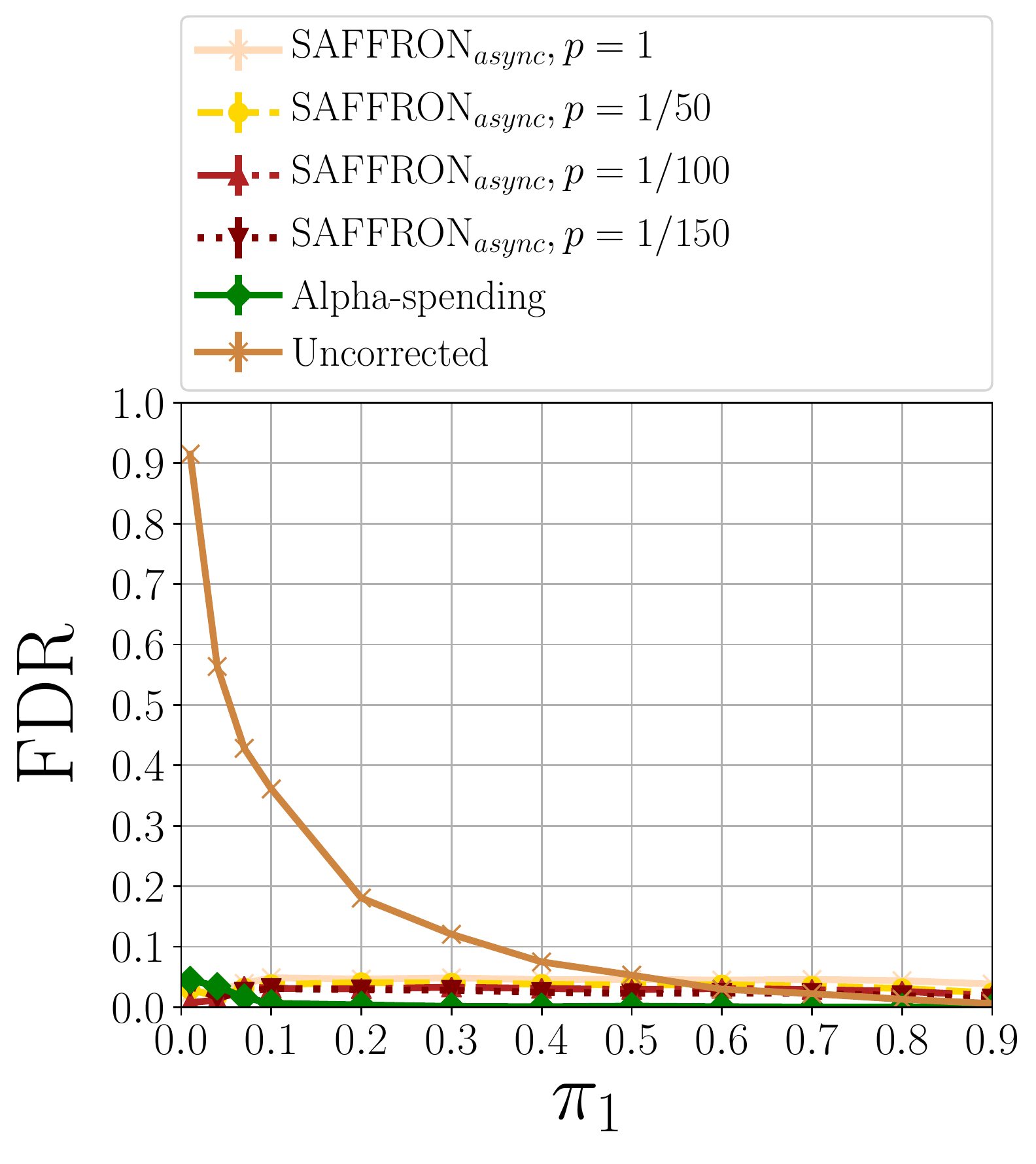}}
\caption{Power and FDR of $\lordasync$ and $\saffasync$ with varying the parameter of asynchrony $p$ of the tests. In all five runs $\lordasync$ and $\saffasync$ have the same parameters ($\{\gamma_j\}_{j=1}^\infty, W_0$). The mean of observations under the alternative is a point mass at $\mu_c=3$.} 
\label{fig:async}
\end{figure}

\begin{figure}[tb]
\centerline{\includegraphics[width=0.25\textwidth]{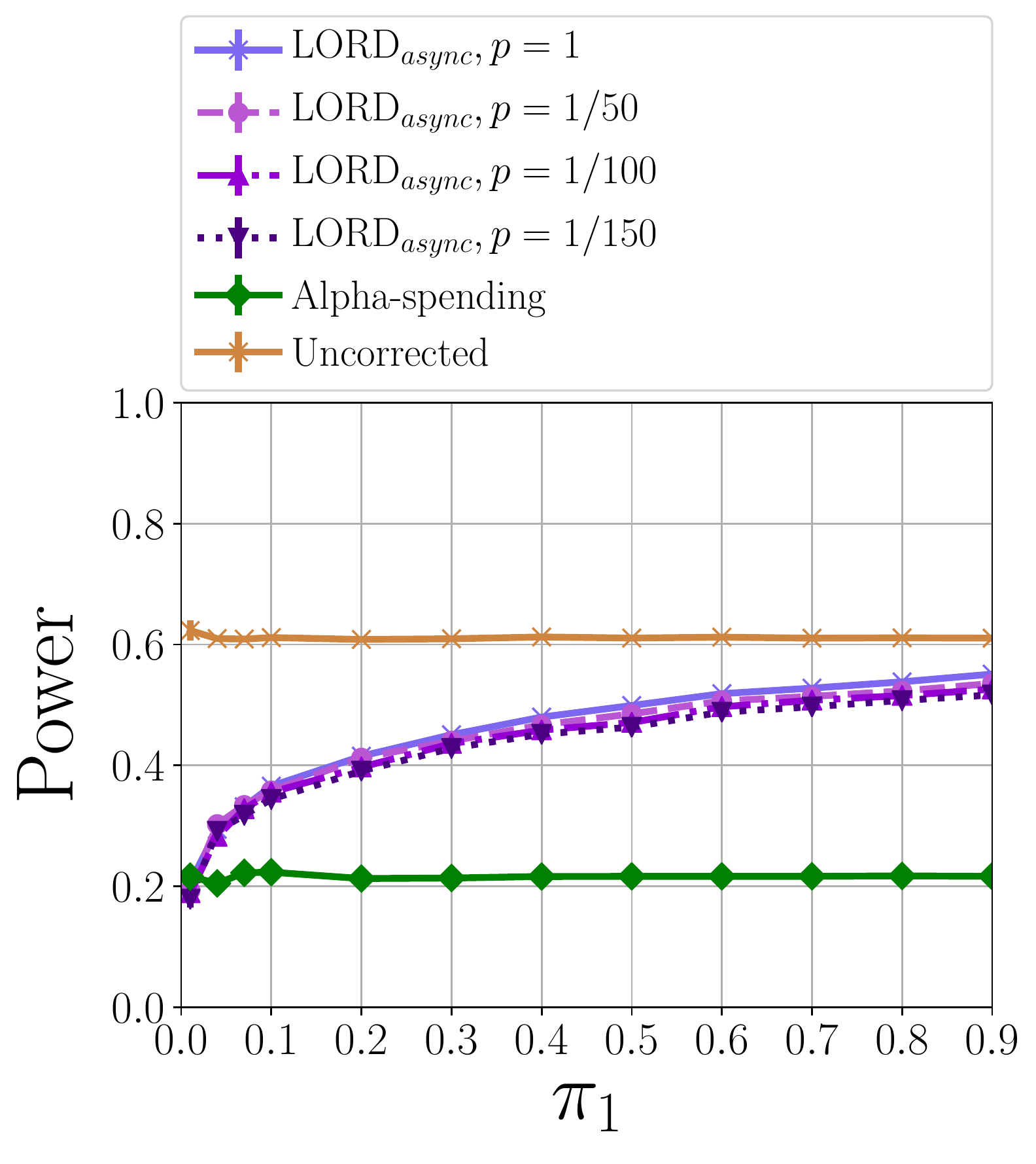}
\includegraphics[width=0.25\textwidth]{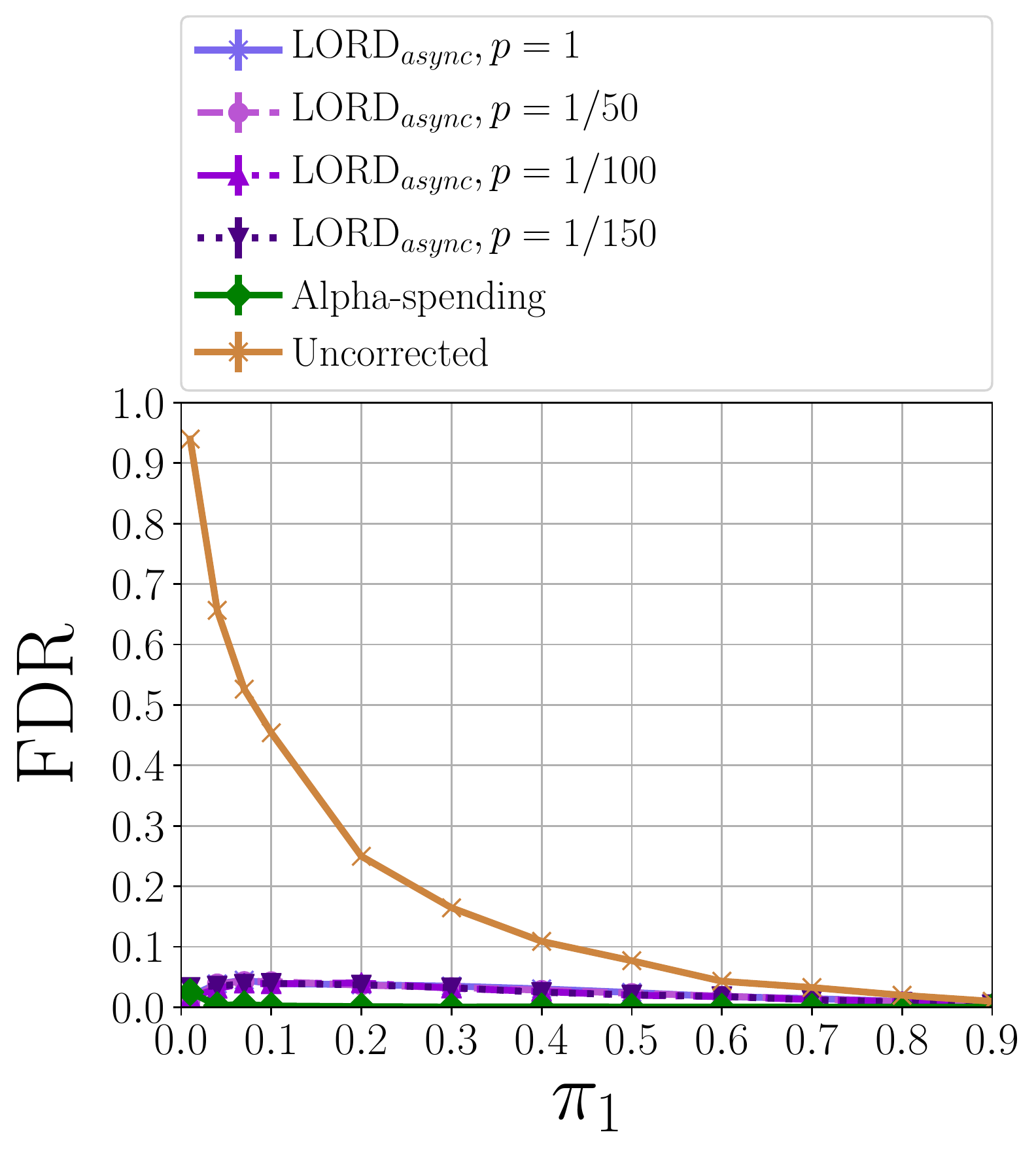}
\includegraphics[width=0.25\textwidth]{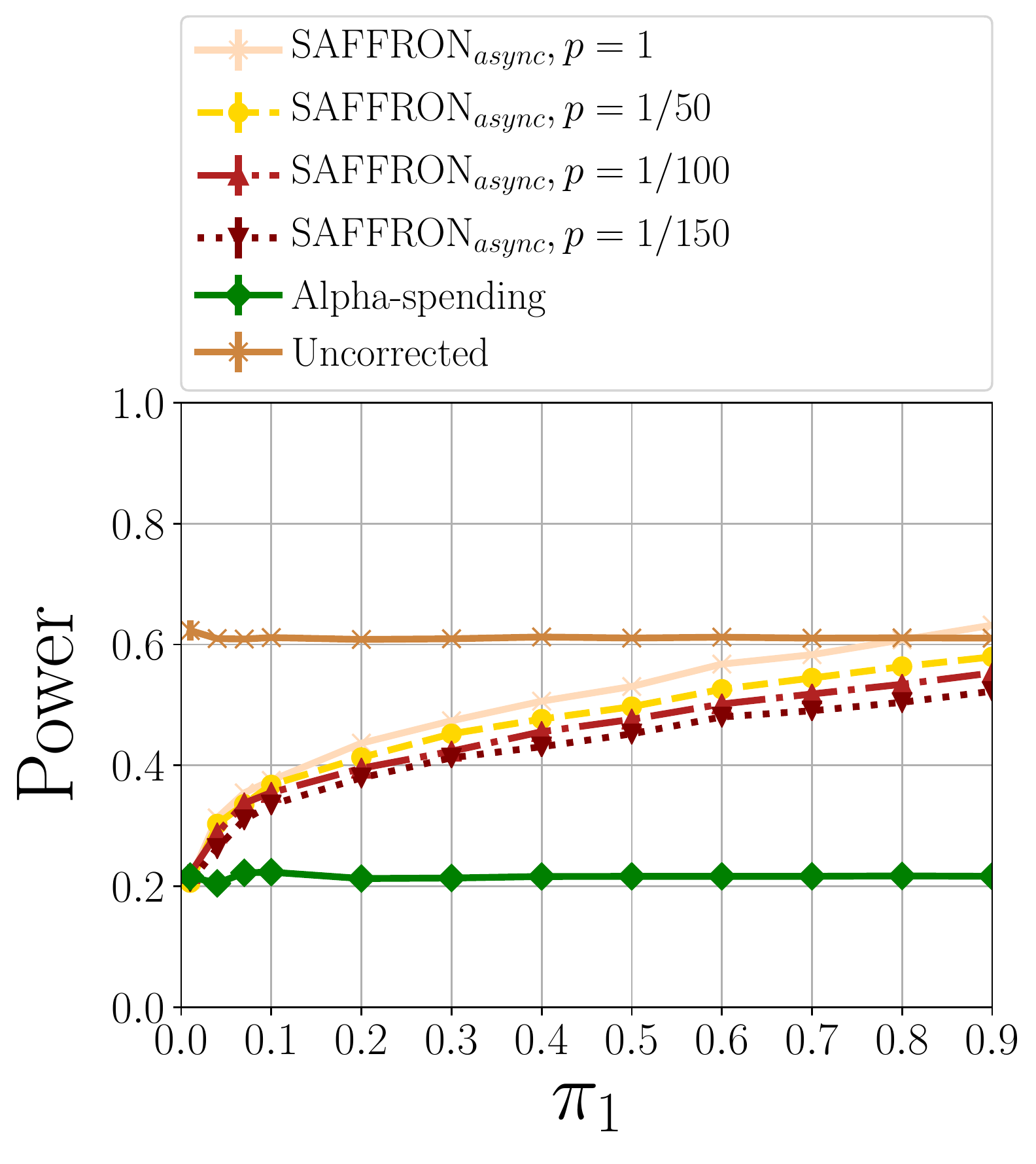}
\includegraphics[width=0.25\textwidth]{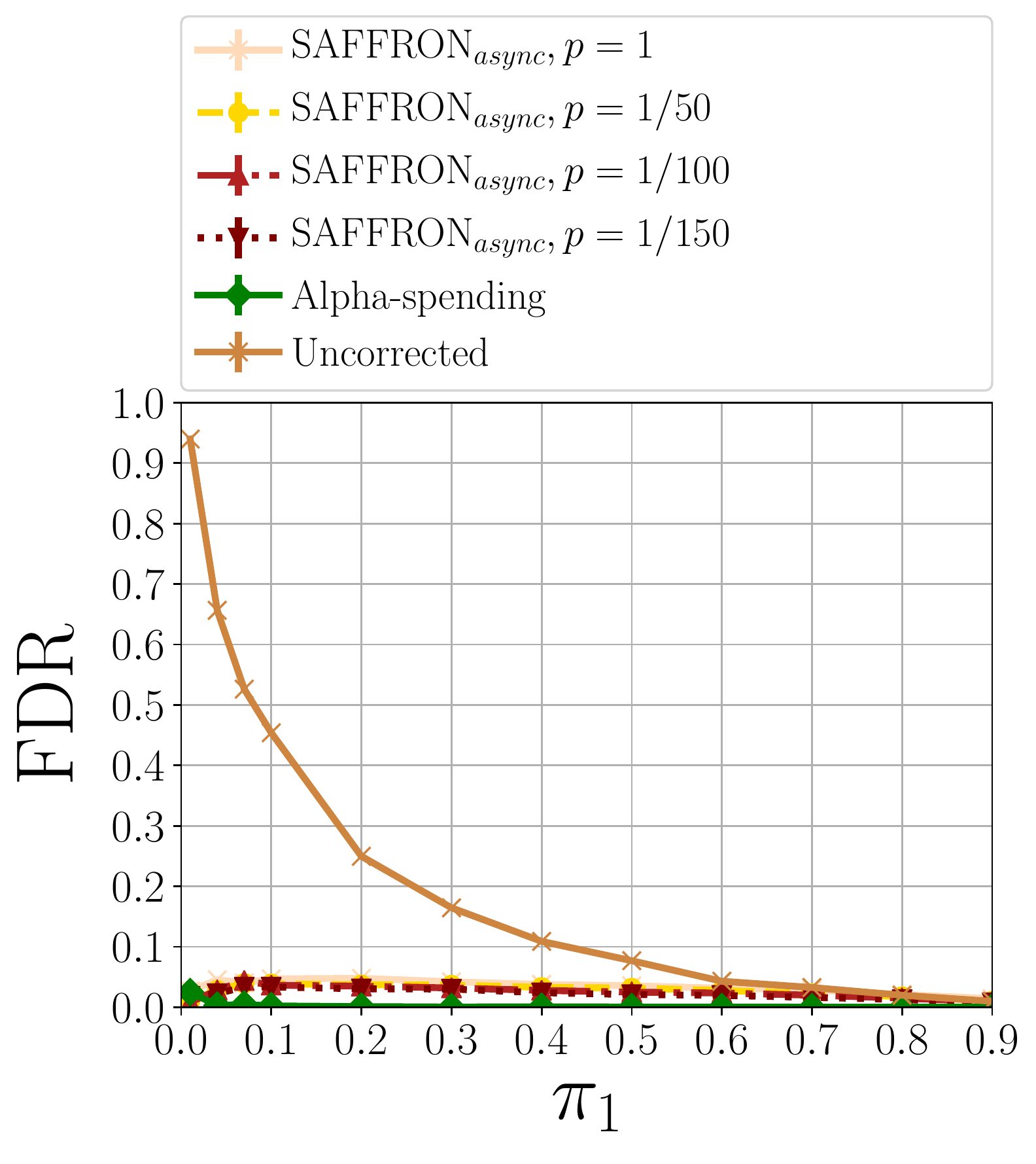}}
\caption{Power and FDR of $\lordasync$ and $\saffasync$ with varying the parameter of asynchrony $p$ of the tests. In all five runs $\lordasync$ and $\saffasync$ have the same parameters ($\{\gamma_j\}_{j=1}^\infty, W_0$). The mean of observations under the alternative is $N(0,2\log(M))$.} 
\label{fig:async2}
\end{figure}

First we show the results of simulated asynchronous tests, in which the $p$-values are independent. At each time step, the test duration is sampled from a geometric distribution with parameter $p$: $E_j\sim j - 1 + \text{Geom}(p)$ for all $j$. This implies that $p=1$ yields the fully synchronous setting, while, as $p$ gets smaller, the expectation of the test duration grows larger, hence the procedure gets more asynchronous, and consequently less powerful. \figref{async} shows numerically how changing $p$ affects the achieved power of $\lordasync$ and $\saffasync$, across different non-null proportions $\pi_1$, when the mean of the alternative is fixed as $\mu_c = 3$. \figref{async2} plots power and FDR of $\lordasync$ and $\saffasync$ against $\pi_1$ for normally distributed means, showing a more gradual change in performance with the increase of asynchrony.

\subsection{Varying the lag of dependence}

\begin{figure}[t]
 \centerline{\includegraphics[width=0.25\textwidth]{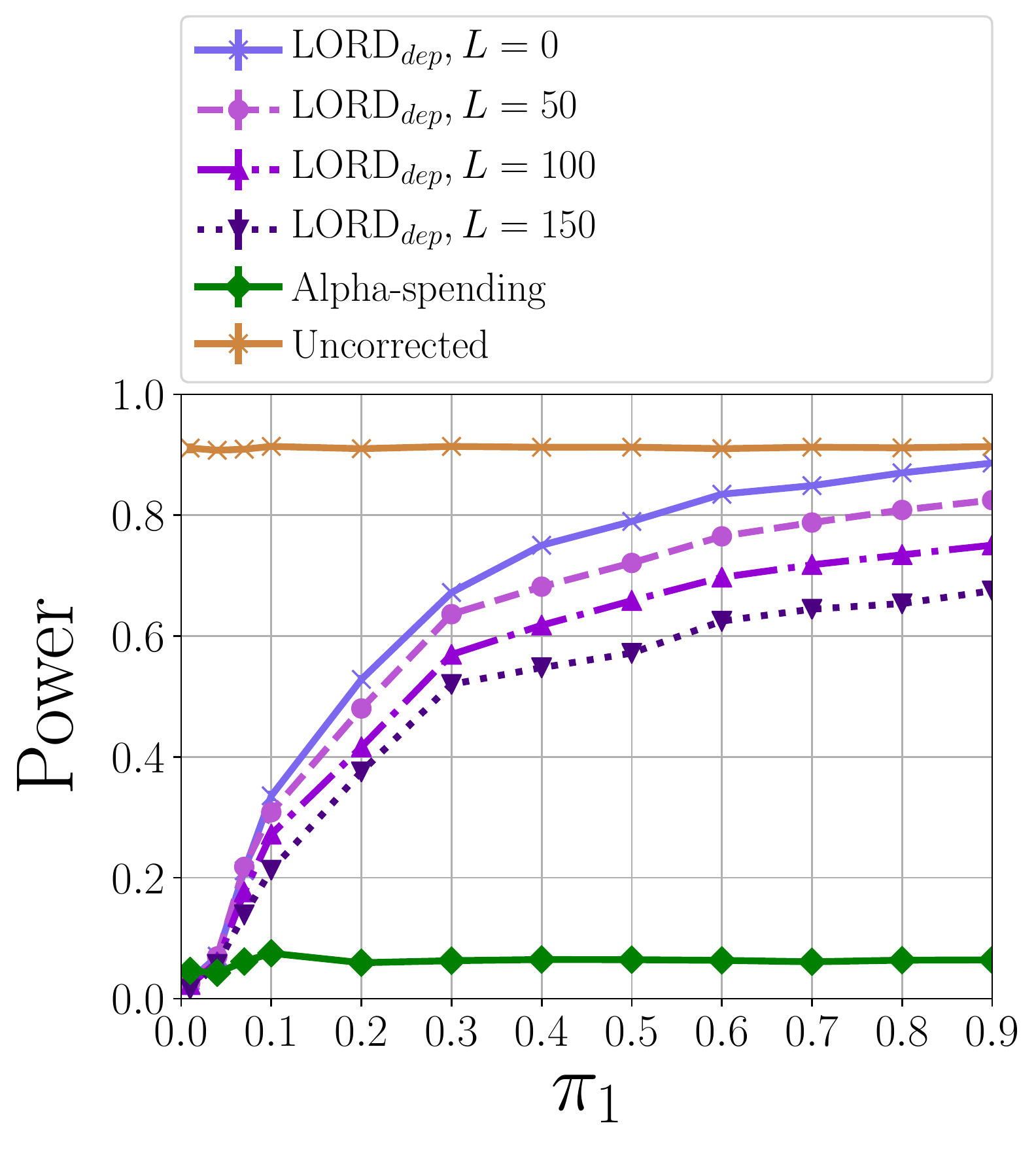}
 \includegraphics[width=0.25\textwidth]{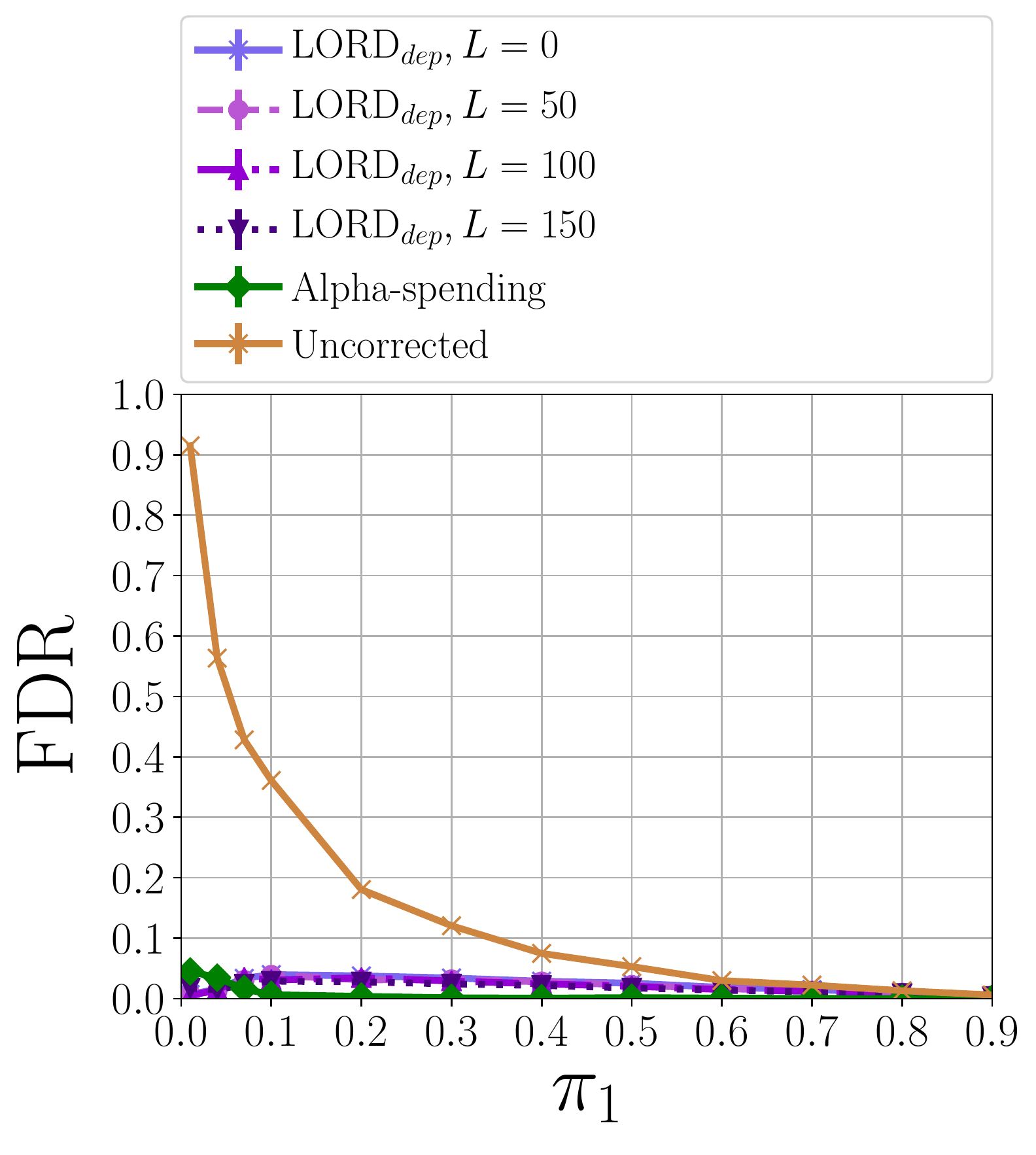}
 \includegraphics[width=0.25\textwidth]{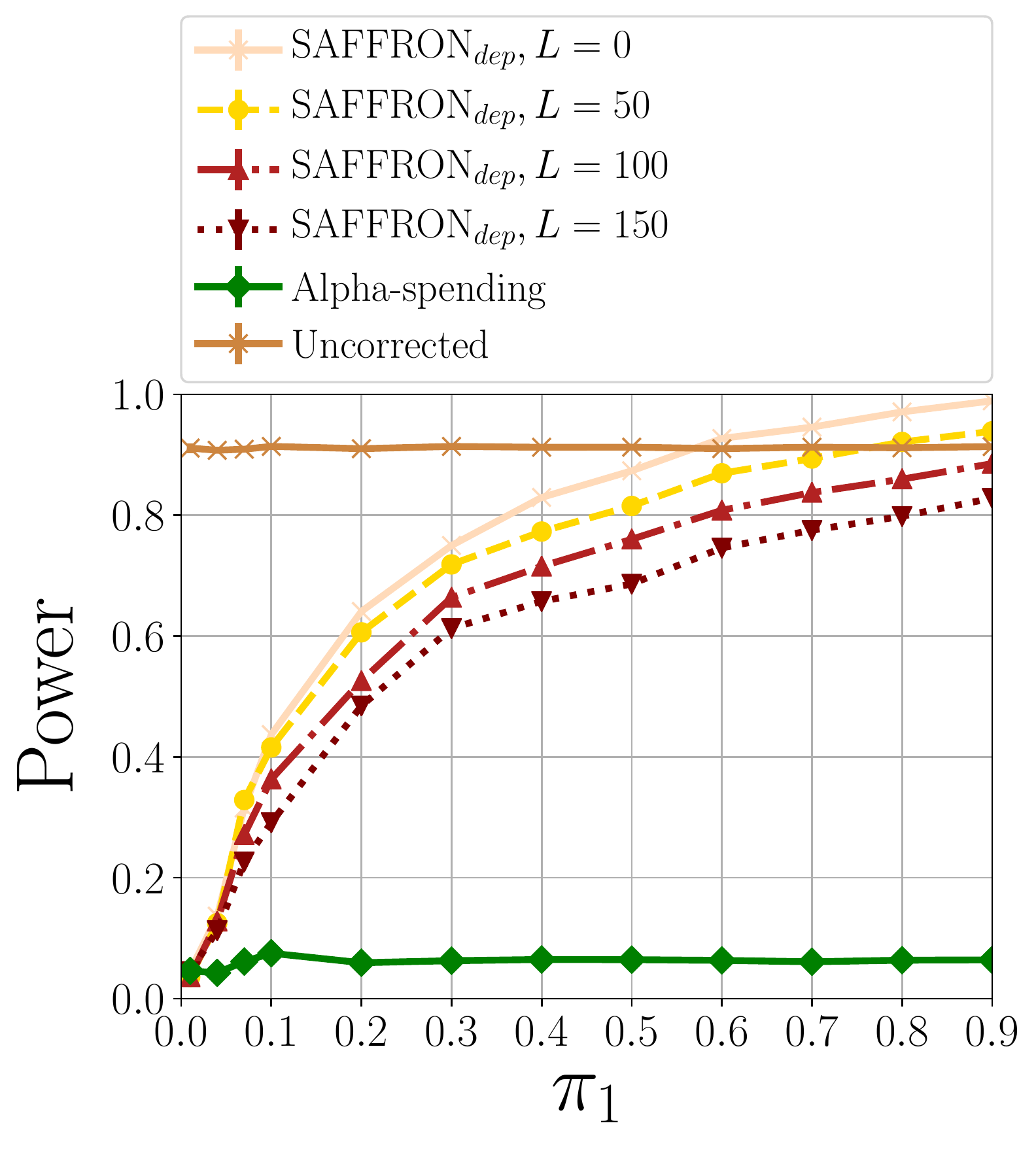}
 \includegraphics[width=0.25\textwidth]{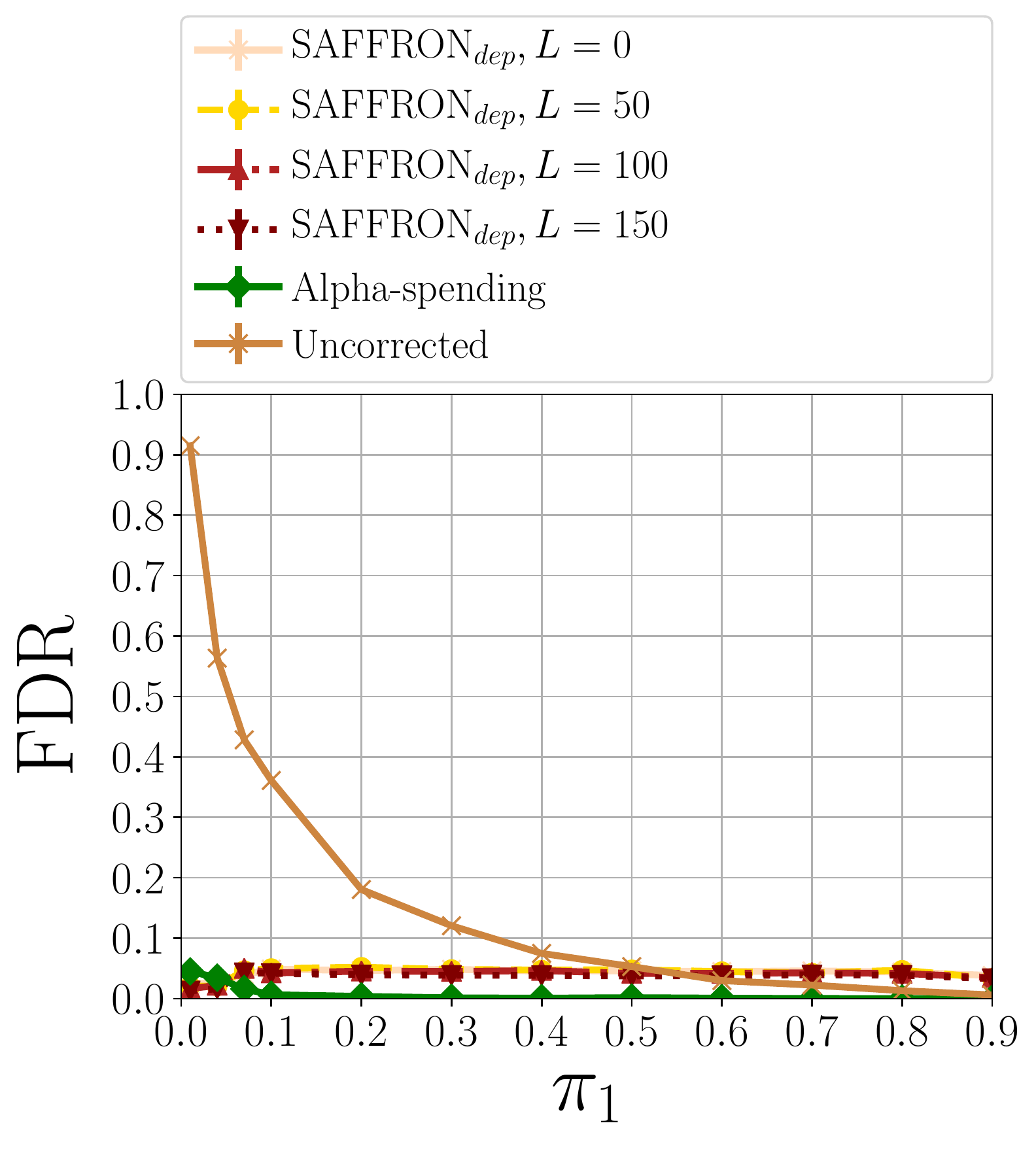}}
 \caption{Power and FDR of $\lordmarkov$ and $\saffmarkov$ with varying the dependence lag $L$ in the $p$-value sequence. In all five runs $\lordmarkov$ and $\saffmarkov$ have the same parameters ($\{\gamma_j\}_{j=1}^\infty, W_0$). The mean of observations under the alternative is a point mass at $\mu_c=3$.} 
 \label{fig:local}
 \end{figure}

\begin{figure}[t]
 \centerline{\includegraphics[width=0.25\textwidth]{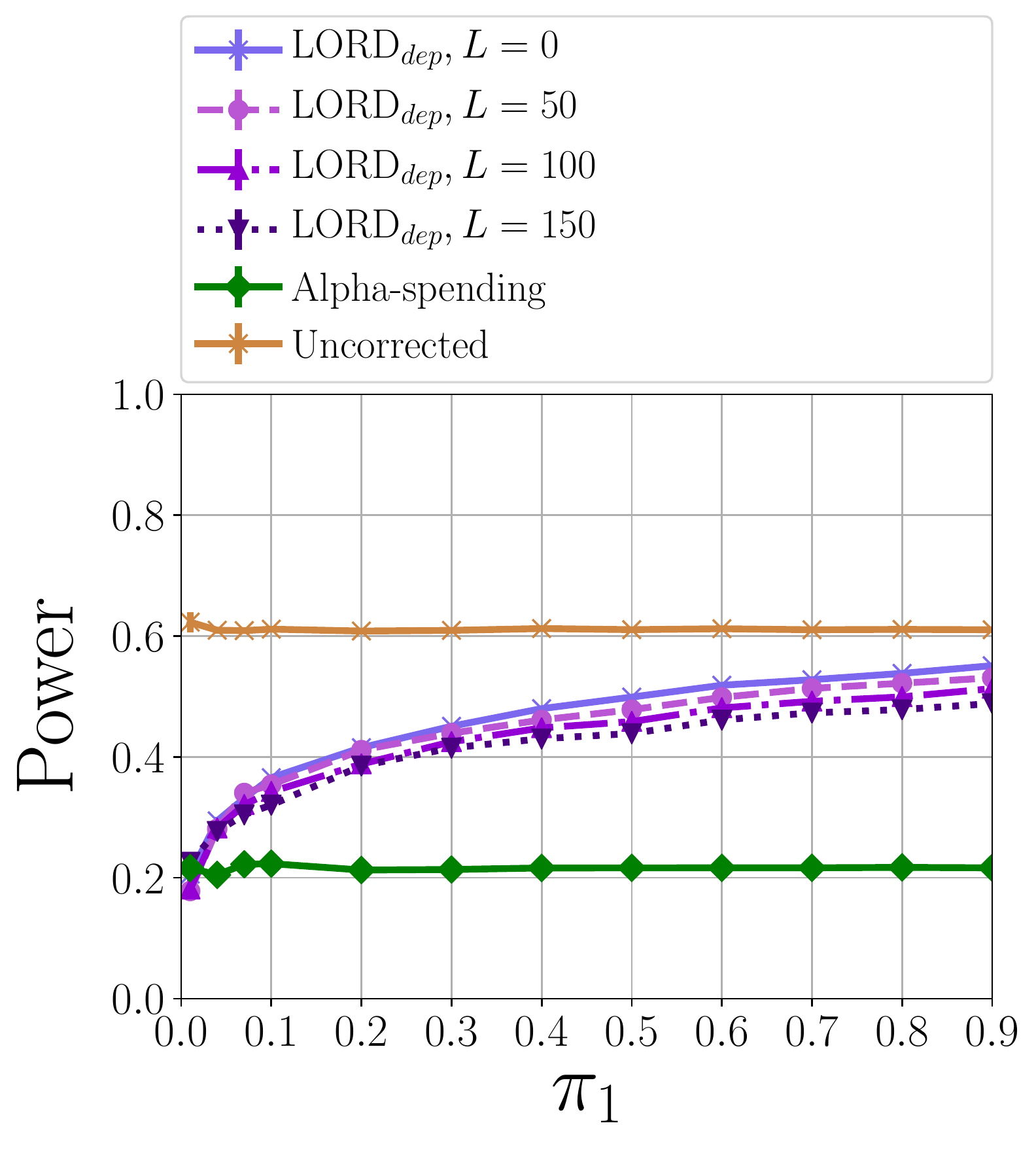}
 \includegraphics[width=0.25\textwidth]{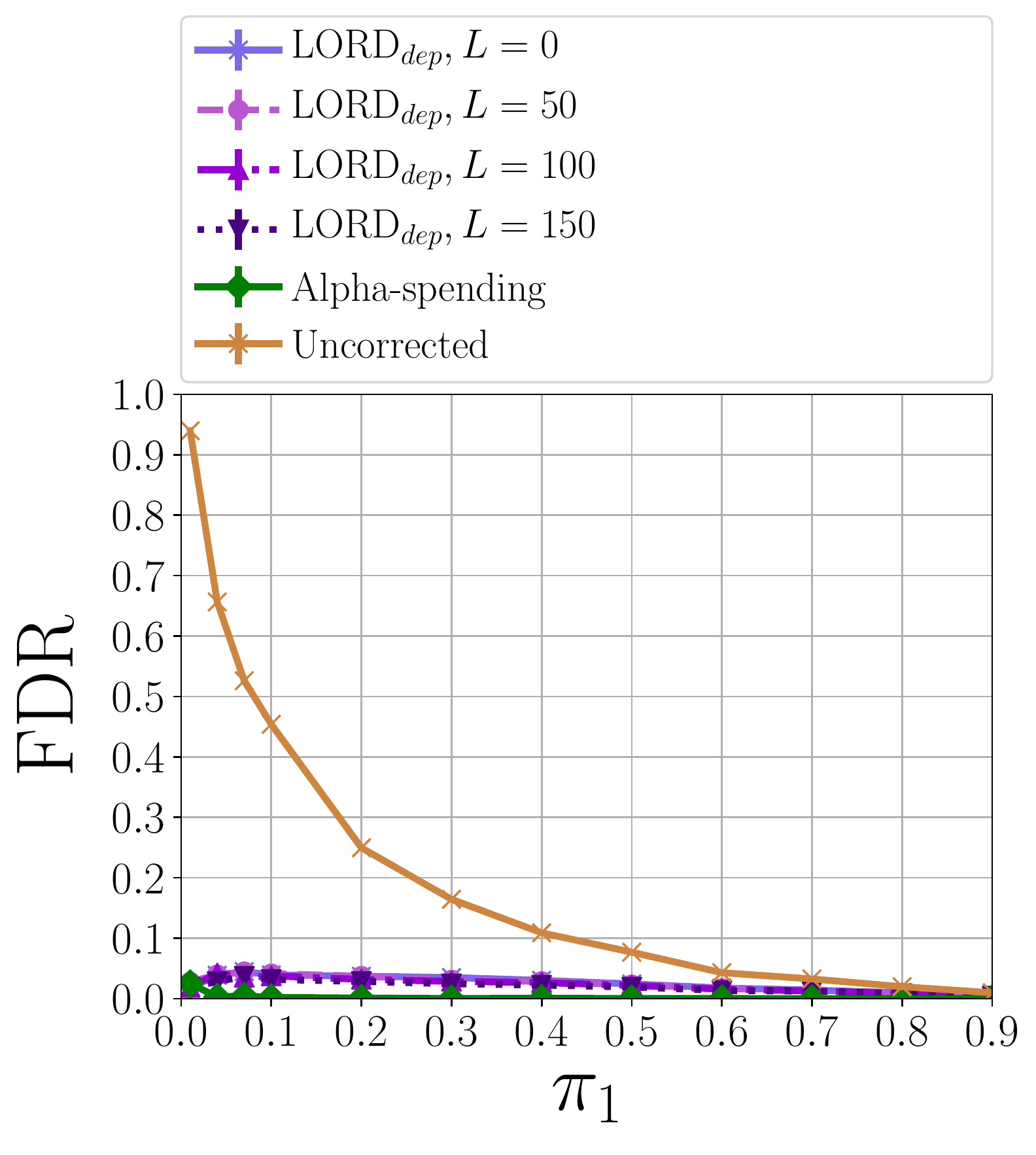}
 \includegraphics[width=0.25\textwidth]{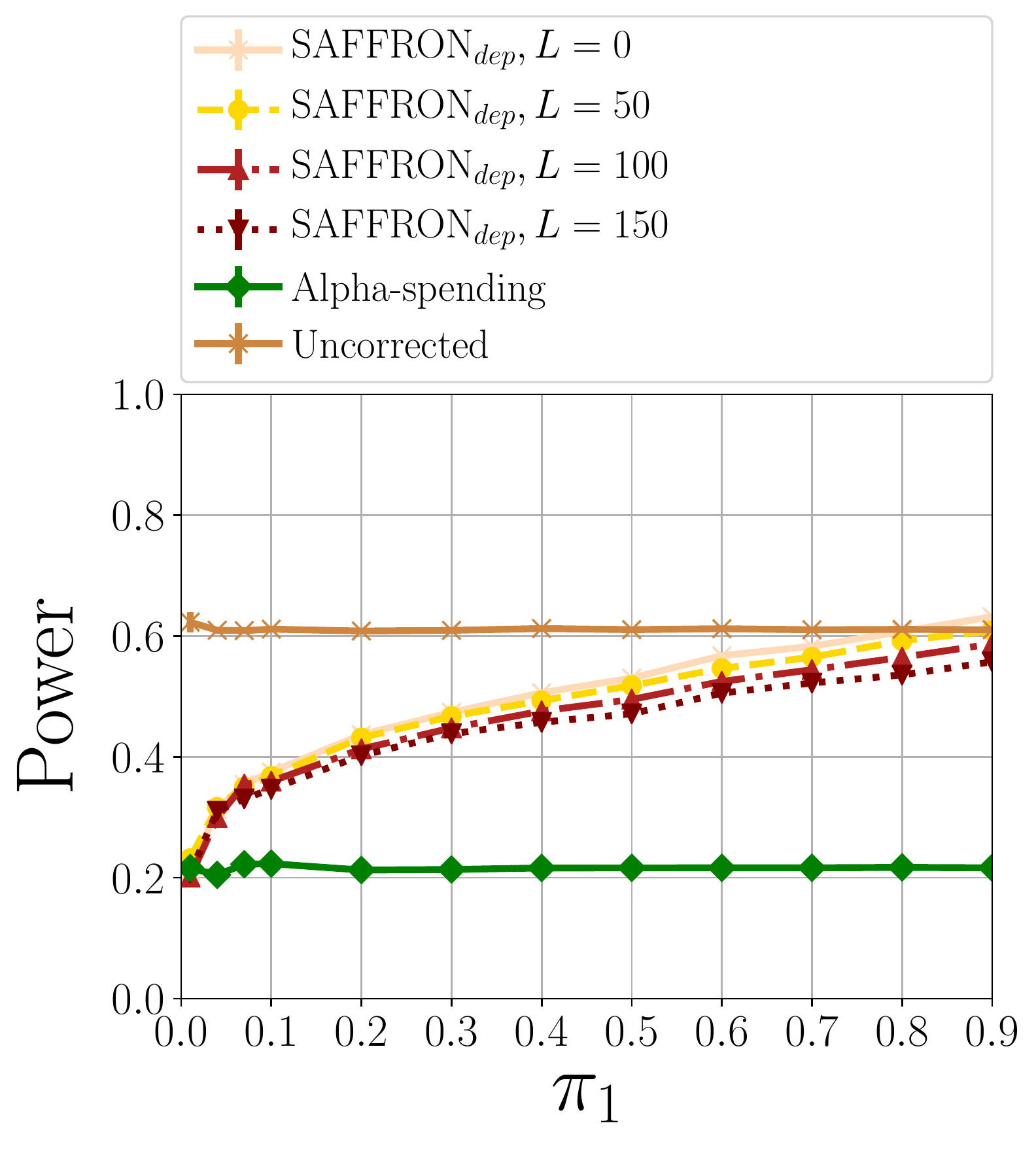}
 \includegraphics[width=0.25\textwidth]{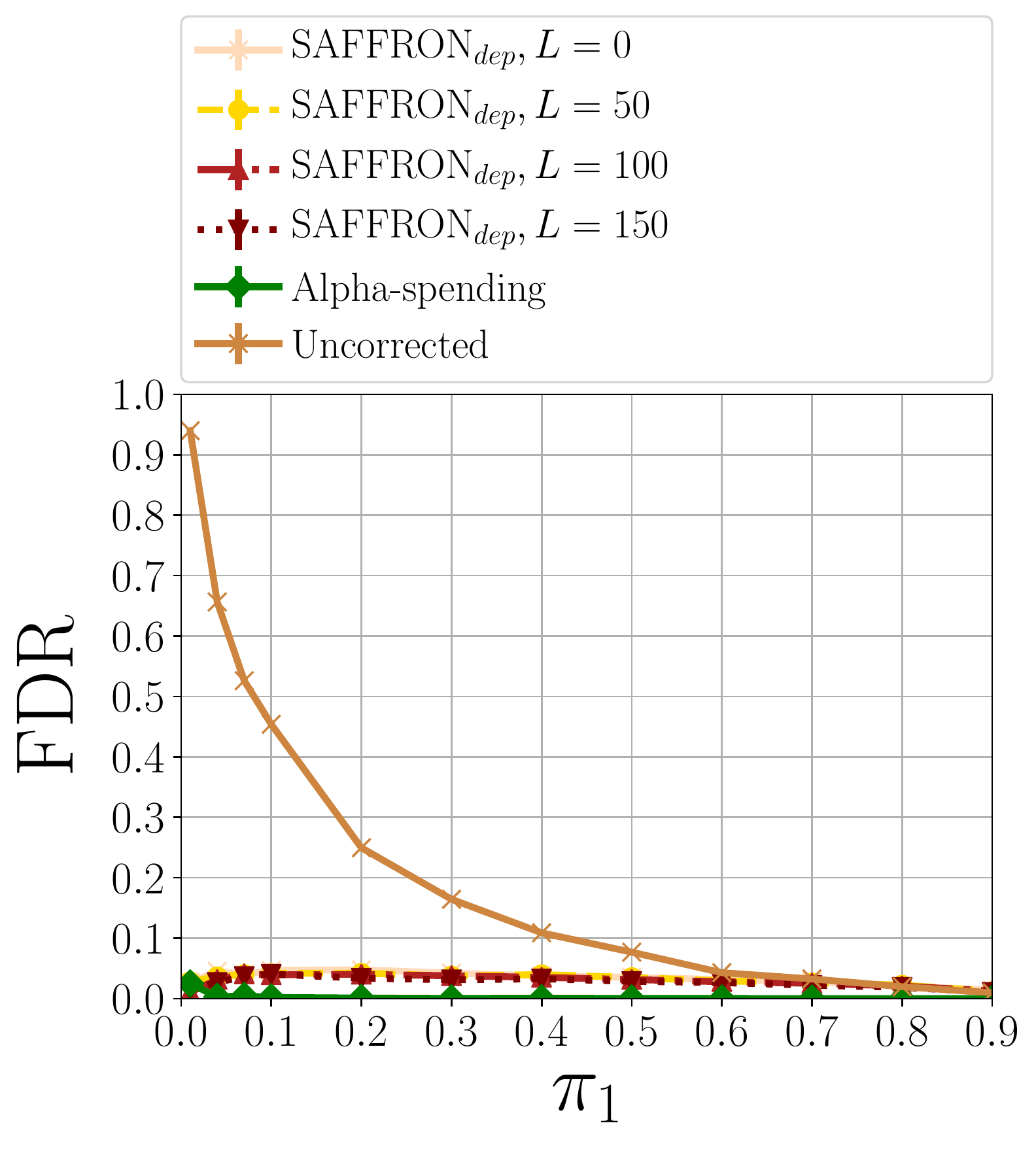}}
 \caption{Power and FDR of $\lordmarkov$ and $\saffmarkov$ with varying the dependence lag $L$ in the $p$-value sequence. In all five runs $\lordmarkov$ and $\saffmarkov$ have the same parameters ($\{\gamma_j\}_{j=1}^\infty, W_0$). The mean of observations under the alternative is $N(0,2\log(M))$.} 
 \label{fig:local2}
 \end{figure}

 The second set of simulations considers synchronous testing of locally dependent $p$-values. We take $L_t$ to be invariant and equal to $L$, which reduces to lagged dependence. We generate an $M$-dimensional vector of Gaussian observations $(Z_1,\dots,Z_M)$, which are marginally distributed according to the model described at the beginning of the section, and have the following $M\times M$ Toeplitz covariance matrix:
 \small
 \begin{equation}
 \label{eqn:toeplitz1}
 \Sigma(M,L,\rho)=
   \begin{bmatrix}
     1 & \rho & \rho^2 & \dots & \rho^L & 0 & \dots & 0 & 0 & 0 \\
     \rho & 1 & \rho & \dots & \rho^{L-1} & \rho^L & \dots & 0 & 0 & 0 \\
     \vdots & & &  \ddots & & & & &\vdots\\
      \vdots & & & & \ddots & & & &\vdots\\
       \vdots & & & &  & \ddots & & &\vdots\\
              \vdots & & & &  &  & \ddots & &\vdots\\
      0 & 0 & 0 & \dots & 0 & 0 & \dots & \rho & 1 & \rho\\
     0 & 0 & 0 & \dots & 0 & 0 & \dots & \rho^2 & \rho & 1
   \end{bmatrix},
 \end{equation}
 \normalsize
  where we set $\rho = 0.5$. \figref{local} compares the power and FDR of $\lordmarkov$ and $\saffmarkov$ under local dependence, when the mean of the observations under the alternative is $\mu_c = 3$ with probability 1. \figref{local2} gives the same comparison when the mean of non-null samples is normally distributed, which yields a slower decrease in performance with increasing the lag.

\subsection{Varying mini-batch sizes}

\begin{figure}[t]
\centerline{\includegraphics[width=0.25\textwidth]{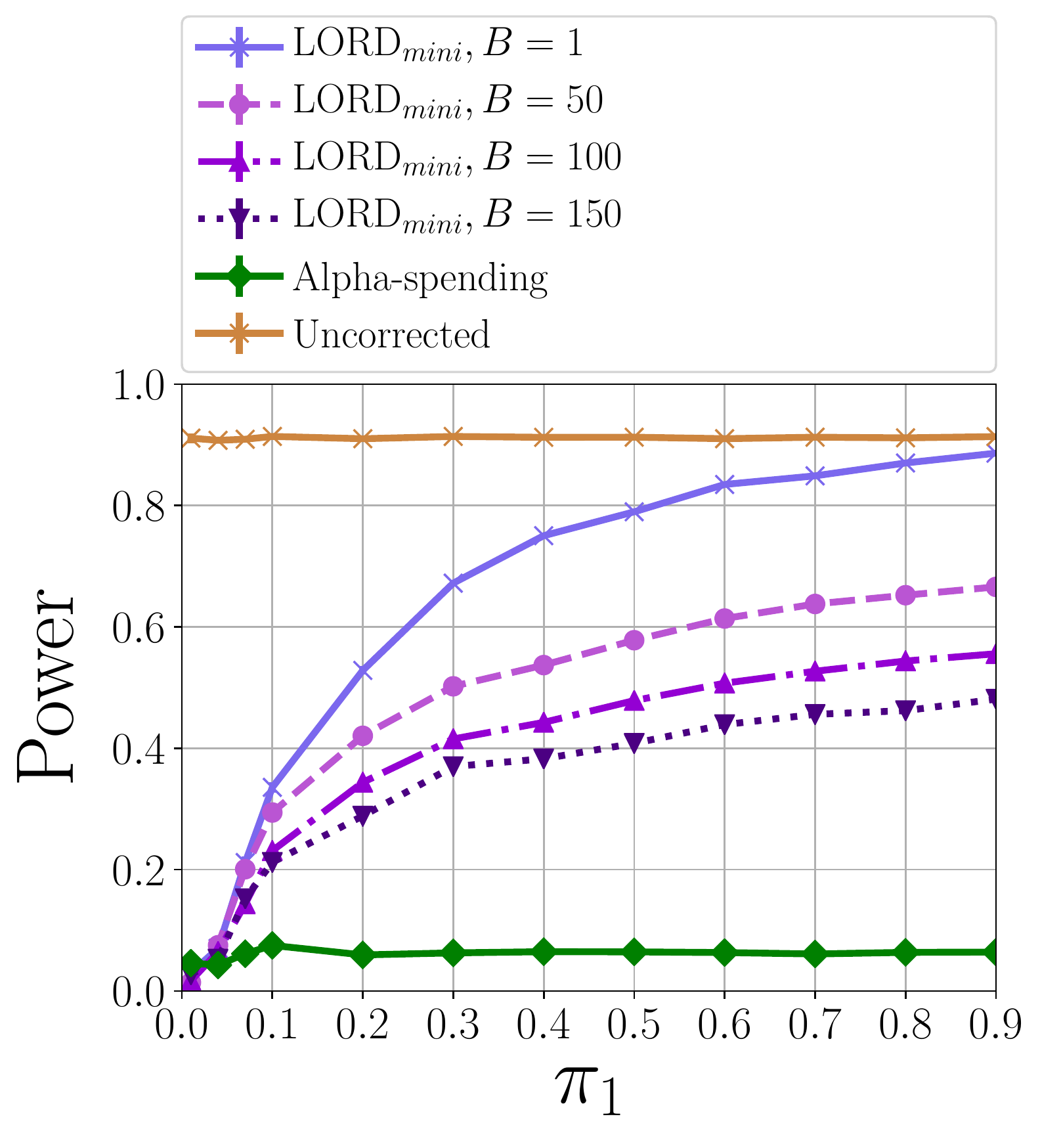}
\includegraphics[width=0.25\textwidth]{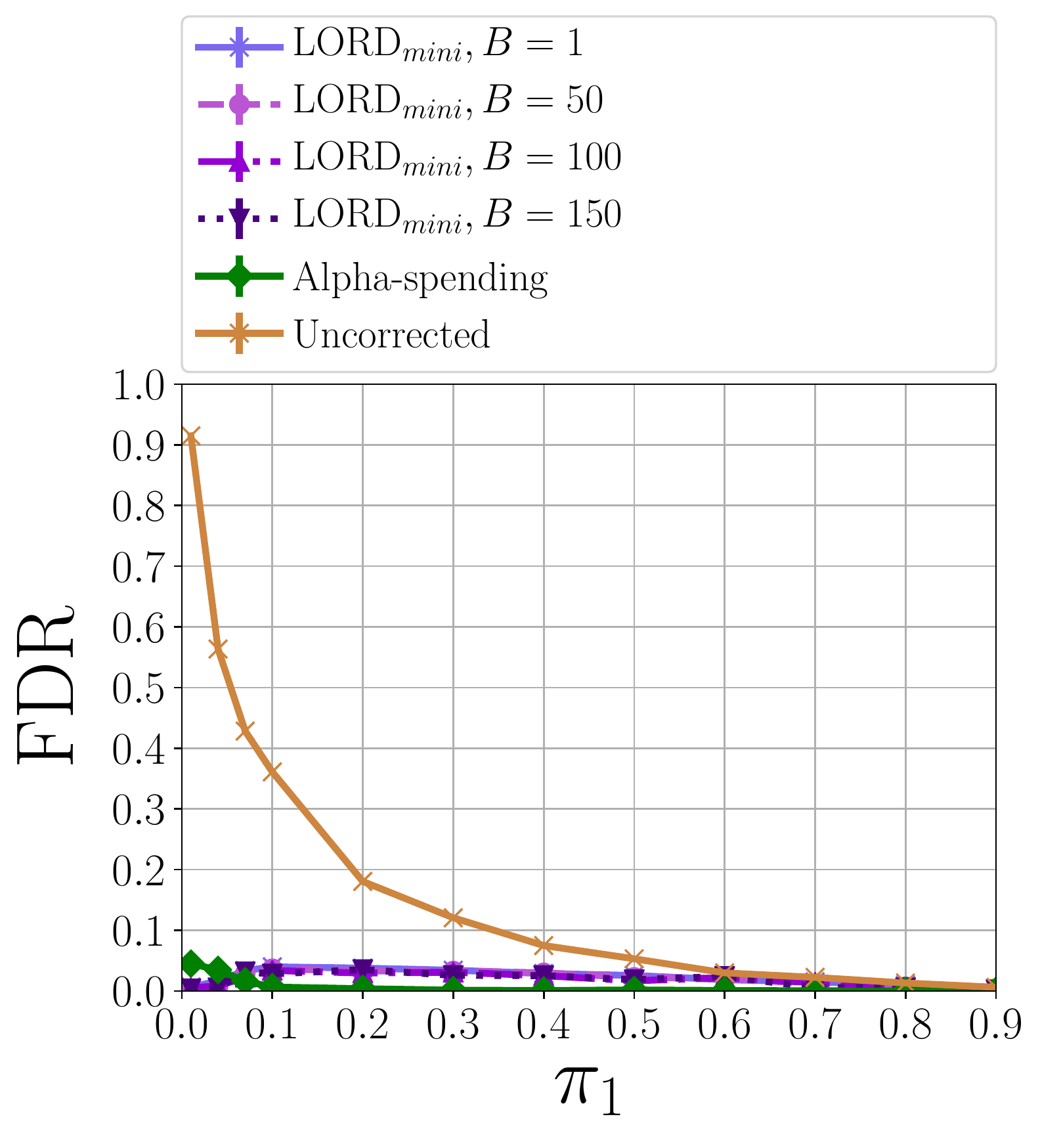}
\includegraphics[width=0.25\textwidth]{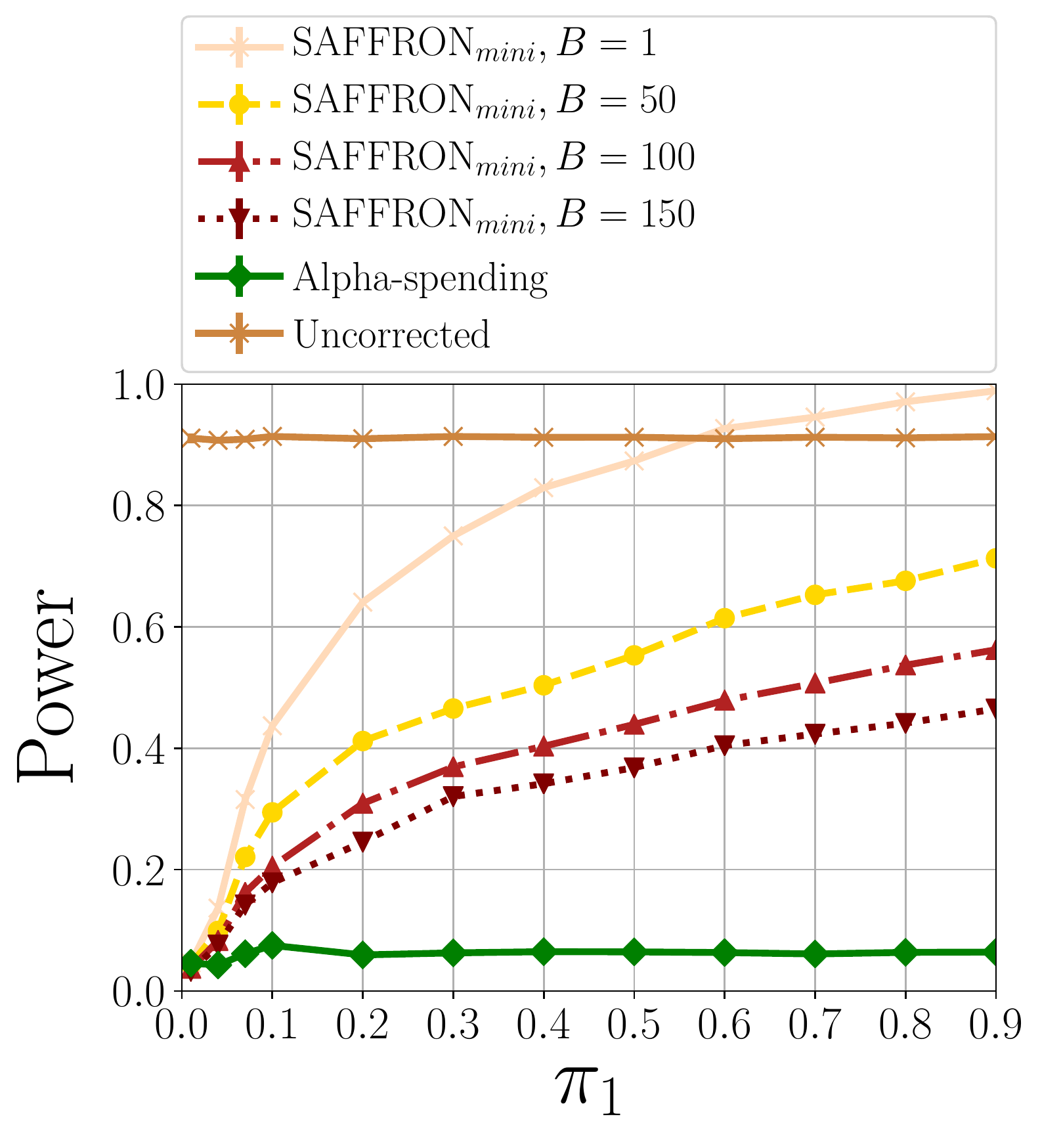}
\includegraphics[width=0.25\textwidth]{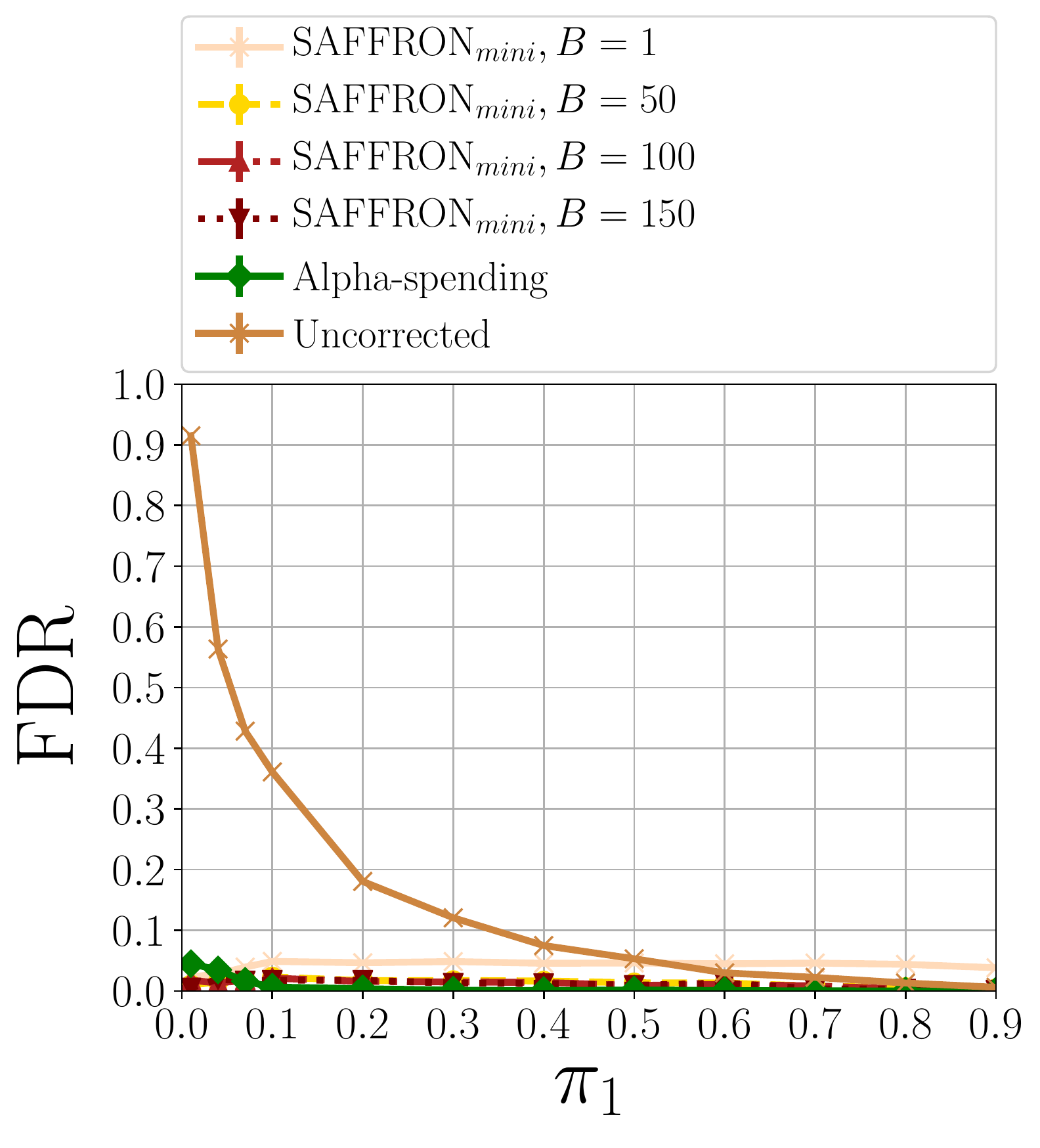}}
\caption{Power and FDR of $\lordmini$ and $\saffmini$ with varying the size of mini-batches. In all five runs $\lordmini$ and $\saffmini$ have the same parameters ($\{\gamma_j\}_{j=1}^\infty, W_0$). The mean of observations under the alternative is a point mass at $\mu_c=3$, and $\rho=0.5$.} 
\label{fig:mini}
\end{figure}

\begin{figure}[t]
\centerline{\includegraphics[width=0.25\textwidth]{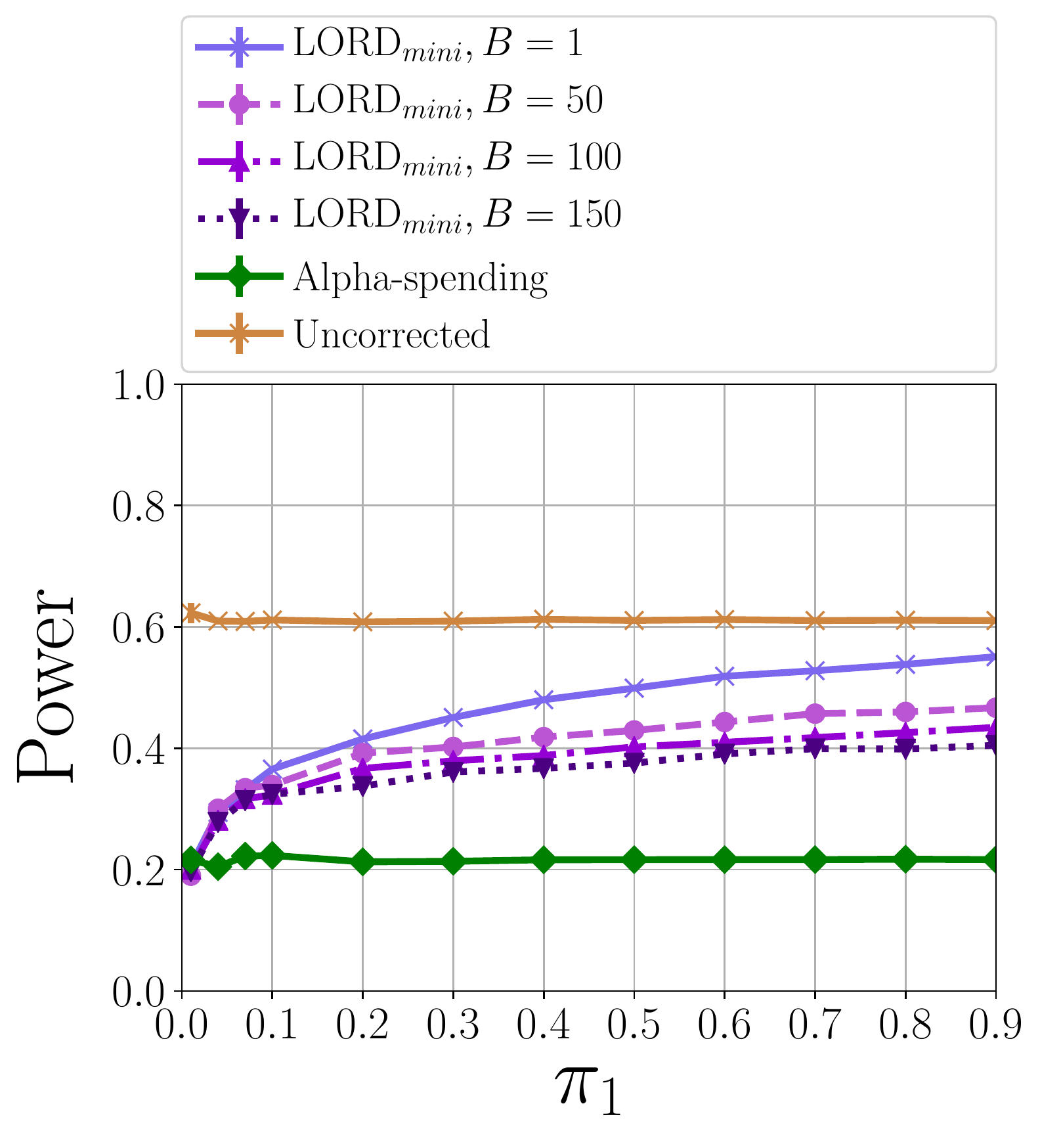}
\includegraphics[width=0.25\textwidth]{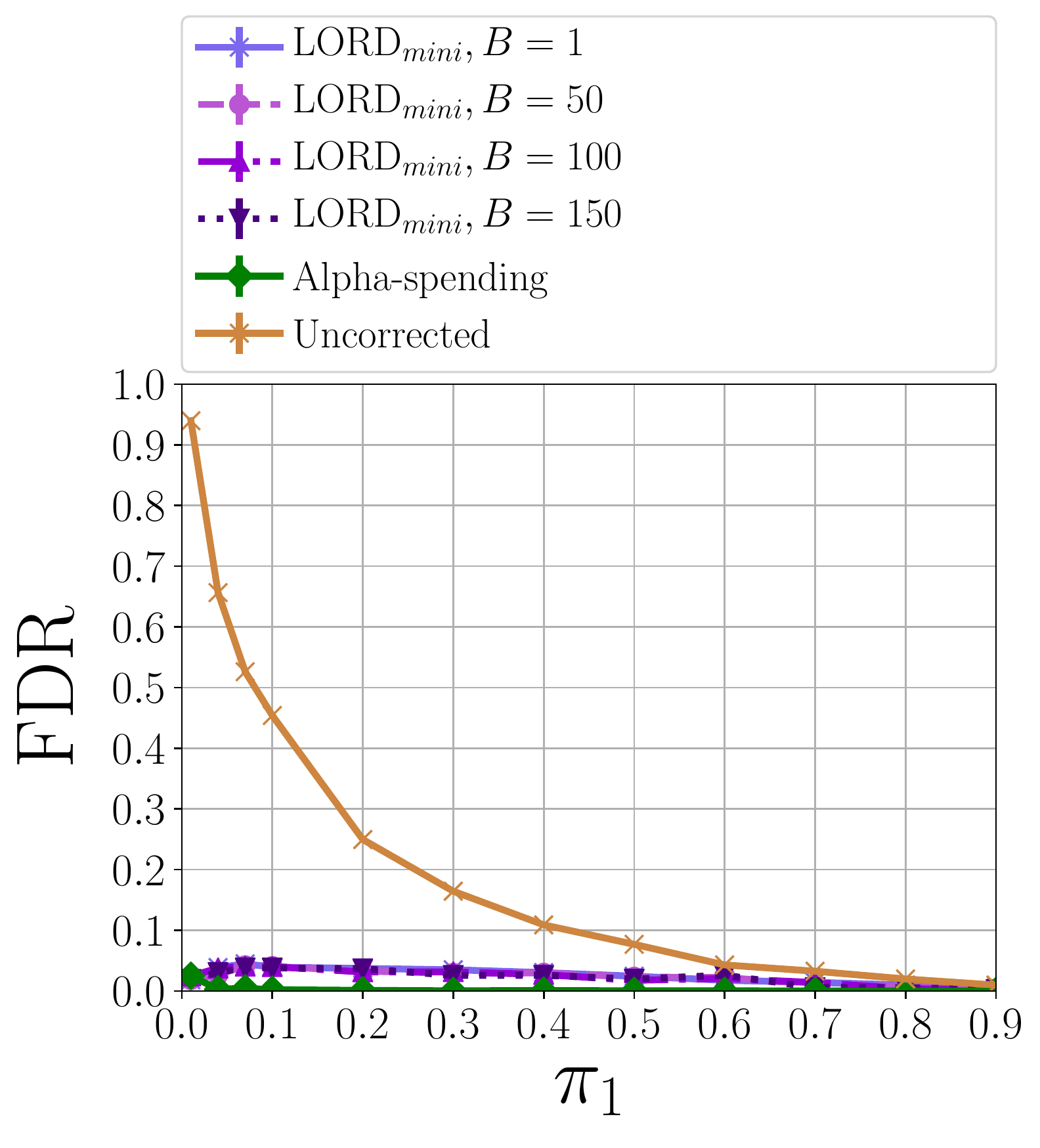}
\includegraphics[width=0.25\textwidth]{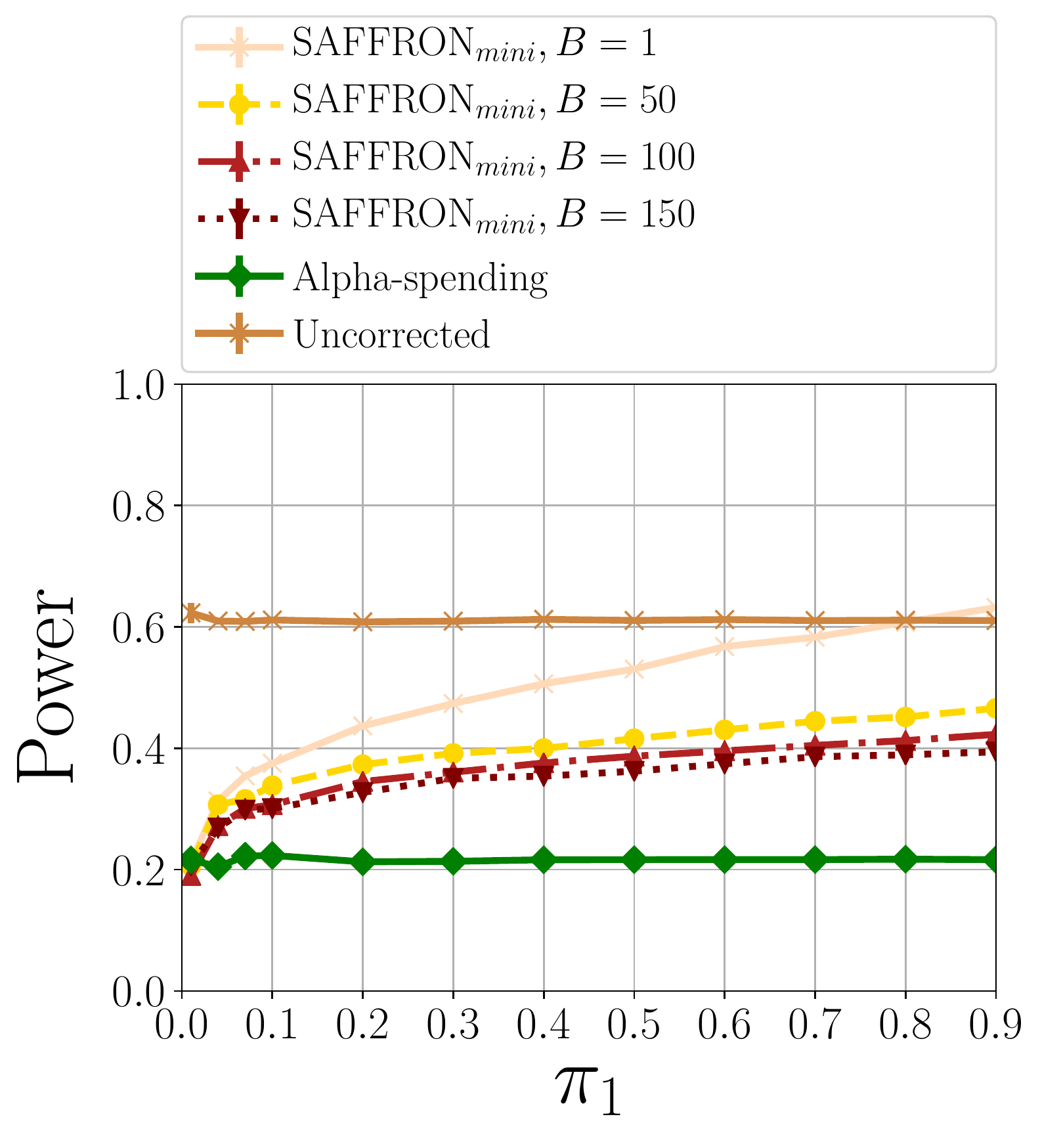}
\includegraphics[width=0.25\textwidth]{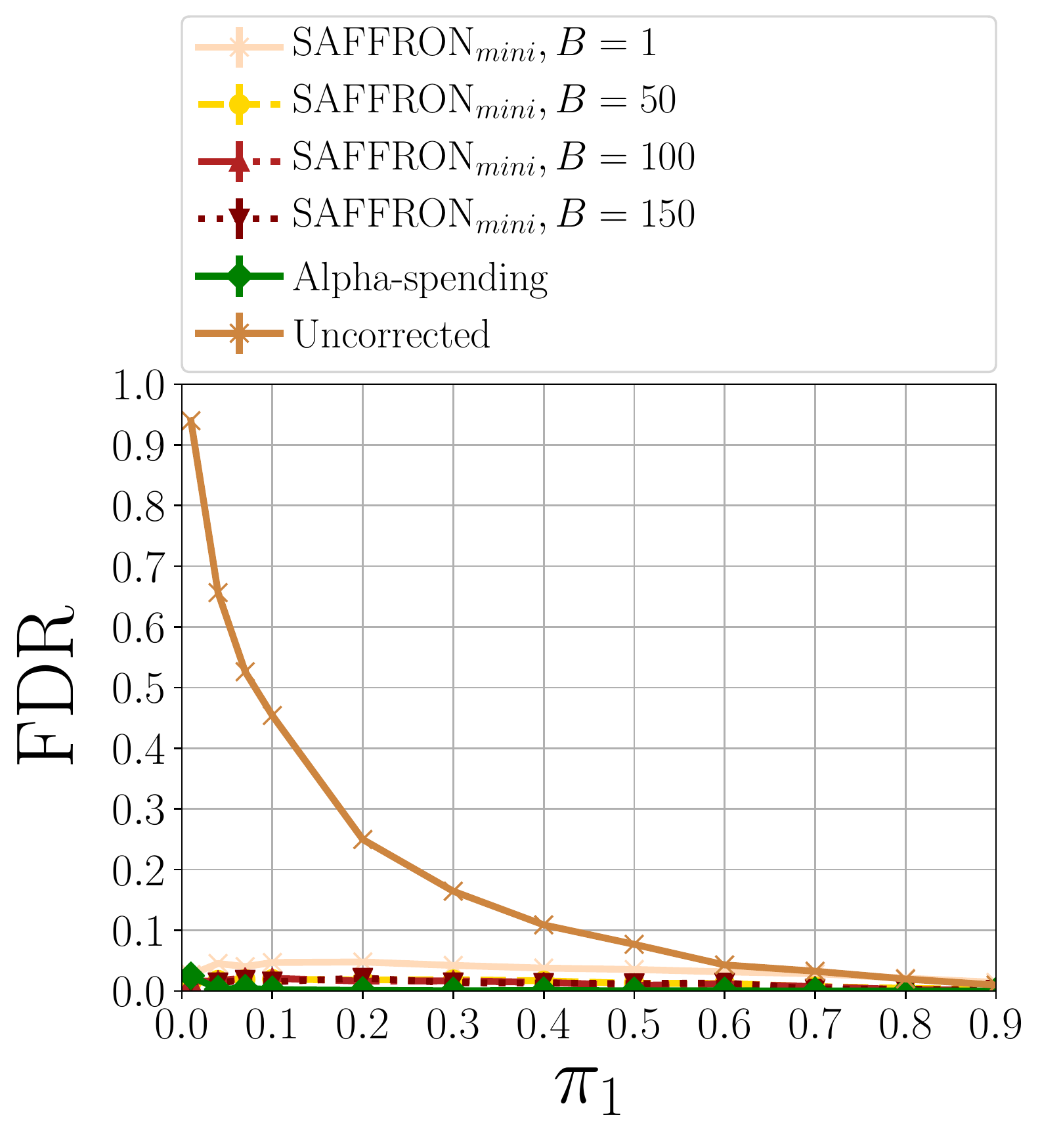}}
\caption{Power and FDR of $\lordmini$ and $\saffmini$ with varying the size of mini-batches. In all five runs $\lordmini$ and $\saffmini$ have the same parameters ($\{\gamma_j\}_{j=1}^\infty, W_0$). The mean of observations under the alternative is $N(0,2\log(M))$, and $\rho=0.5$.} 
\label{fig:mini2}
\end{figure}

Here we analyze the change in performance of $\lordmini$ and $\saffmini$ when the size of mini-batches varies. We fix the batch size $n_b\equiv n$ for all batches $b$. Within each batch tests are performed asynchronously, and all $p$-values within the same batch are dependent. In particular, they follow a multivariate normal distribution, where the marginal distributions are as described at the beginning of this section, and the covariance matrix is the Toeplitz matrix $\Sigma(n,n-1,\rho)$ \eqnref{toeplitz1}, where we fix $\rho = 0.5$.
Dependent $p$-values come in ``blocks'' of size $n$, implying that any two $p$-values belonging to two different batches are independent. \figref{mini} compares the power and FDR of $\lordmini$ and $\saffmini$ for different batch sizes when the mean of the non-null $Z_i$ is a point mass at $\mu_c = 3$, and \figref{mini2} plots the same comparison when the mean of the non-null observations is normally distributed.

\subsection{Comparison with LORD under dependence}

\begin{figure}[t]
\centerline{\includegraphics[width=0.25\textwidth]{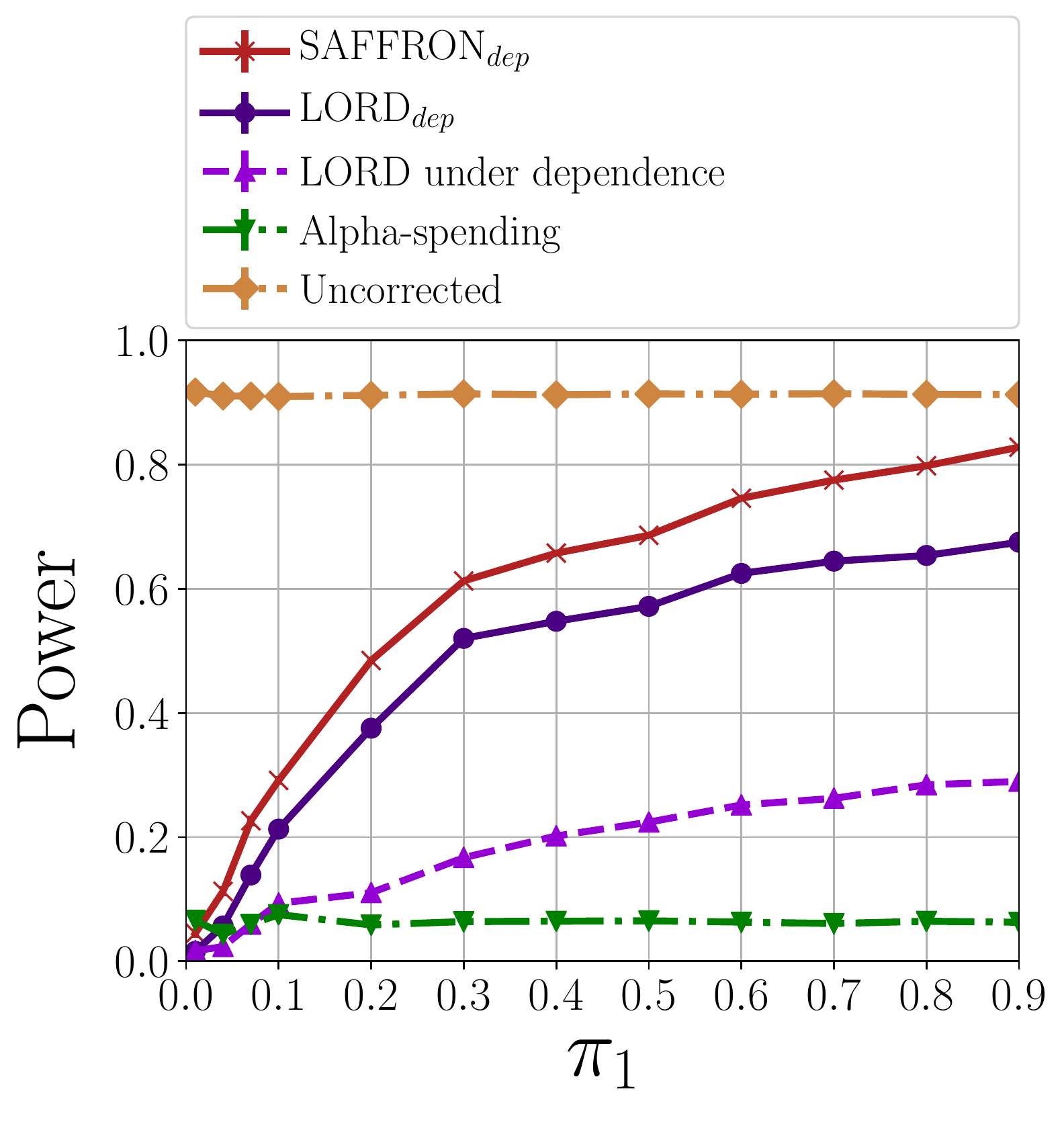}
\includegraphics[width=0.25\textwidth]{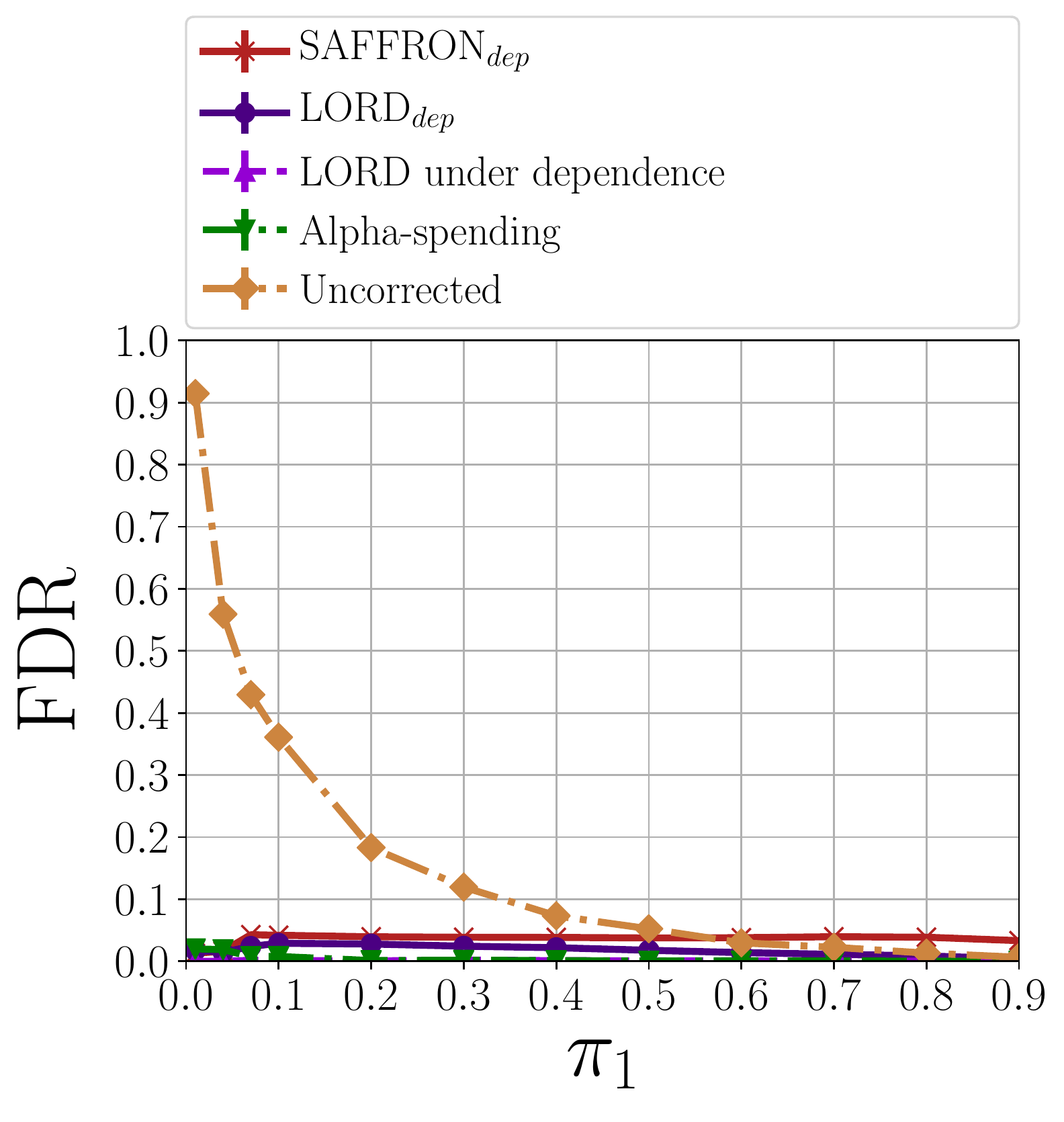}
\includegraphics[width=0.25\textwidth]{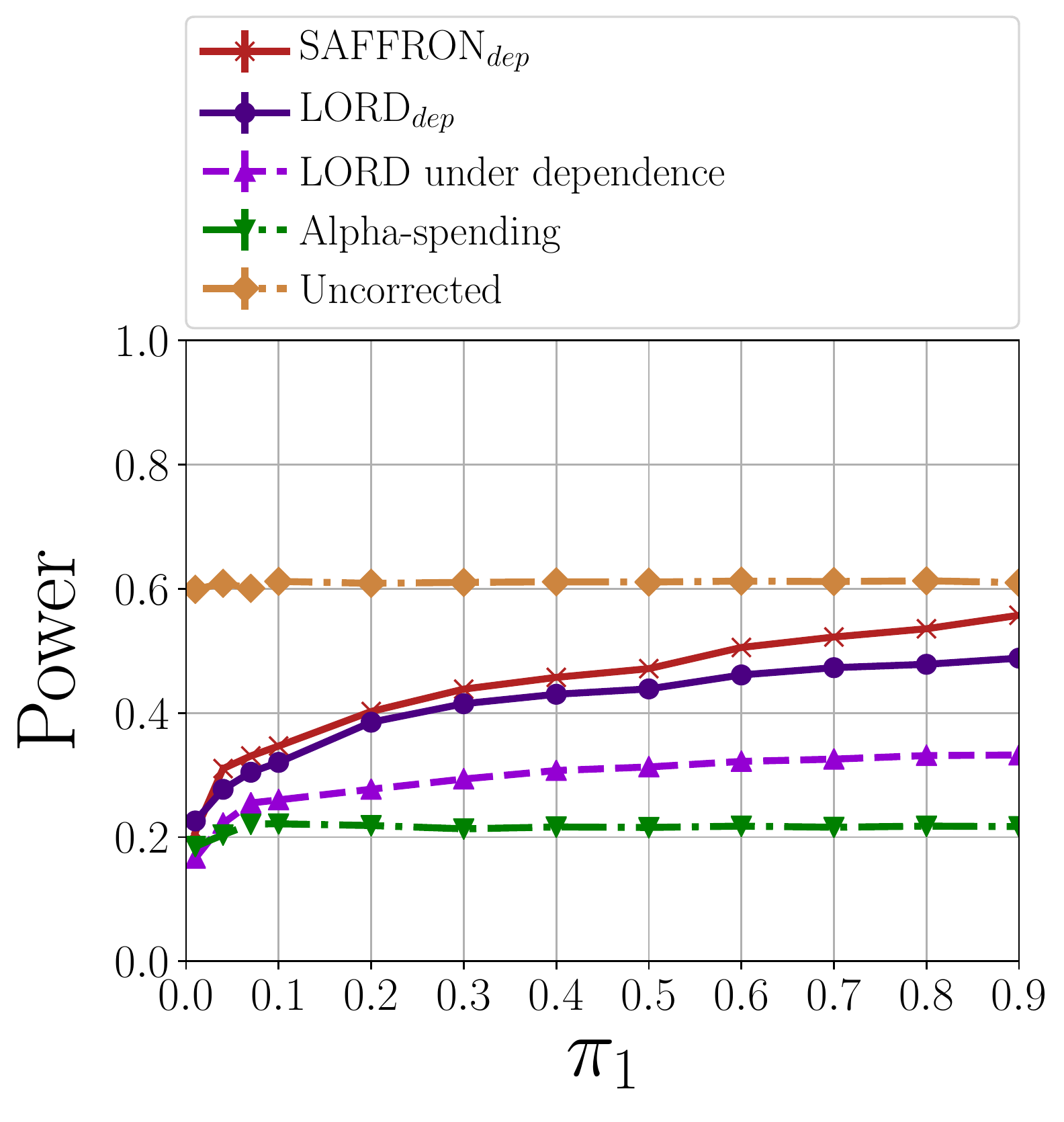}
\includegraphics[width=0.25\textwidth]{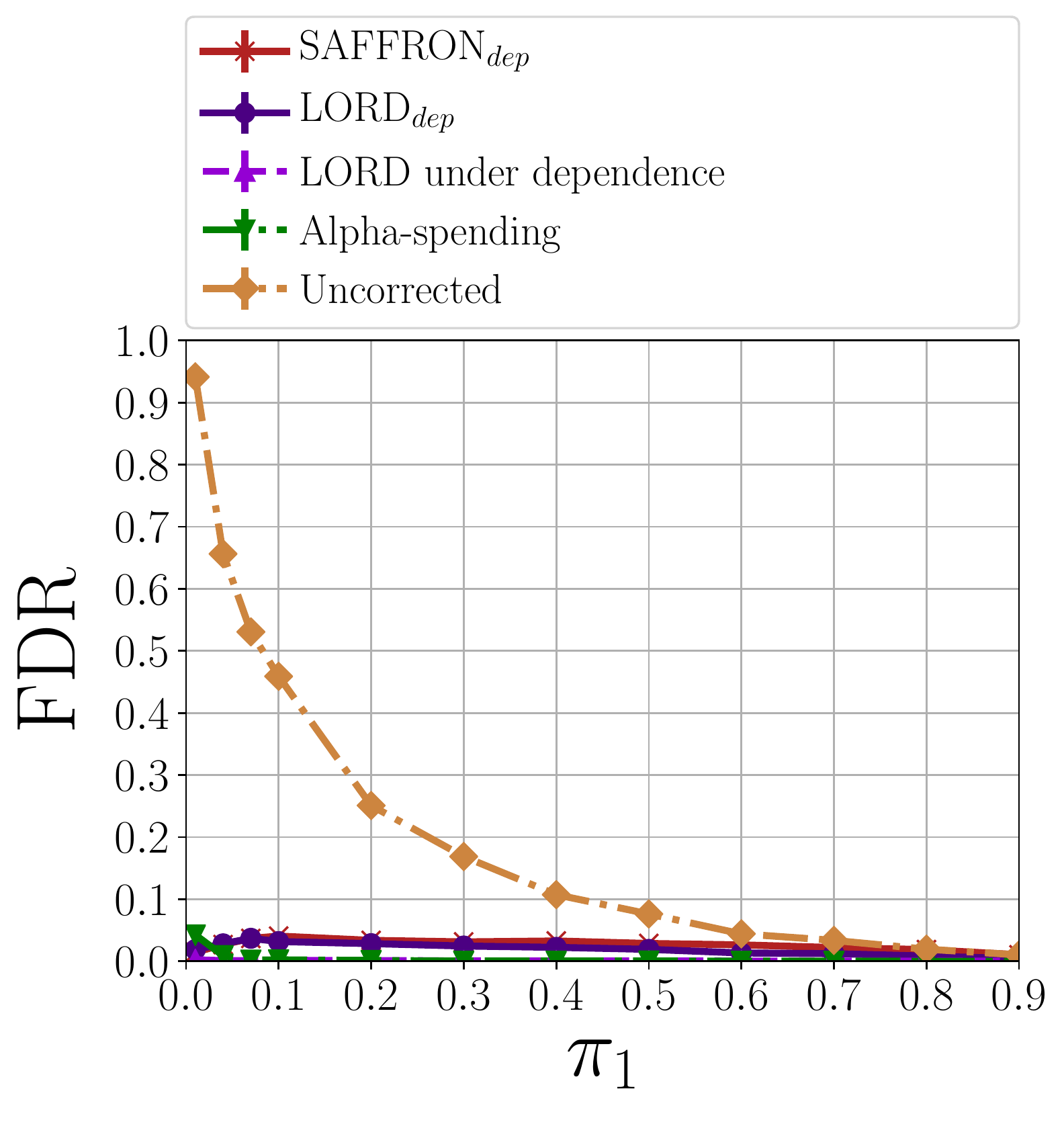}}
\caption{Power and FDR of $\saffmarkov$, $\lordmarkov$, LORD under dependence, and alpha-spending. The decay of test levels in alpha-spending and discount sequence $\{\xi_j\}_{j=1}^\infty$ act according to the sequence $\{\gamma_j\}_{j=1}^\infty$ used for $\saffmarkov$ and $\lordmarkov$. On the left two plots, the mean of observations under the alternative is a point mass at $\mu_c=3$, while on the right two plots, it is distributed as $N(0,2\log(M))$. We fix parameters $\rho=0.5$ and $L=150$.} 
\label{fig:comparison}
\end{figure}

The final set of experiments contrasts $\lordmarkov$ and $\saffmarkov$ to the original LORD algorithm under dependence. The latter controls FDR under arbitrary dependence, however, as mentioned earlier, this entails a similar update to alpha-investing; more precisely, the test levels $\alpha_j^{\text{indep}}$ of LORD under independence have to be discounted by a convergent sequence $\{\xi_j\}_{j=1}^\infty$, resulting in new test levels $\alpha_j\defn\xi_j\alpha_j^{\text{indep}}$, which essentially diminishes the effect of $\alpha_j^{\text{indep}}$ earning extra budget through discoveries. We generate the $p$-value sequence using the same scheme as in Subsection 7.2; they are computed from Gaussian observations with covariance matrix $\Sigma(M,L,\rho)$ \eqnref{toeplitz1}, where we fix $\rho=0.5$ and $L=150$. By construction, this sequence is only locally dependent, which implies that the application of our algorithms comes with provable guarantees. \figref{comparison} compares the power and FDR of $\saffmarkov$, $\lordmarkov$, LORD under dependence, and alpha-spending when the mean of the non-null $Z_i$ is a point mass at $\mu_c = 3$ (left), as well as in the setting with a normally distributed mean under the alternative (right).

\section{Discussion}
\label{sec:summary}

We have presented a unified framework for the design and analysis of online FDR procedures for asynchronous testing, as well as testing locally dependent $p$-values. Our framework reposes on the concept of ``conflict sets,'' and we show the value of this concept for the study of both asynchronous testing and local dependence and for their combination. We derive two specific procedures that make use of conflict sets to yield algorithms that provide online mFDR and FDR control.


Several technical questions remain open for future work. While we have shown strict FDR control of our asynchronous procedures under independence, it is still unclear how to prove their FDR control under local dependence. We believe that it might also be possible to prove FDR control of uncorrected LORD under positive dependence, similarly to how we proved validity of the plain LOND algorithm under positive dependence in Section 6. Finally, it would be of great interest to obtain strict FDR control at stopping times, a problem that remains open even under independence of $p$-values. In mFDR control, this proof relies on a martingale-like argument which decouples the numerator and denominator of the FDP. The expected numerator increments, conditional on the information from past tests, are then controlled by invoking super-uniformity. When false rejection indicators are coupled with the FDP denominator, however, it is less clear how to invoke conditional super-uniformity. This is a non-trivial step even when analyzing FDR at fixed times, as witnessed by many ``super-uniformity lemmas'' in the literature \citep{javanmard2016online, RYWJ17, ramdas2018saffron}.

\newpage

\appendix
\section{Deferred proofs}

\subsection{Proof of \propref{oracleconflict}}
\label{pf:oracleconflict}

Fix a time step $t\in\N$. By this time, exactly $t$ tests have started, and hence at most those $t$ decisions are known. Therefore, by linearity of expectation:
\begin{align*}
\EE{|\V(t)|} &= \EE{\sum_{E_j \leq t, j \in \nulls}
\One{P_j \leq \alpha_j}} \leq  \sum_{j \leq t, j \in \nulls}
\EE{\One{P_j \leq \alpha_j}}.
\end{align*}
Applying the law of iterated expectations by conditioning on $\F^{-\X^{E_j}}$ for each term, we obtain:
\begin{align*}
    \sum_{j \leq t, j \in \nulls} \EE{\One{P_j \leq \alpha_j}} &= \sum_{j \leq t, j \in \nulls} \EE{\EEst{\One{P_j \leq \alpha_j}}{\F^{-\X^{E_j}}}} \leq \sum_{j \leq t, j \in \nulls} \EE{\alpha_j},
\end{align*}
which follows due to measurability of $\alpha_j$ with respect to $\F^{-\X^j}\subseteq \F^{-\X^{E_j}}$, and the super-uniformity assumption. If we assume $\fdp^*(t) \defn \frac{\sum_{j\leq t,j\in\nulls}\alpha_j}{(\sum_{E_j\leq t}R_j)\vee 1} \leq \alpha$, then it follows that:
\begin{align*}
\sum_{j \leq t, j \in \nulls} \EE{\alpha_j}=\EE{\sum_{j \leq t, j \in
    \nulls} \alpha_j}\leq \alpha \EE{\left(\sum_{E_j\leq t}R_j\right)\vee 1}= \alpha \EE{|\cR(t)|\vee 1},
\end{align*}
which follows by linearity of expectation and the assumption on
$\fdp^*(t)$. Rearranging yields the inequality
$\mfdr(t) \defn \frac{\EE{|\V(t)|}}{\EE{|\cR(t)|\vee 1}} \leq
\alpha$, which completes the proof.

\subsection{Proof of \thmref{mfdr-conflict}}
\label{pf:mfdr-conflict}

As stated before, the guarantees for LORD* follow directly from \propref{oracleconflict}, after observing that $\fdp^*_{\text{conf}}(t)\leq \widehat \fdp_{\text{LORD*}}(t)\leq\alpha$ holds almost surely for all $t\in\N$. Therefore, in the rest of this proof, we focus on SAFFRON*.

Fix a time $t$. Then, we have:
\begin{align*}
\EE{|\V(t)|} = \EE{\sum_{E_j \leq t, j \in \nulls}
\One{P_j \leq \alpha_j}} \leq  \sum_{j \leq t, j \in \nulls}
\EE{\One{P_j \leq \alpha_j}},
\end{align*}
where the inequality follows because the set of rejections made by time $t$ could be at most the set $[t]$. Note that $\alpha_j$ and $\lambda_j$ are measurable with respect to $\Ss^{-\X^j}\subseteq \Ss^{-\X^{E_j}}$; therefore, applying iterated expectations by conditioning on $\Ss^{-\X^{E_j}}$ gives:
\begin{align*}
    \sum_{j \leq t, j \in \nulls} \EE{\One{P_j \leq \alpha_j}} \leq \sum_{j \leq t, j \in \nulls} \EE{\alpha_j} \leq \sum_{j \leq t, j \in \nulls} \EE{\alpha_j\frac{\One{P_j>\lambda_j}}{1-\lambda_j}},
\end{align*}
where we apply the super-uniformity assumption. If we assume that \[
\widehat \fdp_{\text{SAFFRON*}}(t) \defn \frac{\sum_{j< t, j\not\in \X^t}\frac{\alpha_j}{1-\lambda_j}\One{P_j>\lambda_j} + \sum_{j\in\X^t\cup\{t\}}\frac{\alpha_j}{1-\lambda_j}}{\left(\sum_{j< t,j\not\in \X^t}R_j\right) \vee 1} \leq \alpha,
\] 
then it follows that:
\begin{align*}
\lefteqn{\sum_{j \leq t, j \in \nulls}  \EE{\alpha_j\frac{\One{P_j>\lambda_j}}{1-\lambda_j}}\leq \sum_{j \leq t} \EE{\alpha_j\frac{\One{P_j>\lambda_j}}{1-\lambda_j}}}\\
& \leq \EE{\sum_{j< t, j\not\in \X^t}\frac{\alpha_j}{1-\lambda_j}\One{P_j>\lambda_j} + \sum_{j\in\X^t\cup\{t\}}\frac{\alpha_j}{1-\lambda_j}}\leq \alpha \EE{\left(\sum_{j< t,j\not\in \X^t}R_j\right)\vee 1}\\
& \leq~ \alpha \EE{|\cR(t)|\vee 1},
\end{align*}
where the first inequality drops the condition $j\in\nulls$, the second one ignores the condition $\One{P_j>\lambda_j}$ for some terms, the third inequality applies the assumption on $\widehat \fdp_{\text{SAFFRON*}}(t)$ and the last inequality uses the fact that $\cR(t)$ contains all past rejections that are no longer conflicting. Rearranging the terms in the previous derivation, we reach the conclusion that $\mfdr(t)\leq \alpha$, which concludes the proof of the theorem.

\subsection{Proof of \thmref{stoppinglord&saff}}
\label{pf:stoppinglord&saff}

We first prove the theorem for LORD*, and then we move on to proving the SAFFRON* guarantees.

\paragraph{LORD*.} For all $t\in\N$, define the process $A(t)$ as:
\small
\begin{align*}
    A(t)&\defn -\sum_{i\leq t,i\in\nulls}\One{E_i\leq t}(\One{P_i\leq\alpha_i} - \alpha_i)= A(t-1) - \sum_{i\leq t,i\in\nulls}\One{E_i=t}(\One{P_i\leq\alpha_i} - \alpha_i),
\end{align*}
\normalsize
where we take $A(0)=0$. Let $H(t) \defn \One{T\geq t}$. Since $T$ is a stopping time, it holds that $\{T\geq t + 1\}=\{T\leq~t\}^c\in \F^{-\X^{t+1}}$, therefore $H(t+1)$ is predictable, that is it is measurable with respect to $\F^{-\X^{t+1}}$. Define the transform $(H \cdot A)$ of $H$ by $A$ as follows:
\begin{align*}
(H\cdot A)(t) &\defn \sum_{m=1}^t H(m)(A(m) - A(m-1))\\
& = \sum_{m=1}^t H(m)\left(- \sum_{i\leq m,i\in\nulls}\One{E_i=m}(\One{P_i\leq\alpha_i} - \alpha_i)\right).
\end{align*}
By taking conditional expectations, we can obtain:
\begin{align*}
\lefteqn{\EEst{(H\cdot A)(t+1)}{\F^{-\X^{t+1}}}}\\
    & = \EEst{(H\cdot A)(t)}{\F^{-\X^{t+1}}} + \EEst{H(t+1)(A(t+1)-A(t))}{\F^{-\X^{t+1}}}\\
    & =\EEst{(H\cdot A)(t)}{\F^{-\X^{t+1}}}\\
    & \ \ + H(t+1)\EEst{- \sum_{i\leq t+1,i\in\nulls}\One{E_i=t+1}(\One{P_i\leq\alpha_i} - \alpha_i)}{\F^{-\X^{t+1}}}\\
    & =\EEst{(H\cdot A)(t)}{\F^{-\X^{E_t}}}\\
    &\ \ + H(t+1)\sum_{i\leq t+1,i\in\nulls}\One{E_i=t+1}\EEst{- (\One{P_i\leq\alpha_i} - \alpha_i)}{\F^{-\X^{t+1}}},
\end{align*}
where the first and last equality follow by linearity of expectation, and the second one uses the predictability of $H(t+1)$. Now we can apply the super-uniformity condition \eqnref{superuniformity-conflict}, since we are summing over null indices: $\EEst{- \One{P_i\leq\alpha_i} + \alpha_i}{\F^{-\X^{E_i}}} \geq -\alpha_i + \alpha_i = 0$.
Therefore, additionally applying the law of iterated expectations, it follows that $\EE{(H\cdot A)(t+1)}\geq\EE{(H\cdot A)(t)}$. Iteratively applying the same argument, we reach the conclusion that, for all $t\in\N$:
\begin{equation}
\label{eqn:stopping}
    \EE{(H\cdot A)(t)}\geq 0.
\end{equation}
So far we have only used the predictability of $H(t)$; observe that, by its definition, and the definition of $A(t)$, $(H\cdot A)(t) = A(T\wedge t) - A(0) = A(T\wedge t)$, and hence by equation \eqnref{stopping}, we obtain $\EE{(H\cdot A)(t)} = \EE{A(T\wedge t)}\geq 0.$

Since $A(T\wedge t)\rightarrow A(T)$ almost surely as $t\rightarrow\infty$, by boundedness of $T$ and dominated convergence we can conclude that $\EE{A(T\wedge t)}\rightarrow \EE{A(T)}$ as $t\rightarrow\infty$. With this we obtain a useful intermediate result:
\begin{equation}
\label{eqn:osres}
    \EE{A(T)}\geq 0.
\end{equation}

Recall that $\cR(t)$ denotes the set of all rejections made by time $t$, and $\V(t)$ denotes the set of false rejections made by time $t$. Consider the following process:
\begin{align*}
B(t)&\defn \alpha (|\cR(t)|\vee 1) - |\V(t)| + \sum_{j\leq t} \alpha_j - \alpha\left(\sum_{j< t,j\not\in\X^t} R_j \vee 1\right) \\
&\geq - |\V(t)| + \sum_{j\leq t} \alpha_j\geq A(t),
\end{align*}
where the final inequality applies the definition of $A(t)$ together with the fact that $\sum_{j\leq t} \alpha_j\geq \sum_{j\leq t} \One{E_j\leq t}\alpha_j$.

Now take a stopping time $T$ such that the conditions of the theorem are satisfied, then:
\begin{align*}
\EE{\alpha (|\cR(T)|\vee 1) - |\V(T)|} &= \EE{B(T) - \sum_{j\leq T}\alpha_j + \alpha\left(\sum_{j<T,j\not\in\X^T} R_j\vee 1\right)}\\
&\geq \EE{B(T)}\geq \EE{A(T)}\geq 0,
\end{align*}
where the first inequality follows by definition of the LORD* FDP estimate, the second one by the relationship already established between $A(t)$ and $B(t)$, and the third inequality applies the intermediate result \eqnref{osres}. Rearranging the terms we have that $\mfdr(T)\leq\alpha$, as desired.

\paragraph{SAFFRON*.} We begin the proof using similar tools as in the LORD* section of the proof. For all $t\in\N$, define the process $A(t)$ as:
\begin{align*}
    A(t)&\defn - \sum_{i\leq t, i\in\nulls}\One{E_i\leq t}\left(\One{P_i\leq\alpha_i} - \One{P_i>\lambda_i}\frac{\alpha_i}{1-\lambda_i}\right)\\
    &= A(t-1) -\sum_{i\leq t,i\in\nulls}\One{E_i=t}\left(\One{P_i\leq\alpha_i} + \One{P_i>\lambda_i}\frac{\alpha_i}{1-\lambda_i}\right),
\end{align*}
where we take $A(0)=0$. Let $H(t) \defn \One{T\geq t}$. Since $T$ is a stopping time, it holds that $\{T\geq t + 1\}=\{T\leq t\}^c\in \F^{-\X^{t+1}}$, therefore $H(t+1)$ is measurable with respect to $\F^{-\X^{t+1}}$. Define the following transform of $H$ by $A$:
\begin{align*}
(H\cdot A)(t) &\defn \sum_{m=1}^t H(m)(A(m) - A(m-1)) \\
&= \sum_{m=1}^t H(m)\left(-\sum_{i\leq m,i\in\nulls}\One{E_i=m}\left(\One{P_i\leq\alpha_i} + \One{P_i>\lambda_i}\frac{\alpha_i}{1-\lambda_i}\right)\right).
\end{align*}
By taking conditional expectations, we can obtain:
\begin{align*}
\lefteqn{
    \EEst{(H\cdot A)(t+1)}{\F^{-\X^{t+1}}}}\\
    & =\EEst{(H\cdot A)(t)}{\F^{-\X^{t+1}}} + \EEst{H(t+1)(A(t+1)-A(t))}{\F^{-\X^{t+1}}}\\
    & =\EEst{(H\cdot A)(t)}{\F^{-\X^{t+1}}}\\
    &\ \ + H(t+1)\EEst{-\sum_{i\leq t+1,i\in\nulls}\One{E_i=t+1}\left(\One{P_i\leq\alpha_i} + \One{P_i>\lambda_i}\frac{\alpha_i}{1-\lambda_i}\right)}{\F^{-\X^{t+1}}}\\
    & =\EEst{(H\cdot A)(t)}{\F^{-\X^{t+1}}}\\
    &\ \ +H(t+1)\sum_{i\leq t+1,i\in\nulls}\One{E_i=t+1}\EEst{-\left(\One{P_i\leq\alpha_i} + \One{P_i>\lambda_i}\frac{\alpha_i}{1-\lambda_i}\right)}{\F^{-\X^{t+1}}},
\end{align*}
where the first equality follows by linearity of expectation and the definition of the transform and the second one uses measurability of $H(t+1)$. The term $-\One{E_i=t+1}(\One{P_i\leq\alpha_i} + \One{P_i>\lambda_i}\frac{\alpha_i}{1-\lambda_i})$ is clearly non-negative when $E_i\neq t+1$. If $E_i = t+1$ however, we can invoke the super-uniformity condition \eqnref{superuniformity-conflict}, since we are summing over null indices:
$$\EEst{-(\One{P_i\leq\alpha_i} + \One{P_i>\lambda_i}\frac{\alpha_i}{1-\lambda_i})}{\F^{-\X^{E_i}}} \geq -\alpha_i + (1-\lambda_i) \frac{\alpha_i}{1-\lambda_i} = 0.$$
Therefore, additionally applying the law of iterated expectations, it follows that $\EE{(H\cdot A)(t+1)}\geq\EE{(H\cdot A)(t)}$. Iteratively applying the same argument, we reach the conclusion that, for all $t\in\N$:
\begin{equation}
\label{eqn:stopping2}
    \EE{(H\cdot A)(t)}\geq 0.
\end{equation}
So far we have only used the predictability of $H(t)$; observe that, by its definition, and the definition of $A(t)$, $(H\cdot A)(t) = A(T\wedge t) - A(0) = A(T\wedge t)$, and hence by equation \eqnref{stopping2}, we obtain $\EE{(H\cdot A)(t)} = \EE{A(T\wedge t)}\geq 0$.

Since $A(T\wedge t)\rightarrow A(T)$ almost surely as $t\rightarrow\infty$, by boundedness of $T$ and dominated convergence we can conclude that $\EE{A(T\wedge t)}\rightarrow \EE{A(T)}$ as $t\rightarrow\infty$. As in the LORD* argument, we reach the result that states:
\begin{equation}
\label{eqn:osres2}
    \EE{A(T)}\geq 0.
\end{equation}

Recall $\cR(t)$, the set of all rejections made by time $t$, and $\V(t)$, the set of false rejections made by time $t$. Consider the following process:
\small
\begin{align*}
B(t) &\defn \alpha (|\cR(t)|\vee 1) - |\V(t)| + \sum_{j< t,j\not\in\X^{t}} \One{P_j>\lambda_j}\frac{\alpha_j}{1-\lambda_j} + \sum_{j\in\X^{t}\cup\{t\}}\frac{\alpha_j}{1-\lambda_j} - \left(\sum_{j< t,j\not\in\X^{t}} R_j\vee 1\right) \alpha\\
&\geq- |\V(t)|  + \sum_{j\leq t} \One{P_j>\lambda_j}\frac{\alpha_j}{1-\lambda_j} \geq A(t),
\end{align*}
\normalsize
where the second inequality applies the definition of $A(t)$ together with $\One{E_j\leq t}\leq 1$.

Now taking a stopping time $T$ that satisfies the conditions of the theorem, we have:
\begin{align*}
\lefteqn{\EE{\alpha (|\cR(t)|\vee 1) - |\V(t)|}}\\
& =\EE{B(T) - \sum_{j< T,j\not\in\X^{T}} \One{P_j>\lambda_j}\frac{\alpha_j}{1-\lambda_j} - \sum_{j\in\X^{T}\cup\{T\}}\frac{\alpha_j}{1-\lambda_j} + \left(\sum_{j< T,j\not\in\X^{T}} R_j\vee 1\right) \alpha}\\
& \geq \EE{B(T)}\geq \EE{A(T)}\geq 0,
\end{align*}
where the first inequality follows by construction of the SAFFRON* empirical FDP estimate, the second inequality uses the proved relationship between $A(t)$ and $B(t)$, and the third inequality applies equation \eqnref{osres2}. Rearranging the terms we have that $\mfdr(T)\leq\alpha$, as desired.

\subsection{Proof of \lemref{superunif}}
\label{pf:superunif}

We begin by focusing on the first inequality. Letting $P_{1:M} = (P_1,\ldots,P_M)$ be the
original vector of $p$-values, we define a ``hallucinated'' vector of
$p$-values $\widetilde{P}^{t\to 1}_{1:M} \defn (\widetilde
P_1,\ldots,\widetilde P_M)$ that equals $P_{1:M}$, except that the
$t$-th component is set to one:
\begin{align*}
  \widetilde P_i = \begin{cases} 1 & \mbox{if $i = t$,} \\
P_i & \mbox{if $i \neq t$.}
  \end{cases}
\end{align*}
Further, denote by $\widetilde E_j$ the finish times of the tests that yield $\widetilde P_j$, and let $\widetilde E_j$ be equal to $E_j$ for all $1\leq j\leq M$. Denote the set of candidates and rejections in the hallucinated sequence at time $i$ by $\widetilde
\C_i$ and $\widetilde \cR_i$, respectively, and let $\widetilde \alpha_i$ be the test level for $\widetilde P_i$.  Also, let $\cR_{1:M} = (\cR_1,\ldots,\cR_M)$ and
$\widetilde{\cR}^{t\to 1}_{1:M} = (\widetilde \cR_1,\ldots, \widetilde
\cR_M)$ denote the vectors of the numbers of rejections using $P_{1:M}$
and $\widetilde{P}^{t\to 1}_{1:M}$, respectively.  Similarly, let $\C_{1:M} =
(\C_1,\ldots,\C_M)$ and $\widetilde{\C}^{t\to 1}_{1:M} = (\widetilde
\C_1,\ldots, \widetilde \C_M)$ denote the vectors of the numbers of candidates using
$P_{1:M}$ and $\widetilde{P}^{t\to 1}_{1:M}$, respectively.

By construction, we have the following properties:
\begin{enumerate}
\item $\widetilde E_j = E_j, \forall j$ implies $\alpha_i = \widetilde \alpha_i$ for all $i \leq E_t$.
\item $\widetilde{\cR}_i = \cR_i$ and $\widetilde{\C}_i = \C_i$ for all $i< E_t$, since the $p$-values from the finished tests and the respective test levels are the same in the original and hallucinated setting.
\item $\widetilde{\cR}_{E_t}\subseteq \cR_{E_t}$ and $\widetilde \C_{E_t}\subseteq \C_{E_t}$, and hence
   $\widetilde{\cR}_i \subseteq \cR_i$ also for all $i > E_t$, due to monotonicity of the test levels $\alpha_i$ and candidacy thresholds $\lambda_i$.
\end{enumerate}
Therefore, on the event $\{P_t > \lambda_t\}$, we have $\cR_{E_t} = \widetilde
\cR_{E_t}$ and $\C_{E_t} = \widetilde \C_{E_t}$, and hence also $\cR_{1:M} =
\widetilde{\cR}^{t\to 1}_{1:M}$ and $\C_{1:M} =
\widetilde{\C}^{t\to 1}_{1:M}$. This allows us to conclude that:
\begin{align*}
\frac{\alpha_t \One{P_t > \lambda_t}}{(1-\lambda_t)g
  (|\cR|_{1:M})} = \frac{\alpha_t \One{P_t >
    \lambda_t}}{(1-\lambda_t)g (|\widetilde{\cR}|^{t\to 1}_{1:M})}.
\end{align*}

Since the null $p$-values are mutually independent and independent of the non-nulls, we conclude that $\widetilde{\cR}^{t\to 1}_{1:M}$ is independent of $P_t$ conditioned on $\F_{\text{async}}^{-\X^{E_t}}$. With this, we can obtain:
\begin{align*}
\EEst{\frac{\alpha_t \One{P_t >
      \lambda_t}}{(1-\lambda_t)g(|\cR|_{1:M})} }{\F_{\text{async}}^{-\X^{E_t}}} =
\EEst{\frac{\alpha_t \One{P_t
      >\lambda_t}}{(1-\lambda_t) g (|\widetilde{\cR}|^{t\to 1}_{1:M})} }{
  \F_{\text{async}}^{-X^{E_t}}}  &\geq \EEst{ \frac{ \alpha_t}{g
    (|\widetilde{\cR}|^{t\to 1}_{1:M})} }{ \F_{\text{async}}^{-\X^{E_t}}} \\
    &\geq
    \EEst{ \frac{ \alpha_t}{g
    (|\cR|_{1:M})} }{ \F_{\text{async}}^{-\X^{E_t}}},
\end{align*}
where the first inequality follows by taking an expectation only with
respect to $P_t$ by invoking the asynchronous super-uniformity property \eqnref{superuniformity-async}, and the second inequality
follows because $g (|\cR|_{1:M}) \geq g(|\widetilde{\cR}|^{t\to 1}_{1:M})$
since $|\cR_i| \geq |\widetilde \cR_i|$ for all $i$ by monotonicity of the test levels and candidacy thresholds. This concludes the proof of the first inequality.

The second inequality uses a similar idea of hallucinating tests with identical finish times, only now the $p$-values that these tests result in are:
\begin{align*}
  \widetilde P_i = \begin{cases} 0 & \mbox{if $i = t$,} \\
P_i & \mbox{if $i \neq t$,}
  \end{cases}
\end{align*}
where $P_i$ are the $p$-values in the original sequence. In a similar fashion, the following observations hold:
\begin{enumerate}
\item $\widetilde E_j = E_j$ implies $\alpha_i = \widetilde \alpha_i$ for all $i \leq E_t$.
 \item $\widetilde{\cR}_i = \cR_i$ and $\widetilde{\C}_i = \C_i$ for all $i < E_t$, since the $p$-values from the finished tests and the respective test levels are the same in the original and hallucinated setting.
\item $\widetilde{\cR}_{E_t}\supseteq \cR_{E_t}$ and $\widetilde \C_{E_t} \supseteq \C_{E_t}$, and hence
   $\widetilde{\cR}_i \supseteq \cR_i$ also for all $i > E_t$, due to monotonicity of the test levels $\alpha_i$.
\end{enumerate}
Then, on the event $\{P_t \leq \alpha_t\}$, we have $\cR_{E_t} = \widetilde
\cR_{E_t}$ and $\C_{E_t} = \widetilde \C_{E_t}$, and hence also $\cR_{1:M} =
\widetilde{\cR}^{t\to 1}_{1:M}$ and $\C_{1:M} =
\widetilde{\C}^{t\to 1}_{1:M}$. From this we conclude that:
\begin{align*}
\frac{\One{P_t \leq \alpha_t}}{g(|\cR|_{1:M})} = \frac{\One{P_t \leq \alpha_t}}{g(|\widetilde{\cR}|^{t\to 1}_{1:M})}.
\end{align*}

As in the first part of the proof, we use the fact that the null $p$-values are mutually independent and independent of the non-nulls, which allows us to conclude that $\widetilde{\cR}^{t\to 1}_{1:M}$ is independent of $P_t$ conditioned on $\F_{\text{async}}^{\X^{E_t}}$. This observation results in the following:
\begin{align*}
\EEst{\frac{\One{P_t \leq \alpha_t}}{g(|\cR|_{1:M})}}{\F_{\text{async}}^{\X^{E_t}}} =
\EEst{\frac{\One{P_t \leq \alpha_t}}{g(|\widetilde{\cR}|^{t\to 1}_{1:M})}}{
  \F_{\text{async}}^{\X^{E_t}}}  &\leq \EEst{ \frac{ \alpha_t}{g
    (|\widetilde{\cR}|^{t\to 1}_{1:M})} }{ \F_{\text{async}}^{\X^{E_t}}}
\\
& \leq \EEst{ \frac{ \alpha_t }{g
    (|\cR|_{1:M})} }{ \F_{\text{async}}^{\X^{E_t}}},
\end{align*}
where the first inequality follows by taking an expectation only with
respect to $P_t$ by invoking the asynchronous super-uniformity property \eqnref{superuniformity-async}, and the second inequality
follows because $g (|\cR|_{1:M}) \leq g(|\widetilde{\cR}|^{t\to 1}_{1:M})$
since $|\cR_i| \leq |\widetilde \cR_i|$ for all $i$ by monotonicity of the
test levels. This concludes the proof of the lemma.

\subsection{Proof of \thmref{fdr-async}}
\label{pf:fdr-async}

\paragraph{$\lordasync$.} Fix a time step $t$. First we show the claim for $\lordasync$, so suppose that 
$$\widehat \fdp_{\lordasync}(t)\defn \frac{\sum_{j\leq t}\alpha_j}{\sum_{j\leq t}\One{P_j\leq\alpha_j, E_j\leq t} \vee 1}\leq\alpha.$$
Then:
\begin{align*}
    \fdr(t) &\defn \EE{\frac{|\V(t)|}{|\cR(t)| \vee 1}}= \EE{\frac{\sum_{j\leq t,j\in\nulls} \One{P_j\leq\alpha_j,E_j\leq t}}{\sum_{j\leq t}\One{P_j\leq \alpha_j,E_j\leq t} \vee 1}}\\
    &\leq \sum_{i\leq t,i\in\nulls} \EE{\frac{ \One{P_i\leq\alpha_i}}{\sum_{j\leq t}\One{P_j\leq \alpha_j,E_j\leq t} \vee 1}},
\end{align*}
where the second equality follows by definition of $\V(t)$ and $\R(t)$, and the inequality drops the condition $E_i\leq t$ from the numerator and applies linearity of expectation. Now we can apply \lemref{superunif} with $g(|\cR|_{1:t}) = (\sum_{i=1}^t |\cR_i|) \vee 1$, together with iterated expectations, to obtain:
\begin{align*}
            \sum_{i\leq t,i\in\nulls} \EE{\frac{ \One{P_i\leq\alpha_i}}{\sum_{j\leq t}\One{P_j\leq \alpha_j,E_j\leq t} \vee 1}} &\leq \sum_{i\leq t,i\in\nulls} \EE{\frac{ \alpha_i}{\sum_{j\leq t}\One{P_j\leq \alpha_j,E_j\leq t} \vee 1}}\\
            &\leq \EE{\widehat \fdp_{\lordasync}(t)}\leq\alpha,
\end{align*}
where the second inequality follows by dropping the condition $i\in\nulls$. This completes the proof for $\lordasync$.

\paragraph{$\saffasync$.} Now we move on to $\saffasync$. Using the same steps as above, for any fixed time $t$, we can conclude the following inequality:
\begin{align*}
\fdr(t) &\leq \sum_{i \leq t, i \in \nulls} \EE{\frac{\alpha_i}{\sum_{j\leq t}\One{P_j\leq \alpha_j,E_j\leq t} \vee 1}}.
\end{align*}
Here we additionally apply the other inequality of \lemref{superunif}, with the same choice $g(|\cR|_{1:t}) = (\sum_{i=1}^t |\cR_i|) \vee 1$, again with iterated expectations:
\begin{align*}
\sum_{i \leq t, i \in \nulls} \EE{\frac{\alpha_i}{\sum_{j\leq t}\One{P_j\leq \alpha_j,E_j\leq t} \vee 1}} &\leq \sum_{i \leq t, i \in \nulls} \EE{\frac{
    \alpha_i\One{P_i >\lambda_i}}{(1-\lambda_i)(\sum_{j\leq t}\One{P_j\leq \alpha_j,E_j\leq t} \vee 1)}}.
\end{align*}
Assuming that the inequality
$$\widehat \fdp_{\saffasync}(t)\defn \frac{\sum_{j\leq t}\frac{\alpha_j}{1-\lambda_j}(\One{P_j>\lambda_j,E_j\leq t} + \One{E_j>t})}{\sum_{j\leq t}\One{P_j\leq \alpha_j,E_j\leq t} \vee 1} \leq \alpha$$
holds, it follows that:
\begin{align*}
\lefteqn{\sum_{i \leq t, i \in \nulls} \EE{\frac{
    \alpha_i\One{P_i >\lambda_i}}{(1-\lambda_i)(\sum_{j\leq t}\One{P_j\leq \alpha_j,E_j\leq t} \vee 1)}}}\\
    & \leq \EE{\frac{\sum_{j\leq t}\frac{\alpha_j}{1-\lambda_j}(\One{P_j>\lambda_j,E_j\leq t} + \One{E_j>t})}{\sum_{j\leq t}\One{P_j\leq \alpha_j,E_j\leq t} \vee 1}}= \EE{\widehat \fdp_{\saffasync}(t)} \leq \alpha,
\end{align*}
where the first inequality follows by dropping the conditions $j\in\nulls$ and $\{P_j>\lambda_j\}$ for some rounds. The second inequality follows by assumption, hence proving the theorem.

\subsection{Proof of \thmref{positivedep}}
\label{pf:positivedep}

For statement (a), we begin by noting that for any $t \in \N$:
\[
\fdr(t) = \EE{\frac{\sum_{i \leq t, i \in \nulls} \One{P_i \leq \alpha_i} }{|\cR(t)|\vee 1}} 
\leq \sum_{i \leq t, i \in \nulls} \EE{\frac{ \One{P_i \leq \alpha_i} }{|\cR(i-1)|\vee 1}}
= \sum_{i \leq t, i \in \nulls} \gamma_i \alpha \EE{\frac{ \One{P_i \leq \alpha_i} }{\alpha_i}} ,
\]
where the first equality follows by definition of \fdr, the sole inequality follows because the number of rejections can only increase with time, and the second equality follows by definition of the LOND rule for  $\alpha_i$.
Lemma 1 from \citet{ramdas2017unified} now asserts that the term in the expectation is bounded by one under PRDS.
Hence, by also noting that $\sum_{i \leq t} \gamma_i \leq 1$ we immediately deduce statement (a).

For statement (b), we follow almost the same sequence of steps to note that:
\begin{align*}
\fdr(t) &= \EE{\frac{\sum_{i \leq t, i \in \nulls} \One{P_i \leq \talpha_i} }{|\cR(t)|\vee 1}} 
\leq \sum_{i \leq t, i \in \nulls} \EE{\frac{ \One{P_i \leq \talpha_i} }{|\cR(i-1)|\vee 1}}\\
&= \sum_{i \leq t, i \in \nulls} \gamma_i \alpha \EE{\frac{ \One{P_i \leq \gamma_i \alpha \beta_i(|\cR(i-1)|\vee 1)} }{\gamma_i \alpha (|\cR(i-1)|\vee 1)}}.
\end{align*}
We now apply Lemma 1 from \citet{ramdas2017unified} with $c=\gamma_i \alpha$ and $f(P)=|\cR(i-1)|\vee 1$ to again assert that the term in the expectation is bounded by one under arbitrary dependence, hence establishing statement (b).


\section{Different instantiations of LORD* and SAFFRON*}

Here we give explicit statements of different instances of LORD* and SAFFRON* described in \secref{async}, \secref{markov} and \secref{minibatch}. All of the following algorithms are special instances of \hyperref[algs1to4]{Algorithms 1-4}, given in \secref{conflict}.

\label{alg:async}
First we state $\lordasync$ and $\saffasync$ explicitly, by taking $\X^t=\X^t_{\text{async}}$ in the statement of LORD* and SAFFRON*. Algorithm 5 and Algorithm 6 state the LORD++ and LOND versions of $\lordasync$, Algorithm 7 states $\saffasync$ for constant candidacy thresholds, i.e. $\{\lambda_j\}\equiv\lambda$, and Algorithm 8 states asynchronous alpha-investing, i.e. $\saffasync$ when $\lambda_j=\alpha_j$. Recall the definition of $r_k$, which in this setting takes the form:
\small
$$r_k=\min\{ i\in [t]: \sum_{j=1}^i R_j\One{E_j\leq i}\geq k\}.$$
\normalsize

\begin{algorithm}[H]
\SetAlgoLined
\footnotesize
\SetKwInOut{Input}{input}
\Input{FDR level $\alpha$, non-negative non-increasing sequence $\{\gamma_j\}_{j=1}^\infty$ such that $\sum_j \gamma_j=1$, initial wealth $W_0\leq\alpha$}
$\alpha_1 = \gamma_1 W_0$\newline
 \For{$t=1,2,\dots$}{
 start $t$-th test with level $\alpha_t$\newline
  $\alpha_{t+1} = \gamma_{t+1} W_0 + \gamma_{t+1-r_1} (\alpha - W_0) +  \left(\sum_{j\geq 2}\gamma_{t+1-r_j}\right)\alpha$
 }
 \caption{The asynchronous LORD++ algorithm as a version of $\lordasync$}
\end{algorithm}

\begin{algorithm}[H]
\SetAlgoLined
\footnotesize
\SetKwInOut{Input}{input}
\Input{FDR level $\alpha$, non-negative non-increasing sequence $\{\gamma_j\}_{j=1}^\infty$ such that $\sum_j \gamma_j=1$}
$W_0 = \alpha$\newline
$\alpha_1 = \gamma_1 W_0$\newline
 \For{$t=1,2,\dots$}{
 start $t$-th test with level $\alpha_t$\newline
  $\alpha_{t+1} = \alpha\gamma_{t+1}\left((\sum_{j=1}^t \One{P_j\leq\alpha_j, E_j \leq t})\vee 1\right)$
 }
 \caption{The asynchronous LOND algorithm as a version of $\lordasync$}
\end{algorithm}

\begin{algorithm}[H]
\SetAlgoLined
\footnotesize
\SetKwInOut{Input}{input}
\Input{FDR level $\alpha$, non-negative non-increasing sequence $\{\gamma_j\}_{j=1}^\infty$ such that $\sum_j \gamma_j=1$, candidate threshold $\lambda\in(0,1)$, initial wealth $W_0\leq \alpha$}
$\alpha_1 = (1-\lambda)\gamma_1 W_0$\newline
 \For{$t=1,2,\dots$}{
 start $t$-th test with level $\alpha_t$\newline
  $\alpha_{t+1} = \min\left\{\lambda,(1-\lambda)\left(W_0\gamma_{t+1-C^\#_{0+}} + (\alpha - W_0)\gamma_{t+1-r_1-C^\#_{1+}} + \sum_{j\geq 2} \alpha\gamma_{t+1-r_j-C^\#_{j+}}\right)\right\}$,\\
  where $C^\#_{j+}=\sum_{i=r_j+1}^t |\C_i|$

 }
 \caption{The $\saffasync$ algorithm for constant $\lambda$}
\end{algorithm}


\begin{algorithm}[H]
\SetAlgoLined
\footnotesize
\SetKwInOut{Input}{input}
\Input{FDR level $\alpha$, non-negative non-increasing sequence $\{\gamma_j\}_{j=1}^\infty$ such that $\sum_j \gamma_j=1$, initial wealth $W_0\leq \alpha$}
\ $s_1 = \gamma_1 W_0$\newline
$\alpha_1 = s_1/(1+s_1)$\newline
 \For{$t=1,2,\dots$}{
 start $t$-th test with level $\alpha_t$\newline
  $s_{t+1} = W_0\gamma_{t+1-R^\#_0+} + (\alpha - W_0)\gamma_{t+1-r_1-R^\#_{1+}} + \sum_{j\geq 2} \alpha\gamma_{t+1-r_j-R^\#_{j+}}$, where $R^\#_{j+}=\sum_{i=r_j+1}^t |\cR_i|$\newline
$\alpha_{t+1} = s_{t+1}/(1+s_{t+1})$
 }
 \caption{The asynchronous alpha-investing algorithm as a special case of $\saffasync$}
\end{algorithm}

\label{alg:local}
Below we give explicit statements of $\lordmarkov$ and $\saffmarkov$ as special cases of LORD* and SAFFRON*. Algorithm 9 and Algorithm 10 state LORD++ and LOND under local dependence, both as instances of $\lordmarkov$. Algorithm 11 states $\saffmarkov$ for the constant sequence $\{\lambda_j\}\equiv\lambda$, and Algorithm 12 states alpha-investing under local dependence, which is a particular instance of $\saffmarkov$ obtained by taking $\lambda_j = \alpha_j$. The definition of $r_k$ under local dependence simplify to:
$$r_k = \min \{ i\in[t]: \sum_{j=1}^{i-L_{i+1}} R_j \geq k\}.$$

\begin{algorithm}[H]
\SetAlgoLined
\footnotesize
\SetKwInOut{Input}{input}
\Input{FDR level $\alpha$, non-negative non-increasing sequence $\{\gamma_j\}_{j=1}^\infty$ such that $\sum_j \gamma_j=1$, initial wealth $W_0\leq\alpha$}
$\alpha_1 = \gamma_1 W_0$\newline
 \For{$t=1,2,\dots$}{
 run $t$-th test with level $\alpha_t$\newline
  $\alpha_{t+1} = \gamma_{t+1} W_0 + \gamma_{t+1-r_1} (\alpha - W_0) + \left(\sum_{j=2}^\infty\gamma_{t+1-r_j}\right)\alpha$
 }
 \caption{The LORD++ algorithm under local dependence as a version of $\lordmarkov$}
\end{algorithm}

\begin{algorithm}[H]
\SetAlgoLined
\footnotesize
\SetKwInOut{Input}{input}
\Input{FDR level $\alpha$, non-negative non-increasing sequence $\{\gamma_j\}_{j=1}^\infty$ such that $\sum_j \gamma_j=1$}
$W_0=\alpha$\newline
$\alpha_1 = \gamma_1 W_0$\newline
 \For{$t=1,2,\dots$}{
 run $t$-th test with level $\alpha_t$\newline
  $\alpha_{t+1} = \alpha \gamma_{t+1} \left((\sum_{i=1}^{t-L_{t+1}}R_i)\vee 1\right)$
 }
 \caption{The LOND algorithm under local dependence as a version of $\lordmarkov$}
\end{algorithm}

\begin{algorithm}[H]
\SetAlgoLined
\footnotesize
\SetKwInOut{Input}{input}
\Input{FDR level $\alpha$, non-negative non-increasing sequence $\{\gamma_j\}_{j=1}^\infty$ such that $\sum_j \gamma_j=1$, candidate threshold $\lambda\in(0,1)$, initial wealth $W_0\leq \alpha$}
$\alpha_1 = \gamma_1 W_0$\newline
 \For{$t=1,2,\dots$}{
 run $t$-th test with level $\alpha_t$\newline
  $\alpha_{t+1} = \min\left\{\lambda,(1-\lambda)\left(W_0\gamma_{t+1-C_0+} + (\alpha - W_0)\gamma_{t+1-r_1-C_{1+}} + \alpha(\sum_{j\geq 2} \gamma_{t+1-r_j-C_{j+}})\right)\right\}$, \\
  where $C_{j+}=\sum_{i=r_j+1}^{t-L_{t+1}} C_i$

 }
 \caption{The $\saffmarkov$ algorithm for constant $\lambda$}
\end{algorithm}


\begin{algorithm}[H]
\SetAlgoLined
\footnotesize
\SetKwInOut{Input}{input}
\Input{FDR level $\alpha$, non-negative non-increasing sequence $\{\gamma_j\}_{j=1}^\infty$ such that $\sum_j \gamma_j=1$, initial wealth $W_0\leq \alpha$}
\ $s_1 = \gamma_1 W_0$\newline
$\alpha_1 = s_1/(1 + s_1)$\newline
 \For{$t=1,2,\dots$}{
 run $t$-th test with level $\alpha_t$\newline
  $s_{t+1} = W_0\gamma_{t+1-R_0+} + (\alpha - W_0)\gamma_{t+1-r_1-R_{1+}} + \sum_{j\geq 2} \alpha\gamma_{t+1-r_j-R_{j+}}$, where $R_{j+}=\sum_{i=r_j+1}^{t-L_{t+1}} R_i$\newline
  $\alpha_{t+1} = s_{t+1}/(1 + s_{t+1})$
 }
 \caption{The alpha-investing algorithm under local dependence as a special case of $\saffmarkov$}
\end{algorithm}

\label{alg:mini}
Algorithms 13 and 14 describe the mini-batch versions of LORD++ and LOND respectively, both as cases of $\lordmini$. Algorithm 15 is a variant of $\saffmini$ with $\lambda_j$ chosen constant and equal to some $\lambda\in(0,1)$, and Algorithm 16 is the alpha-investing version of $\saffmini$, in which $\lambda_j=\alpha_j$. In this setting, the definition of $r_k$ is slightly tweaked in order to satisfy the convention of double indexing; $r_k$ refers to the \emph{batch} in which the $k$-th non-conflicting rejection occurs:
$$r_k\defn\min\{ i\in [b-1]: \sum_{j=1}^i |\cR_j|\geq k\}.$$

\begin{algorithm}[H]
\SetAlgoLined
\footnotesize
\SetKwInOut{Input}{input}
\Input{FDR level $\alpha$, non-negative non-increasing sequence $\{\gamma_j\}_{j=1}^\infty$ such that $\sum_j \gamma_j=1$, initial wealth $W_{1,0}\leq\alpha$}
$\alpha_{1,1} = \gamma_1 W_{1,0}$\newline
\For{$b=1,2,\dots$}{
\If{$b> 1$}{
 $W_{b,0} = W_{b-1,n} + \alpha |\cR_{b-1}| - W_{1,0}\One{r_1 = b-1}$
}
 \For{$t=1,2,\dots,n_b$}{
 start $t$-th test in the $b$-th batch with level $\alpha_{b,t}$\newline
  $\alpha_{b,t+1} = \gamma_{\sum_{i=1}^{b-1}n_i+t+1} W_{1,0} + \gamma_{\sum_{i=1}^{b-1}n_i + t+1-\sum_{i=1}^{r_1}n_i} (\alpha - W_{1,0}) +  \left(\sum_{j=2}^\infty\gamma_{\sum_{i=1}^{b-1}n_i + t+1-\sum_{i=1}^{r_j}n_i}\right)\alpha$
 }
 }
 \caption{The mini-batch LORD++ algorithm as a version of $\lordmini$}
\end{algorithm}

\begin{algorithm}[H]
\SetAlgoLined
\footnotesize
\SetKwInOut{Input}{input}
\Input{FDR level $\alpha$, non-negative non-increasing sequence $\{\gamma_j\}_{j=1}^\infty$ such that $\sum_j \gamma_j=1$}
$\alpha_{1,1} = \gamma_1 \alpha$\newline
\For{$b=1,2,\dots$}{
 \For{$t=1,2,\dots,n_b$}{
 start $t$-th test in the $b$-th batch with level $\alpha_{b,t}$\newline
  $\alpha_{b,t+1} = \alpha \gamma_{\sum_{i=1}^{b-1}n_i+t+1} \left((\sum_{j=1}^{b-1}|\cR_j|)\vee 1\right)$
 }
 }
 \caption{The mini-batch LOND algorithm as a version of $\lordmini$}
\end{algorithm}


\begin{algorithm}[H]
\SetAlgoLined
\footnotesize
\SetKwInOut{Input}{input}
\Input{FDR level $\alpha$, non-negative non-increasing sequence $\{\gamma_j\}_{j=1}^\infty$ such that $\sum_j \gamma_j=1$, initial wealth $W_{1,0}\leq\alpha$, constant $\lambda$}
$\alpha_{1,1} = (1-\lambda)\gamma_1 W_{1,0}$\newline
\For{$b=1,2,\dots$}{
 \For{$t=1,2,\dots,n$}{
 start $t$-th test in the $b$-th batch with level $\alpha_{b,t}$\newline
  $\alpha_{b,t+1} = (1-\lambda)(\gamma_{\sum_{i=1}^{b-1}n_i-|\C_0^{+}|+t+1} W_{1,0} + \gamma_{\sum_{i=1}^{b-1}n_i -|\C^{+}_1| + t+1-\sum_{i=1}^{r_1}n_i} (\alpha - W_{1,0}) +  \left(\sum_{j=2}^\infty\gamma_{\sum_{i=1}^{b-1}n_i -|\C^{+}_j| + t+1-\sum_{i=1}^{r_1}n_i}\right)\alpha)$, ~~ where $|\C^{+}_j| = \sum_{j=r_j+1}^{b-1} |\C_j|$
 }
 }
 \caption{The $\saffmini$ algorithm for constant $\lambda$}
\end{algorithm}


\begin{algorithm}[H]
\SetAlgoLined
\footnotesize
\SetKwInOut{Input}{input}
\Input{FDR level $\alpha$, non-negative non-increasing sequence $\{\gamma_j\}_{j=1}^\infty$ such that $\sum_j \gamma_j=1$, initial wealth $W_{1,0}\leq \alpha$}
\ \ $s_{1,1} = \gamma_1 W_{1,0}$\newline
$\alpha_{1,1} = s_{1,1}/(1 + s_{1,1})$\newline
\For{$b=1,2,\dots$}{
 \For{$t=1,2,\dots,n$}{
 start $t$-th test in the $b$-th batch with level $\alpha_{b,t}$\newline
  $s_{b,t+1} = \gamma_{\sum_{i=1}^{b-1}n_i-|\C^{+}_0|+t+1} W_{1,0} + \gamma_{\sum_{i=1}^{b-1}n_i -|\C^{+}_1| + t+1-\sum_{i=1}^{r_1}n_i} (\alpha - W_{1,0}) +  \left(\sum_{j=2}^\infty\gamma_{\sum_{i=1}^{b-1}n_i -|\C^{+}_j| + t+1-\sum_{i=1}^{r_j}n_i}\right)\alpha$, ~~ where $|\C^{+}_j| = \sum_{j=r_j+1}^{b-1} |\C_j|$\newline
  $\alpha_{b,t+1} = s_{b,t+1}/(1 + s_{b,t+1})$
 }
 }
 \caption{The mini-batch alpha-investing as a special case of $\saffmini$}
\end{algorithm}


\section{Experiments on real data with local dependence}
\label{sec:realdata}

We perform an additional case study on a high-throughput phenotypic data set from the International Mouse Phenotyping Consortium (IMPC) data repository. This database documents gene knockout experiments on mice and annotates every protein coding gene by exploring the impact of the gene knockout on the resulting phenotype. This is an example of a continuously growing data set, as the family of hypotheses and the new knockouts grow with time. \citet{karp2017prevalence} tested the role of genotype and the role of sex as a modifier of genotype effect. In this section, we focus on the $p$-values resulting from the analysis of genotype effects.

This set of $p$-values exhibits local dependence---the same set of mice is used to test multiple hypotheses adjacent on the time horizon, while $p$-values computed at sufficiently distant time points are statistically independent.

We use a subset of the database organized by \citet{robertson2019onlinefdr}. The data are publicly available at \texttt{https://zenodo.org/record/2396572}. Hypotheses belonging to the same batch have the same experimental ID. For the sake of computational efficiency, we only analyze the hypotheses whose experimental ID is in the interval $[36700, 37000)$. This results in 4275 distinct hypotheses and their corresponding $p$-values, split into 172 batches of varying sizes. For different target FDR levels, we report the number of discoveries made by $\saffmarkov$, $\lordmarkov$, and alpha-spending. Since we expect a small number of truly relevant genes, for $\saffmarkov$ we set $\lambda = 0.1$; all other parameters for all three algorithms are as in \secref{sims}. We cannot evaluate the FDR and power due to lack of ground truth, however we know that theoretically all three procedures control the FDR at the target level.

\begin{figure}[H]
\centerline{\includegraphics[width=0.5\textwidth]{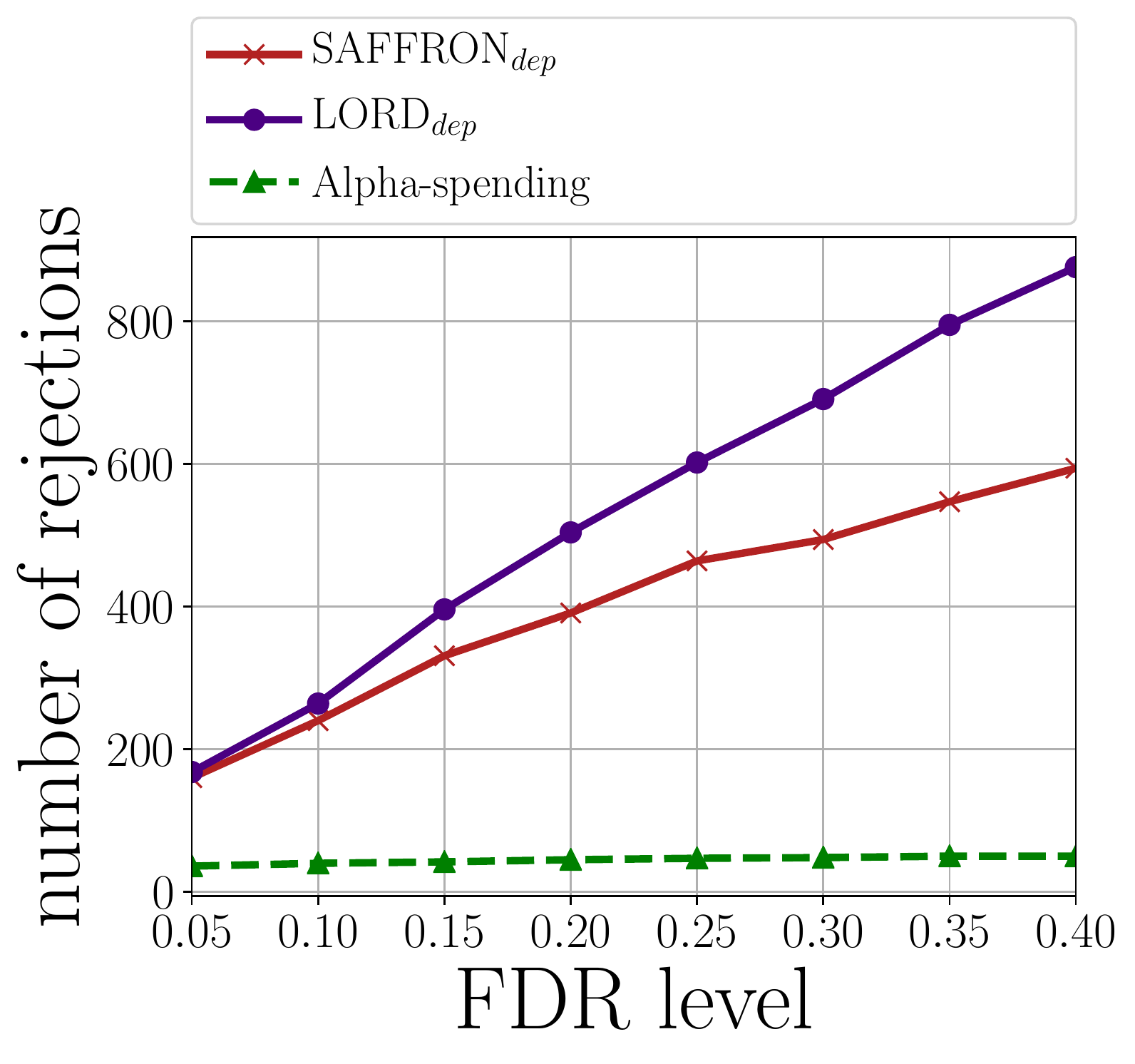}}
\caption{Number of rejections made by $\saffmarkov$, $\lordmarkov$, and alpha-spending, for different target FDR levels.} 
\label{fig:realdata}
\end{figure}


\section{Positive regression dependency on a subset (PRDS)}
\label{prds}

For convenience, here we briefly review the definition of positive regression dependency on a subset (PRDS).

\begin{definition}
Let $\mathcal{D}\subseteq[0,1]^n$ be any non-decreasing set, meaning that $x\in\mathcal{D}$ implies  $y\in\mathcal{D}$, for all $y$ such that $y_i\geq x_i$ for all $i\in[n]$. We say that a vector of $p$-values $P=(P_1,\dots,P_n)$ satisfies positive dependence (PRDS) if for any null index $i\in\nulls$ and arbitrary non-decreasing $\mathcal{D}\subseteq[0,1]^n$, the function $t\rightarrow \PPst{P\in\mathcal{D}}{P_i\leq t}$ is non-decreasing over $t\in(0,1]$.
\end{definition}

Clearly, independent $p$-values satisfy PRDS. Another important example is given for Gaussian observations. Suppose $Z=(Z_1,\dots,Z_n)$ is a multivariate Gaussian with covariance matrix $\Sigma$, and let $P=(\Phi(Z_1,),\dots,\Phi(Z_n))$ be a vector of $p$-values, where $\Phi$ is the standard Gaussian CDF. Then, $P$ satisfies PRDS if and only if, for all $i\in\nulls$ and $j\in[n]$, $\Sigma_{ij}\geq 0$.

\section{Examining the difference between mFDR and FDR}

In \secref{sims}, we plotted strict $\fdr$ estimates, obtained by averaging the false discovery proportion over 200 independent trials; on the other hand, the main guarantees of this paper apply to $\mfdr$ control. For this reason, here we provide the plot of both $\mfdr$ and $\fdr$ estimates, for all experiments in \secref{sims}. We estimate $\mfdr$ by computing the ratio of the average number of false discoveries and the average total number of discoveries.

\begin{figure}[H]
\centerline{\includegraphics[width=0.24\textwidth]{lordasyncfdr.pdf}
\includegraphics[width=0.24\textwidth]{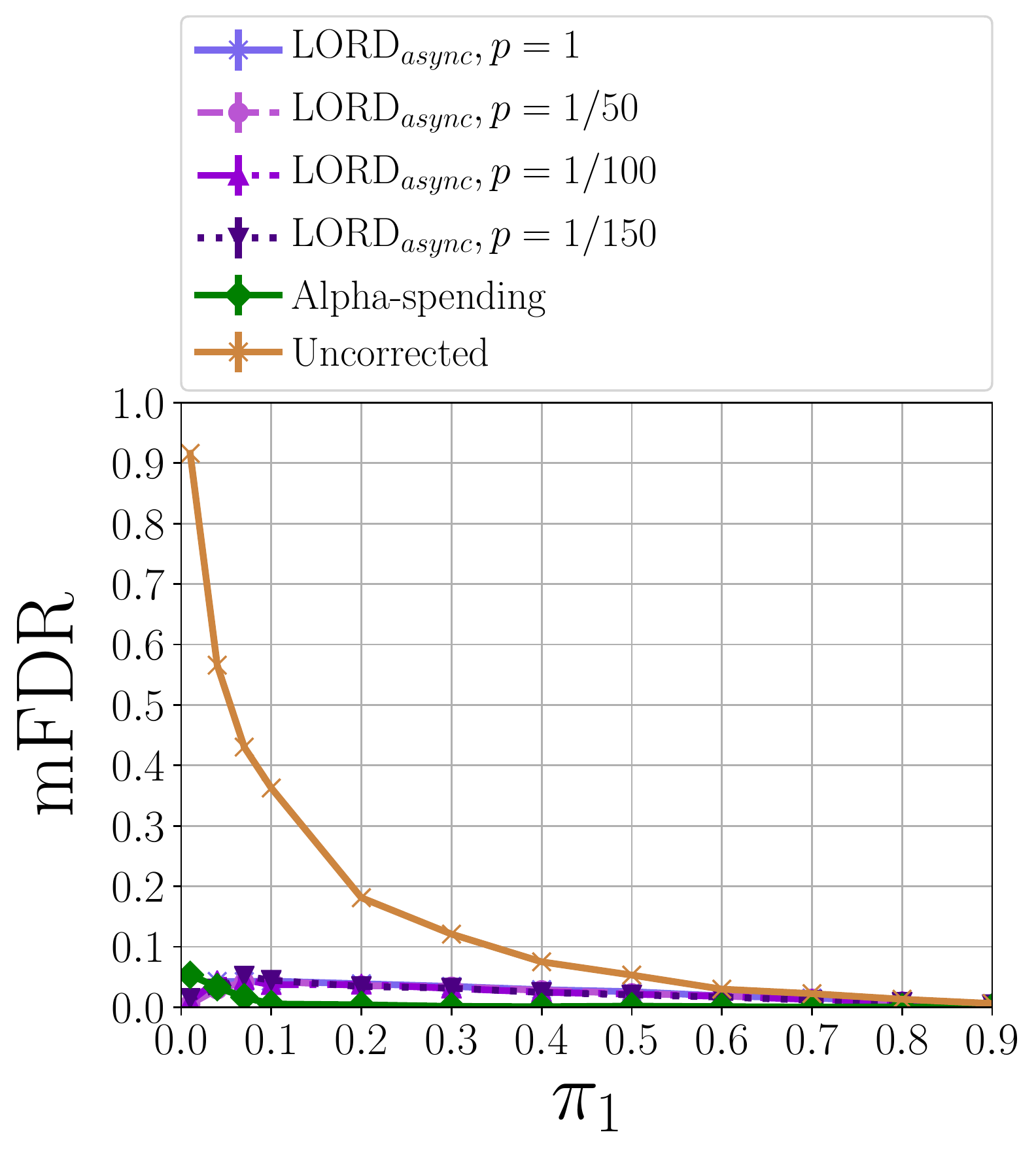}
\includegraphics[width=0.24\textwidth]{lordasyncfdr2.pdf}
\includegraphics[width=0.24\textwidth]{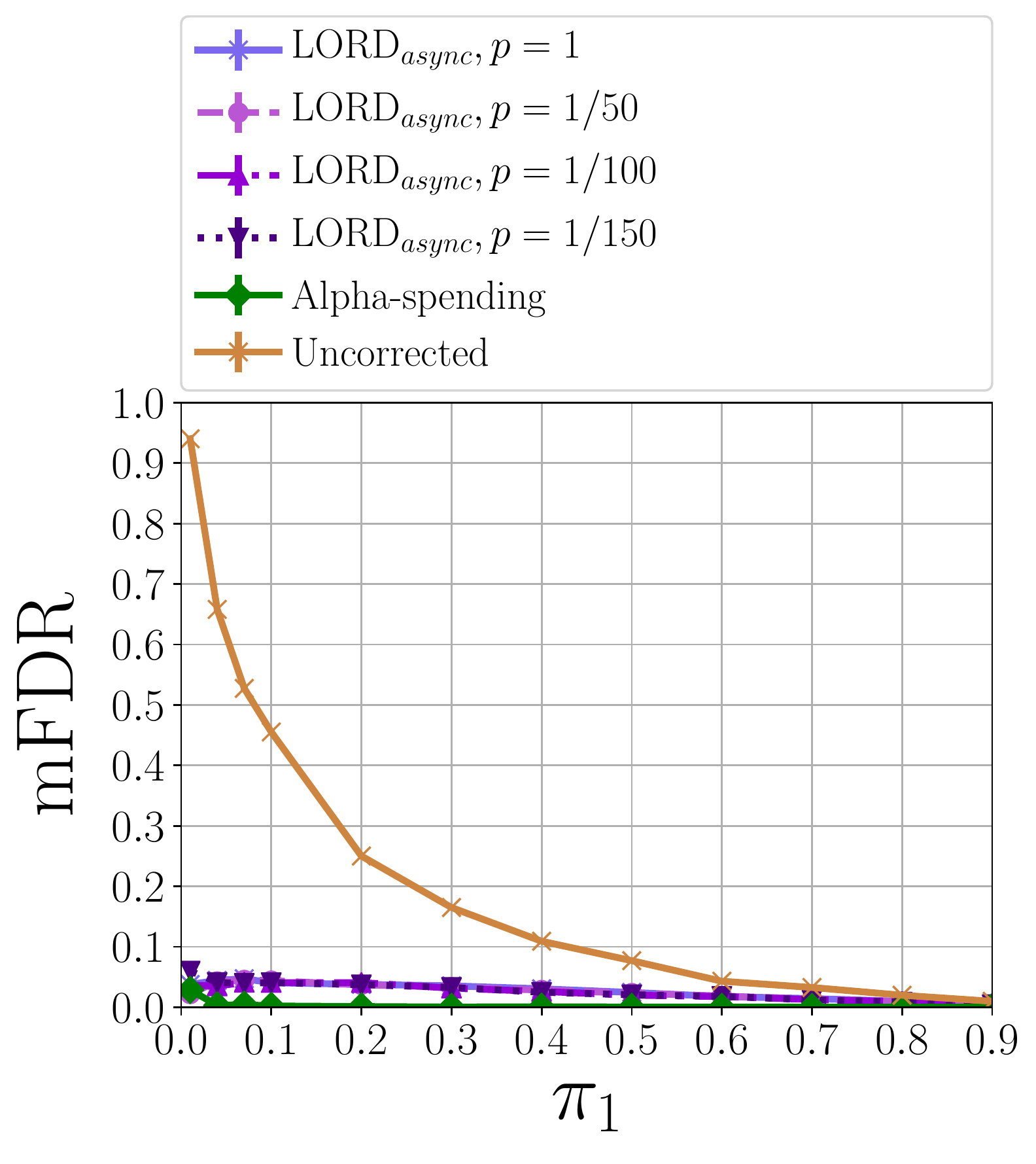}}
\centerline{\includegraphics[width=0.24\textwidth]{saffasyncfdr.pdf}
\includegraphics[width=0.24\textwidth]{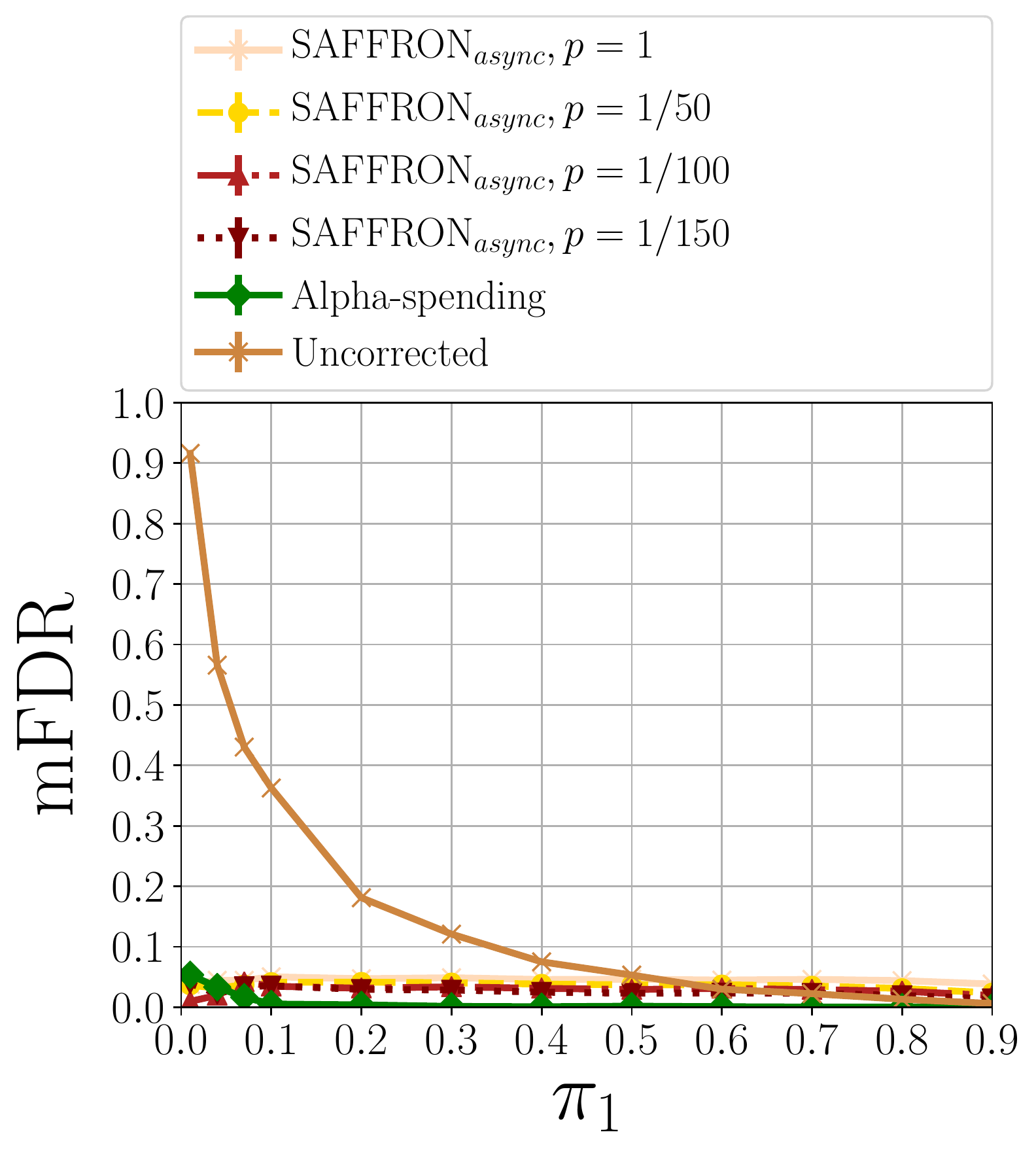}
\includegraphics[width=0.24\textwidth]{saffasyncfdr2.pdf}
\includegraphics[width=0.24\textwidth]{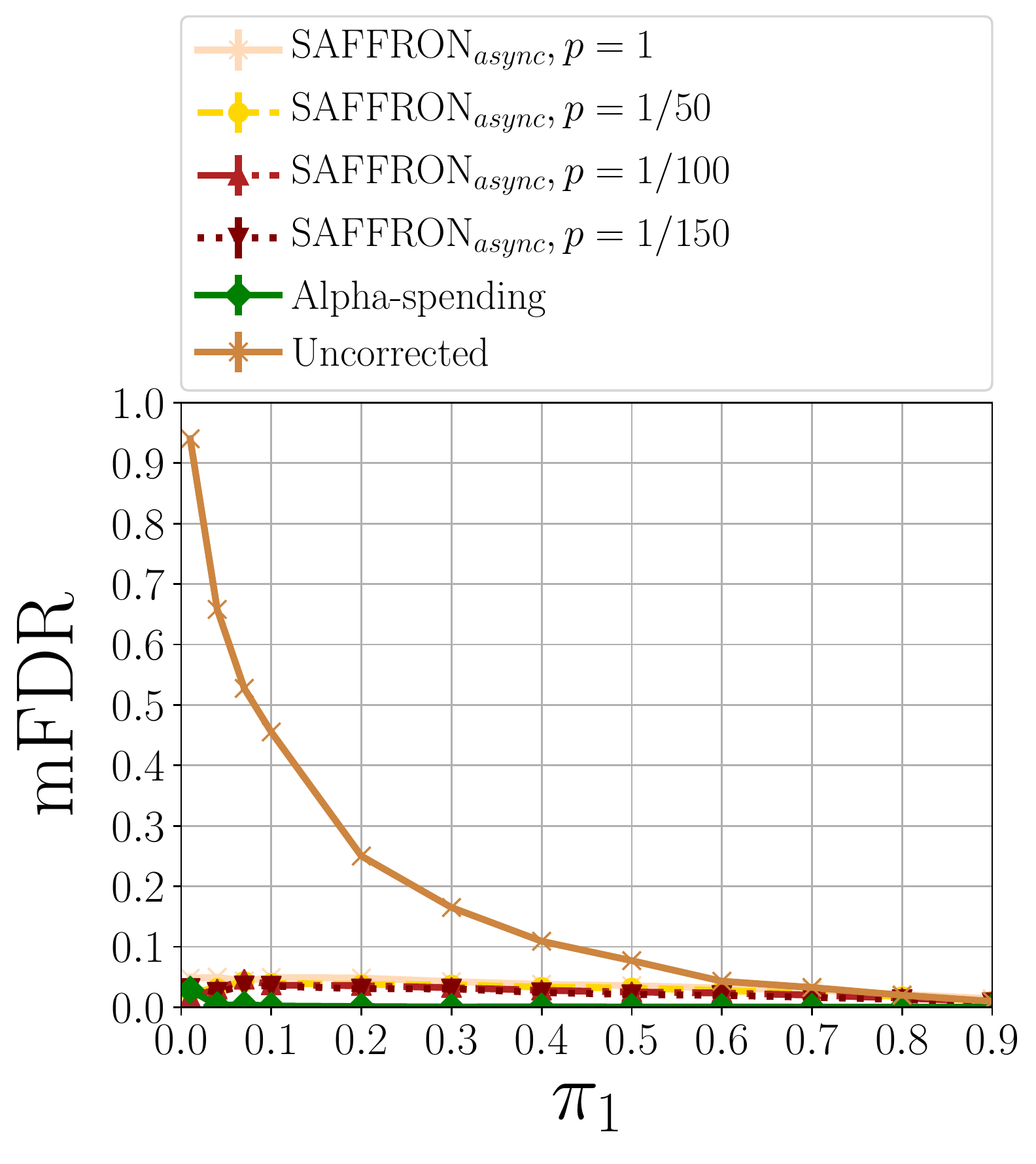}}
\caption{The left plots reproduce FDR from \figref{async} and \figref{async2}, while the right plots show mFDR for the same experiments.} 
\end{figure}

\begin{figure}[H]
\centerline{\includegraphics[width=0.24\textwidth]{lordlocfdr.pdf}
\includegraphics[width=0.24\textwidth]{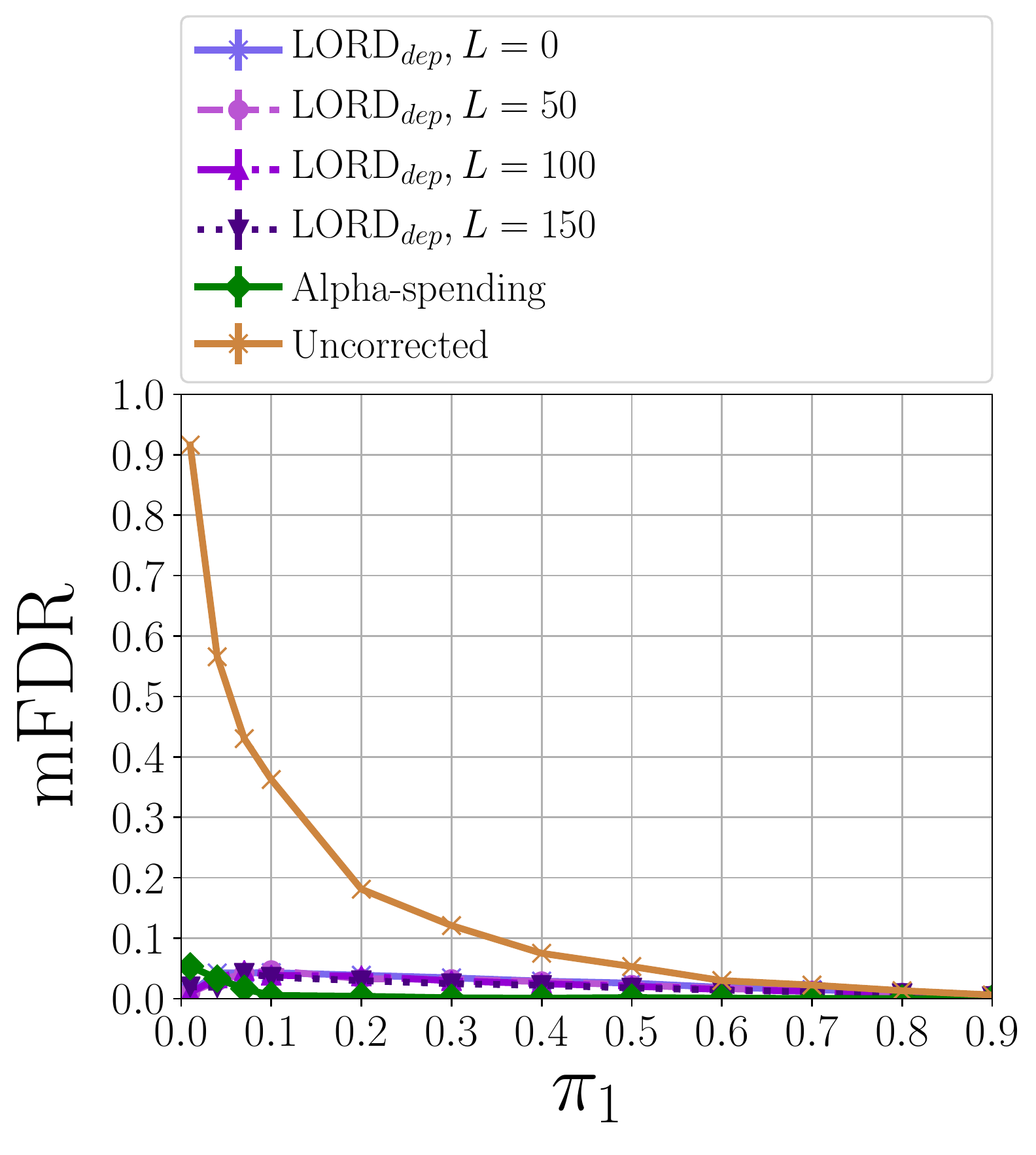}
\includegraphics[width=0.24\textwidth]{lordlocfdr2.pdf}
\includegraphics[width=0.24\textwidth]{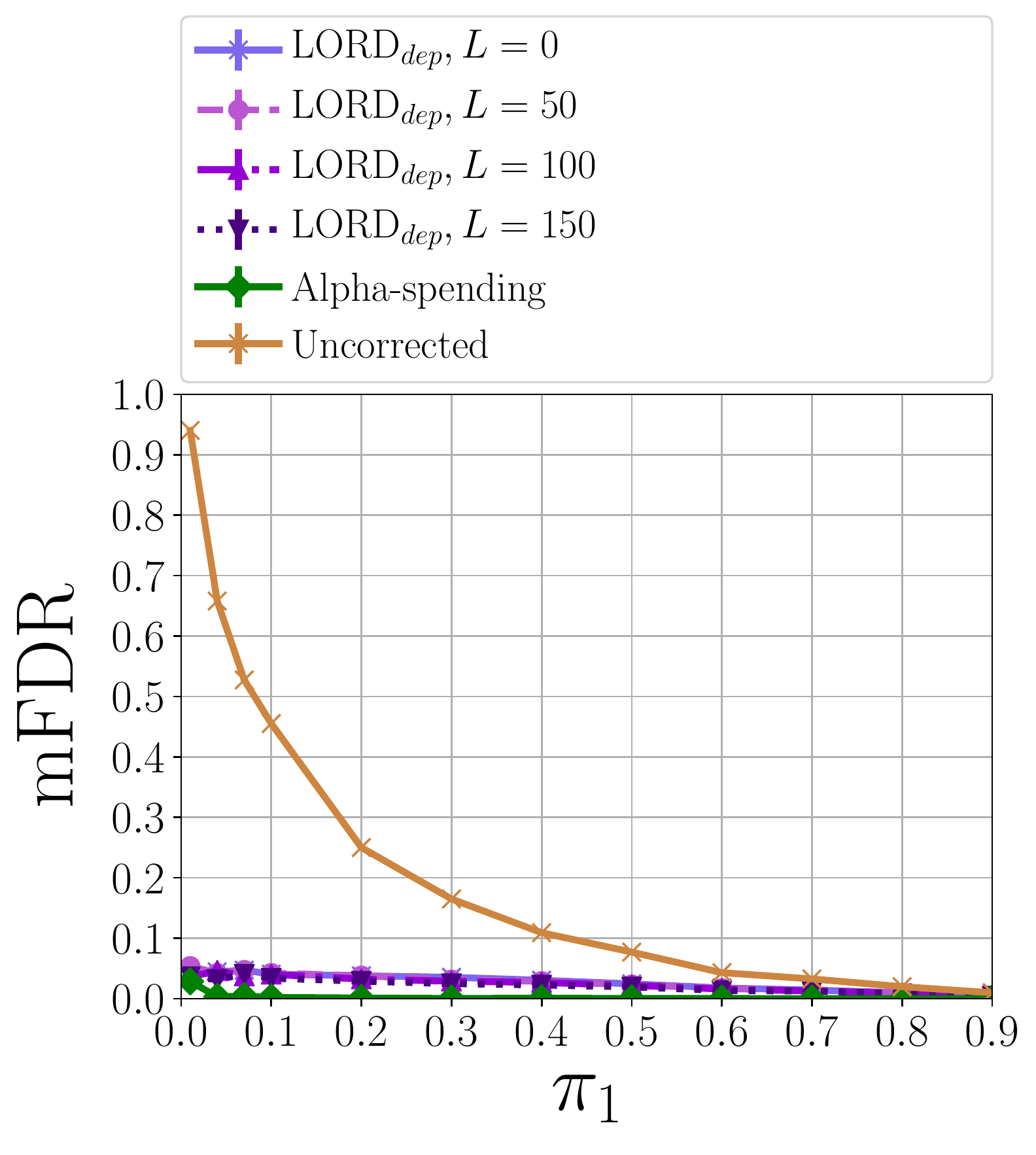}}
\centerline{\includegraphics[width=0.24\textwidth]{safflocfdr.pdf}
\includegraphics[width=0.24\textwidth]{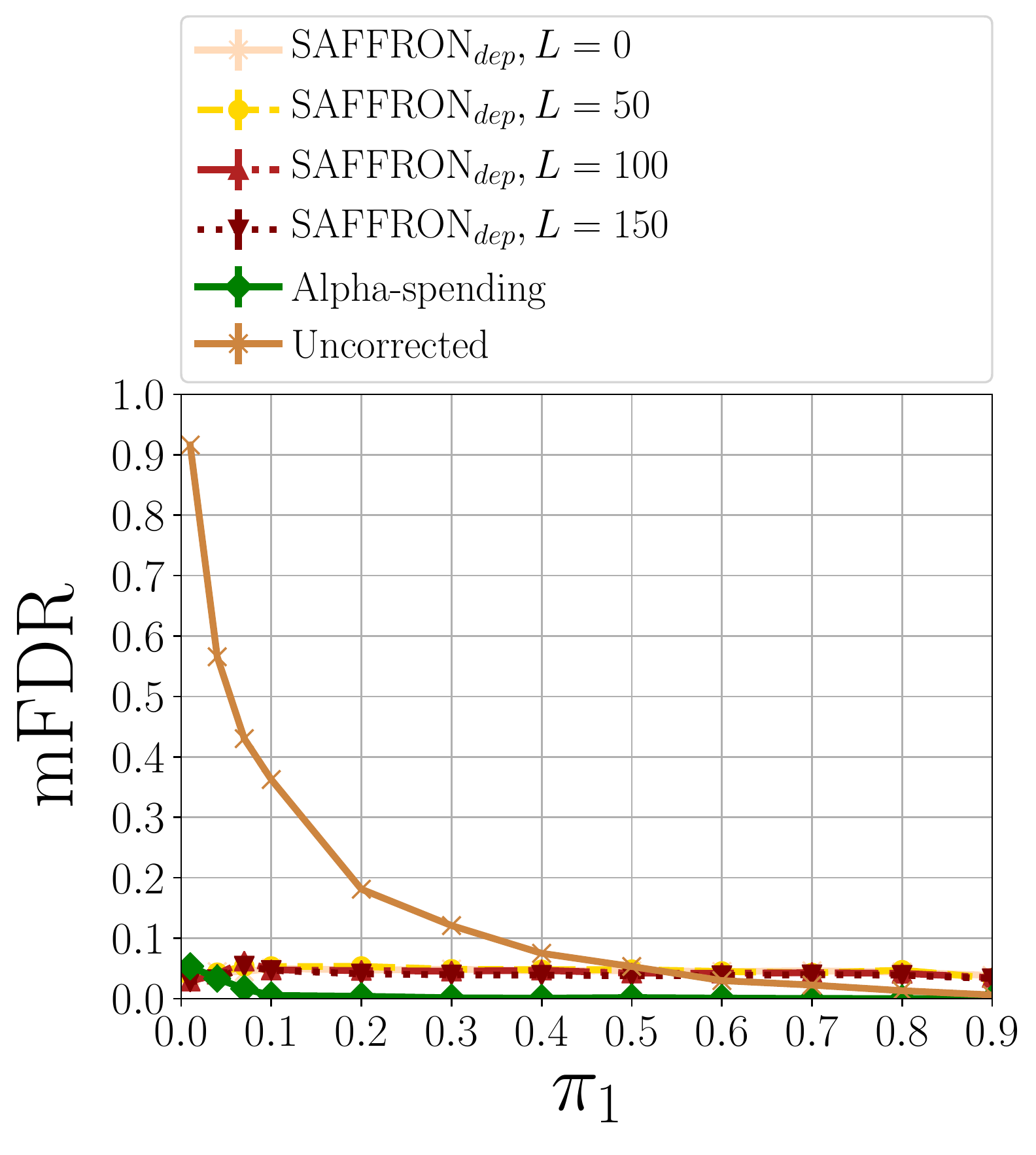}
\includegraphics[width=0.24\textwidth]{safflocfdr2.pdf}
\includegraphics[width=0.24\textwidth]{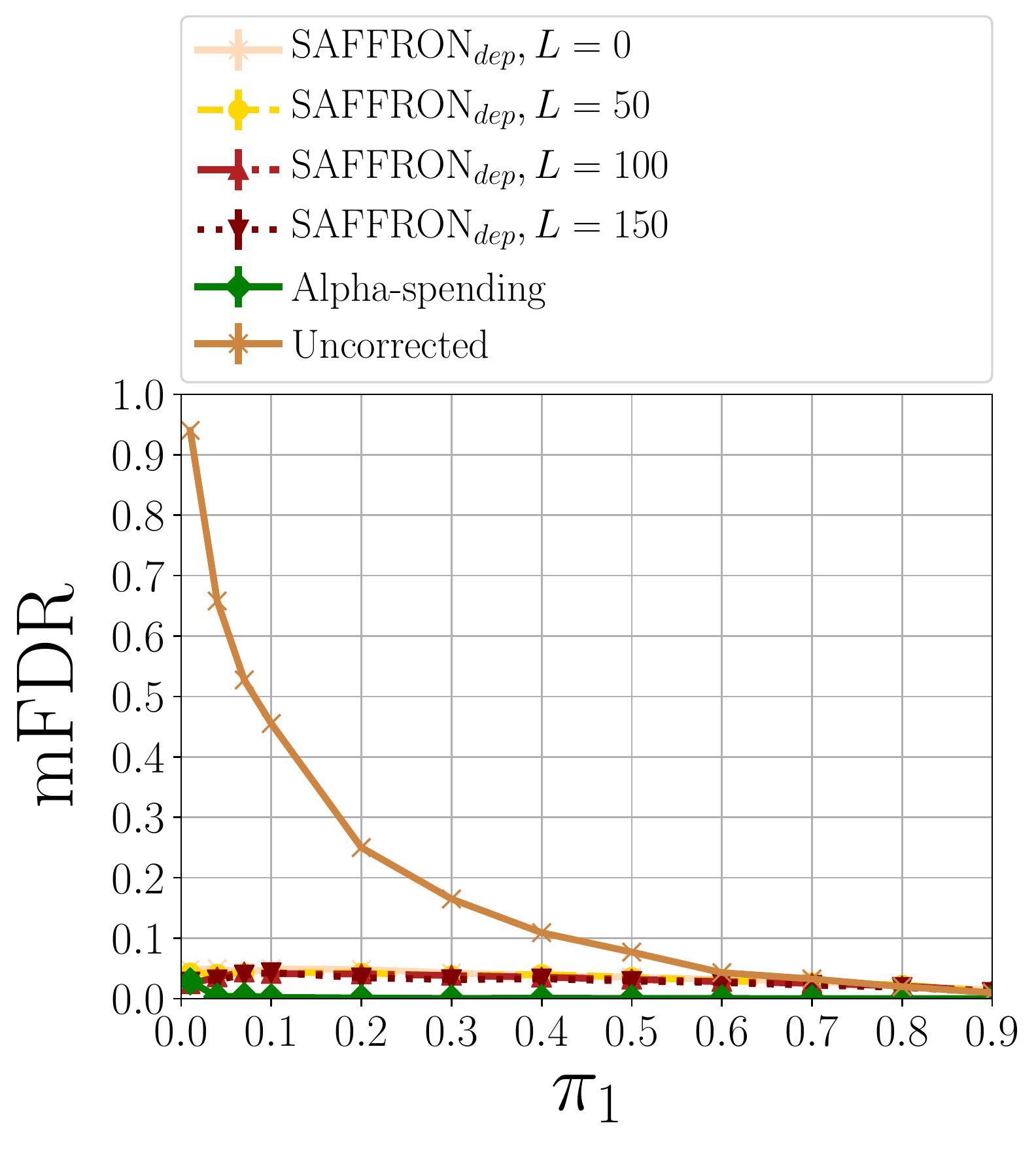}}
\caption{The left plots reproduce FDR from \figref{local} and \figref{local2}, while the right plots show mFDR for the same experiments.} 
\end{figure}

\begin{figure}[H]
\centerline{\includegraphics[width=0.24\textwidth]{lordminifdr.pdf}
\includegraphics[width=0.24\textwidth]{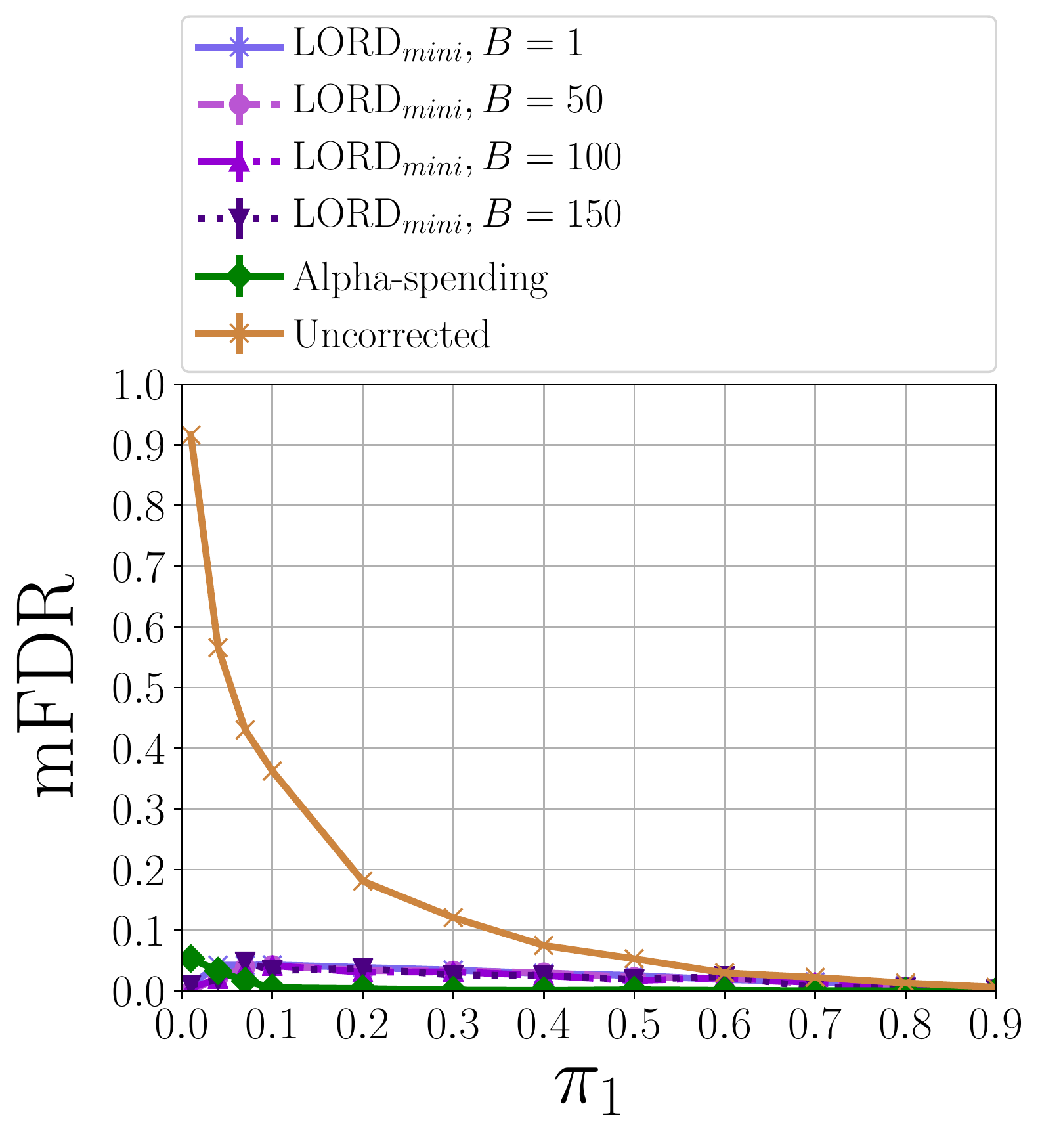}
\includegraphics[width=0.24\textwidth]{lordminifdr2.pdf}
\includegraphics[width=0.24\textwidth]{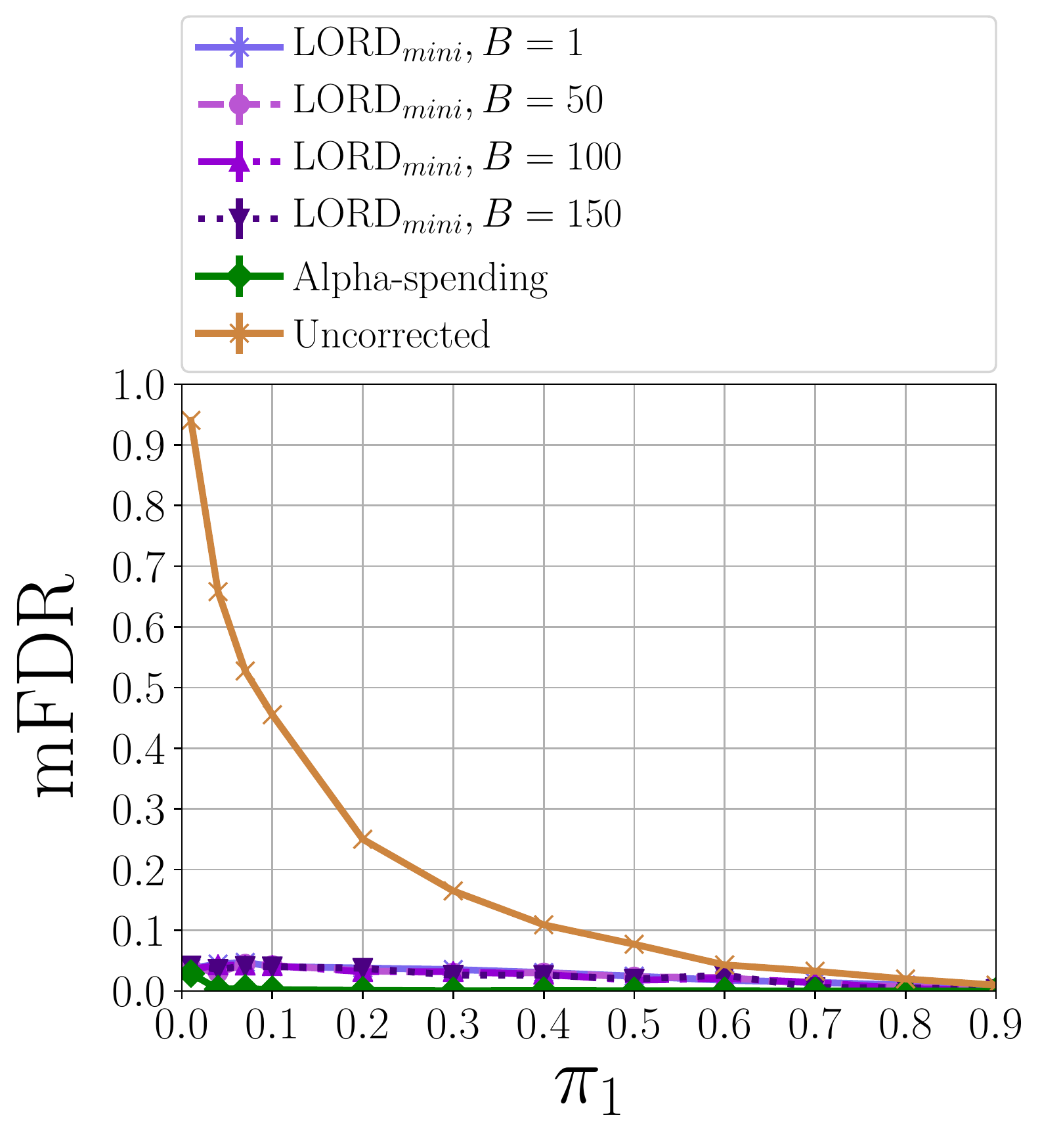}}
\centerline{\includegraphics[width=0.24\textwidth]{saffminifdr.pdf}
\includegraphics[width=0.24\textwidth]{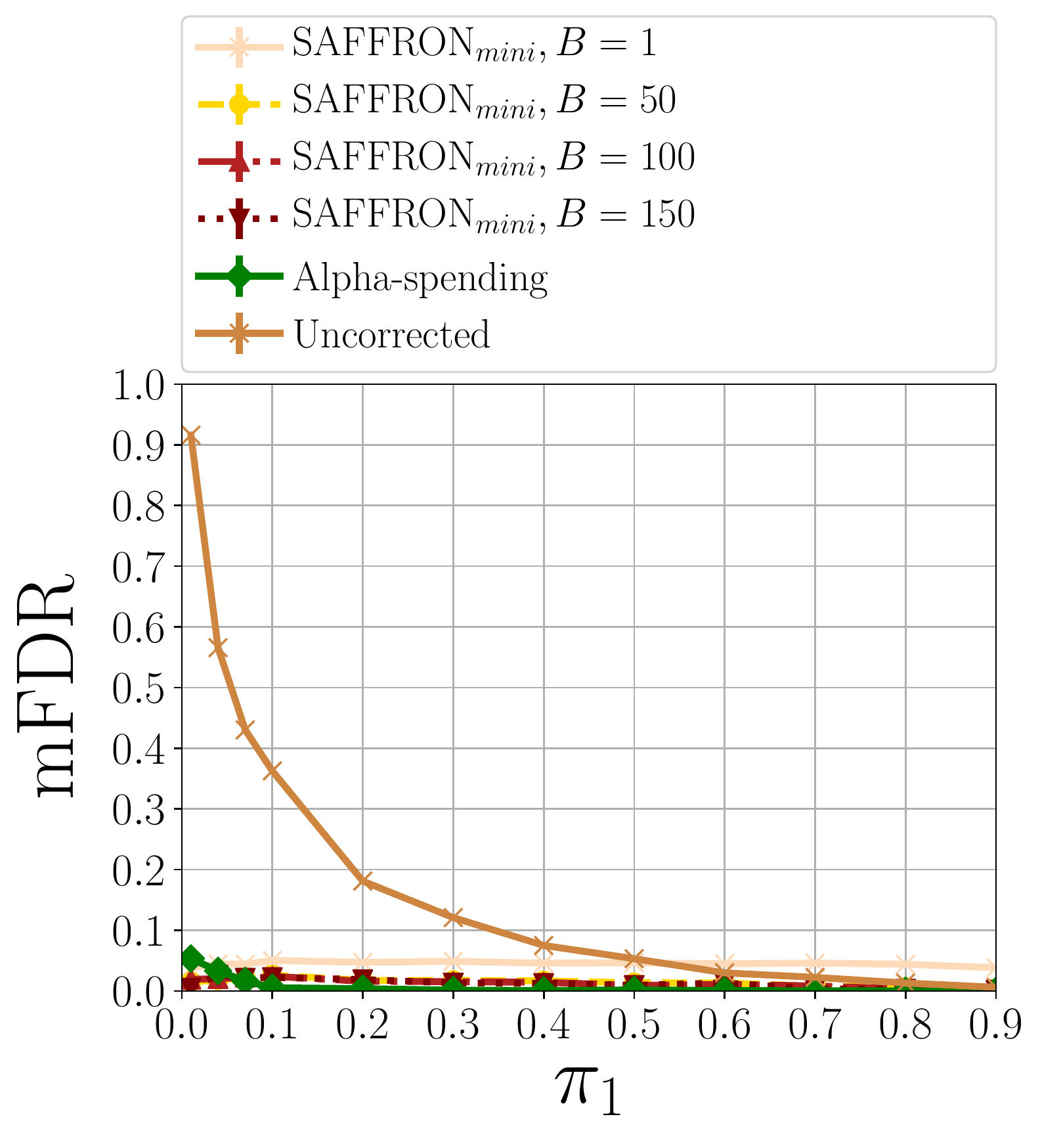}
\includegraphics[width=0.24\textwidth]{saffminifdr2.pdf}
\includegraphics[width=0.24\textwidth]{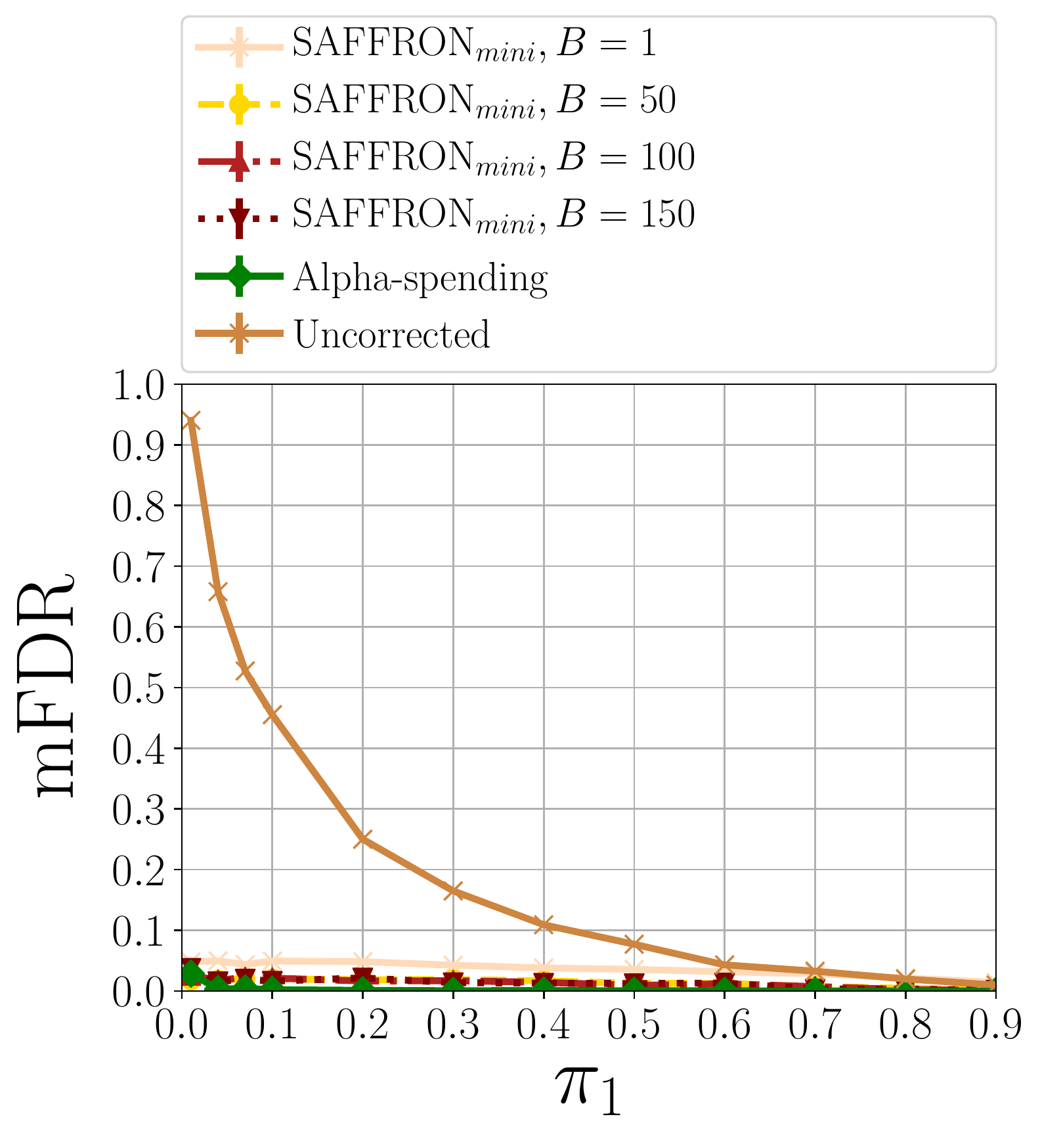}}
\caption{The left plots reproduce FDR from \figref{mini} and \figref{mini2}, while the right plots show mFDR for the same experiments.} 
\end{figure}

\begin{figure}[H]
\centerline{\includegraphics[width=0.24\textwidth]{comparisonfdr.pdf}
\includegraphics[width=0.24\textwidth]{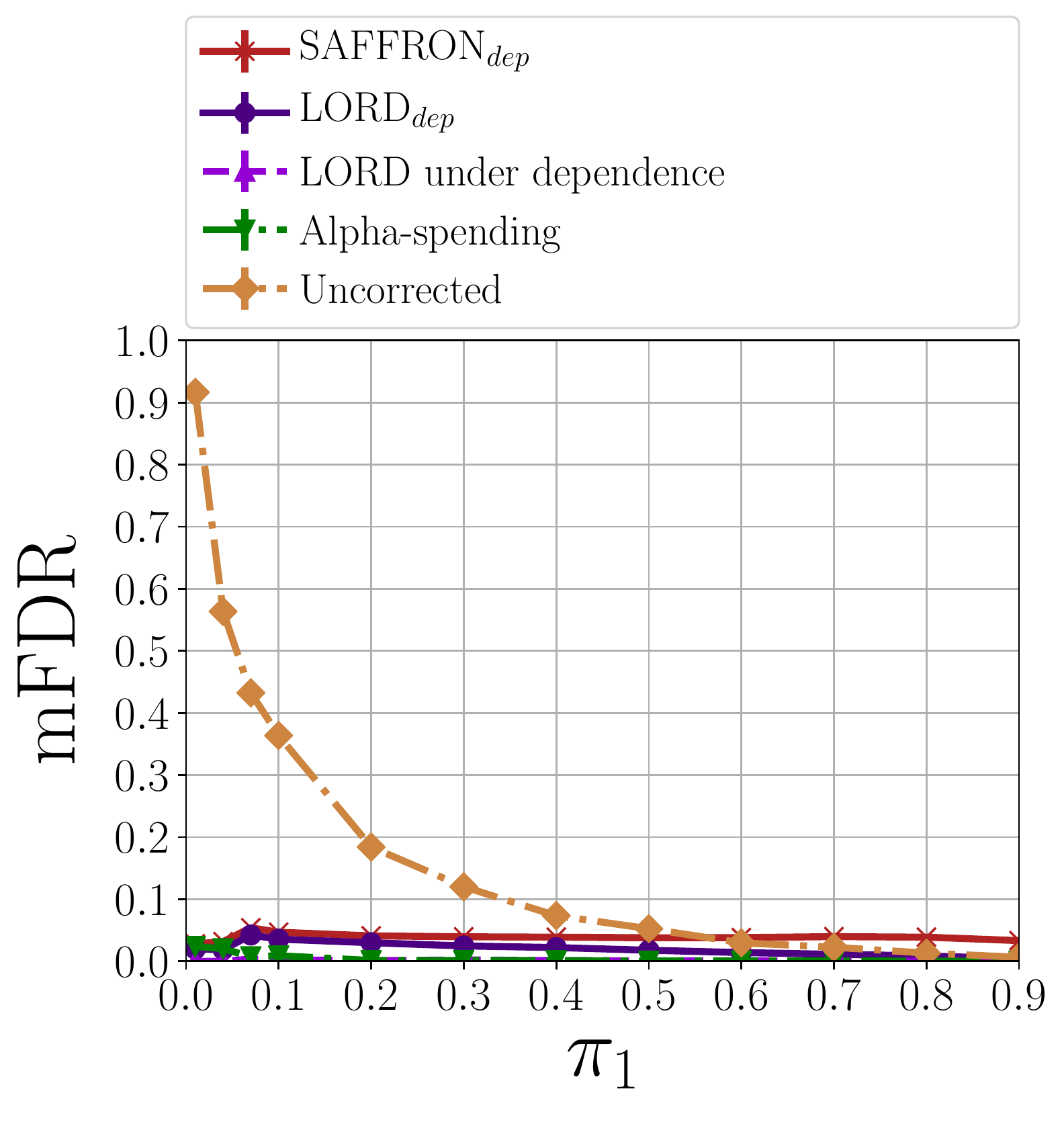}\includegraphics[width=0.24\textwidth]{comparisonfdr2.pdf}
\includegraphics[width=0.24\textwidth]{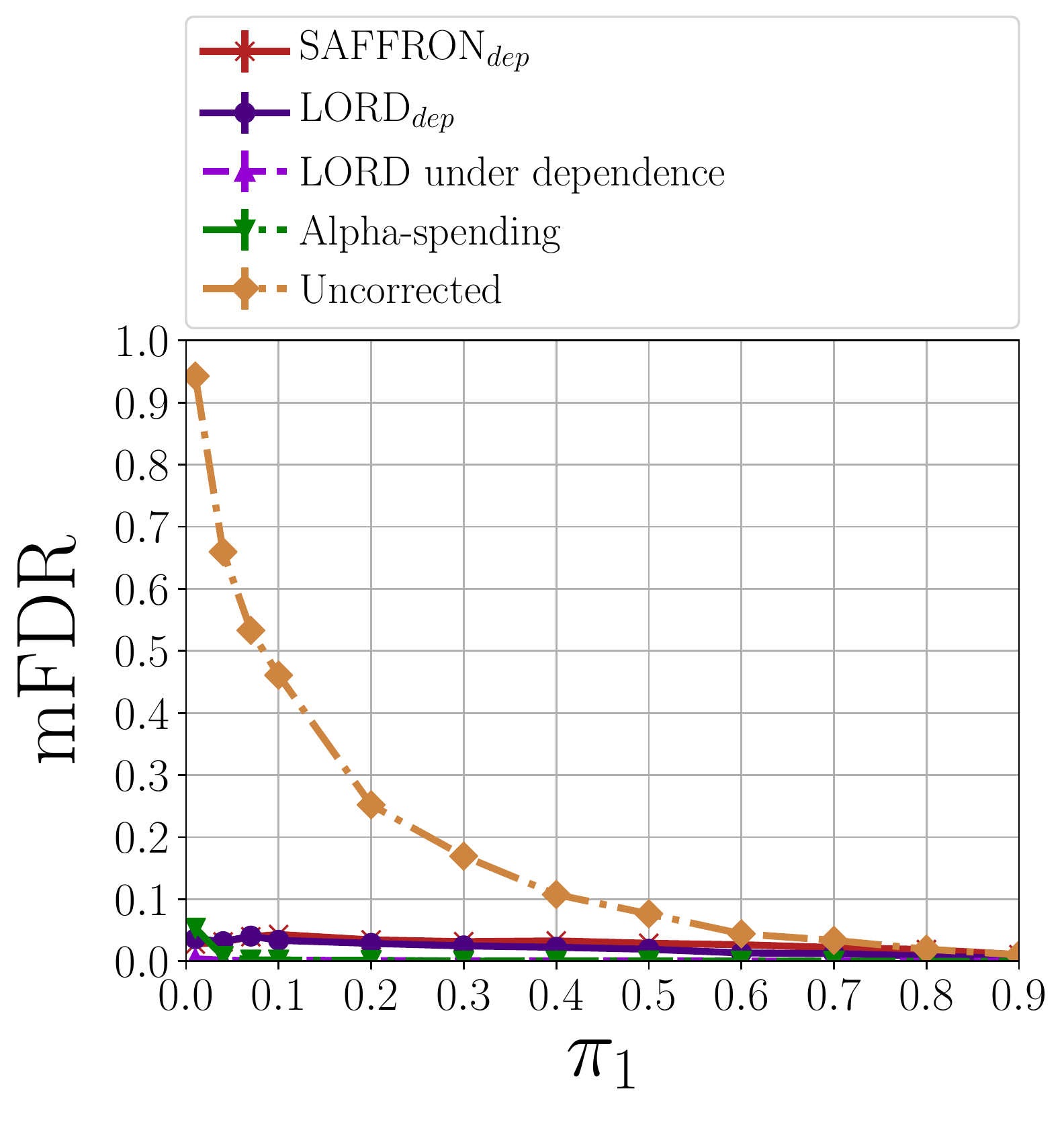}}
\caption{The left plots reproduce FDR from \figref{comparison}, while the right plots show mFDR for the same experiments.} 
\end{figure}


\end{document}